\documentclass[aps,prb,twocolumn,groupedaddress]{revtex4-1}
\usepackage[dvips]{graphicx}
\usepackage{braket}
\begin{document}

\renewcommand{\vec}[1]{\mathbf{#1}}




\title{Elastic constants and thermodynamical quantities for crystal lattices from many-body perturbation theory}


\author{Ville J. H\"{a}rk\"{o}nen}
\email[]{ville.j.harkonen@jyu.fi}
\affiliation{Department of Chemistry, University of Jyv\"{a}skyl\"{a}, PO Box 35, FI-40014, Finland}


\date{\today}

\begin{abstract}
The method of many body Green's functions is used to derive algebraic expressions for the different elastic and thermodynamical quantities such as the free energy, internal energy, entropy, heat capacity, elastic constants (adiabatic and isothermal) and the coefficient of thermal expansion. The perturbation expansion is developed up to third-order and diagrams corresponding to the equations are represented. The present results extend the existing ones by giving expressions for the elastic constants of arbitrary order and terms which are higher-order in the interatomic force constants that have been obtained earlier. The perturbation expansion in terms of arbitrary macroscopical parameters is considered and the similarity of expansions with respect to different parameters is emphasized. A physical interpretation of the harmonic phonon eigenvectors is considered. To the author’s knowledge, no such similar interpretation of the harmonic phonon eigenvectors has been given in the literature before.
\end{abstract}


\maketitle

\section{Introduction}
\label{cha:Introduction}
Some elements of the dynamical theory of crystal lattices applied in its present form have not changed much since the book of Born and Huang \cite{Born-Huang-DynamicalTheory-1954}. Until the past few decades, it was not possible to calculate the central quantities of the theory by using \textit{ab initio} methods, in general. The development of computers and computational methods has made it possible to calculate the second-order interatomic force constants (IFCs) and thus phonon eigenvalues and eigenvectors within the harmonic approximation by using, for example, the density functional perturbation theory (DFPT) for arbitrary crystal structures \cite{Baroni-RevModPhys.73.515-DFTP-2001,QE-2009}. These quantities have a rather central role in the calculation of different physical quantities and cast a foundation, for instance, for the perturbation theoretic calculation of the thermodynamical quantities of crystal lattices.

There has also been development in the calculation of anharmonic IFCs. The third-order IFCs have been calculated for some structures by using DFPT \cite{Deinzer-PhysRevB.67.144304-2003,Ward-PhysRevB.80.125203-Tcond-2009,Paulatto-PhysRevB.91.054304_lifetimes-2015} and in some cases, the fourth-order IFCs have been obtained by using self-consistent methods \cite{Paulatto-PhysRevB.91.054304_lifetimes-2015,Tadano-SCF-Phonon-PhysRevB.92.054301-2015} (no open access DFPT or self-consistent code available). However, by using the super cell method, one is able to calculate estimates for the third-order IFCs for arbitrary structures. This is established, for example, in an open access code used to calculate lattice thermal conductivity values for arbitrary crystal structures \cite{Li-ThirdOrderPy_PhysRevB.86.174307-2012,Li-Gaussian_PhysRevB.85.195436-2012,Li-shengbte-2014}.

The Helmholtz free energy of a lattice can be written in terms of the lattice Hamiltonian, from which the various thermodynamical quantities may be derived from. Such quantities are the internal energy, heat capacity, entropy, elastic constants (adiabatic and  isothermal) and the coefficient of thermal expansion (CTE). For example, the stress and elastic constants are relatively important quantities when the thermal expansion is studied. In particular, the so-called negative thermal expansion (NTE) is a subject of rather intensive study \cite{Biernacki-NTE-PhysRevLett.63.290-1989,Evans-Mary-chem.mat.9602959.NTE-1996,Mary-05041996.science.ZrW2O8.NTE-1996,Evans-1997311-NTE-1998,Ernst-NTEPDOS-1998,Mittal-NTE-PhysRevLett.86.4692-2001,Barrera.Bruno.Barron-0953-8984-17-4-R03.jour.phys.cond.mat.NTE-rew-2005,Goodwin-ColossalNTE-2008,Miller-NTE-Review-2009,Dove-NTEreview-2016}. The NTE is mostly an anharmonic phenomenon and is thus dependent on the anharmonic IFCs. Usually, the calculational study of the CTE is made by using the quasiharmonic approximation (QHA) \cite{Xu-CTE-PhysRevB.43.5024-1991,Pavone-DiamondLatDyn-PhysRevB.48.3156-1993,Wei-ThermodynamicalPropSi-PhysRevB.50.14587-1994,Karch-SiCarbLatDyn-PhysRevB.50.17054-1994,Xie-CTE-PhysRevB.60.9444-1999,Zhao-QHA-2006,Zakharchenko-CTE-PhysRevLett.102.046808-2009,Norouzzadeh-CTEpaper-2016} which takes into an account the lowest-order of anharmonicity. When the anharmonicity becomes stronger the lowest-order description may not be sufficient to describe the NTE and other phenomenon properly \cite{Zakharchenko-CTE-PhysRevLett.102.046808-2009}. As mentioned, the developments of the computational methods probably allow, sooner or later, more detailed study of the CTE (and NTE) and may give some useful information about the phenomena.

Born and Huang  represented a perturbation theoretic approach to determine the Helmholtz free energy from which one may calculate the aforementioned thermodynamical and elastic quantities \cite{Born-Huang-DynamicalTheory-1954}. Later on, there have been rather many works considering the calculation of free energy and other thermodynamical quantities \cite{Maradudin-AnharmonicThermodynamics-1961,Leibfried-SolidStatePhysics-1961,Maradudin-ThermalExpFreqShifts-1962}, one approach being used is the method of many-body Green’s functions \cite{Zubarev-GreensFunc0038-5670-3-3-R02-1960,Kadanoff-Baym-Quant.Stat.Mech-1962,Cowley-TheLatticeDynamicsOfAnAnharmonicCrystal-1963,Cowley-AnharmonicCrystals-1968,Shukla-AnharmonicHelmholtz-1971,Shukla-AnharmonicFreeEnergy-PhysRevB.3.4055-1971,Shukla-FreeEnergy-PhysRevB.10.3660-1974,Barron-DynamicalPropertiesOfSolids-1974}. For example, Cowley has derived expressions for the second-order isothermal elastic constants and thermodynamical quantities (some third-order terms were represented for isothermal elastic constants) \cite{Cowley-TheLatticeDynamicsOfAnAnharmonicCrystal-1963}. Similar work has been represented by Barron and Klein where some third-order terms for the second-order isothermal elastic constants were given in diagrams. Shukla and Cowley have calculated expressions for the free energy by using the anharmonic Hamiltonian up to fourth-order \cite{Shukla-AnharmonicFreeEnergy-PhysRevB.3.4055-1971,Shukla-FreeEnergy-PhysRevB.10.3660-1974}. In that time, the calculation of the higher-order IFCs was not possible in the case of arbitrary crystal structures, which might be one reason why most of the third-order terms for the elastic constants \textit{et cetera} were not explicitly shown. While the isothermal and adiabatic stress together with the second-order isothermal elastic constants are useful quantities in the study of CTE, the third and higher-order elastic constants can be used in the study of physical acoustics and non-linear elasticity of crystals \cite{Mason-PhysicalAcoustics-1971,Graham-SolidsUnderHighPressure-1993}. There is already some \textit{ab initio} computational studies, where the elastic constants up to fourth-order have been obtained for realistic materials \cite{Wang-ElasticConst-PhysRevB.79.224102-2009}. However, in the methods used in Refs. \cite{Lopuszynski-ElasticCOnt-PhysRevB.76.045202-2007,Wang-ElasticConst-PhysRevB.79.224102-2009,Golesorkhtabar-ElaStic-20131861-2013}, for example, only the static lattice contributions are taken into account, while the approach used in the present work includes also the so-called vibrational contributions.

It might be possible that in the near future, the calculation of third or even higher-order IFCs becomes possible in the same manner that is done at the present in the case of the second-order IFCs. This gives a reason to derive the algebraic expressions for higher-order terms in a systematic manner in so that these expressions may be used in the numerical calculations, whenever the systematic calculation of the higher-order IFCs becomes possible. The purpose of the present work is to write the explicit expressions for different thermodynamical and elastic quantities to be used in the actual calculations by using the lattice dynamical Hamiltonian and Green's function method. If the anharmonic forces are sufficiently strong, the perturbation theory with a few lowest-order terms may not describe the system properly and one probably have to use some more rigorous methods, such as self-consistent theory for phonons described, for instance, in Refs. \cite{Koehler-SFC-HarmonicAppr-PhysRevLett.17.89-1966,Gotze-TransportTheorQuantumCryst-1969,Werthamer-SFCPhonon-PhysRevB.1.572-1970,Gotze-DynamicalPropertiesOfSolids-1974}. It is also mentioned that recently, a treatment considering corrections to the harmonic thermodynamical quantities has been established in Ref. \cite{Allen-QuasiparticleTheory-PhysRevB.92.064106-2015}.

This paper is organized as follows. In Sec. \ref{LatticeDynamics}, the lattice Hamiltonian for strained and unstrained crystal is given. Also, the Hamiltonian for a wider class of parameters is shown. Section \ref{cha:ThermodynamicalRelations} considers the thermodynamical properties and definitions of the thermal expansion coefficient, elastic constants (isothermal and adiabatic) \textit{et cetera} are given. The perturbation expansion used to derive the results of this work is introduced in Sec. \ref{cha:PerturbationExpansionForThermodynamicalQuantities}. The results from first to third-order in perturbation are given in Secs. \ref{cha:FirstOrderResults}-\ref{cha:ThirdOrderResults} and are compared with the results obtained earlier by other researchers.

\section{Lattice dynamics}
\label{LatticeDynamics}

\subsection{Hamiltonian}
\label{Hamiltonian}
The theory of lattice dynamics has been described, for example, in Refs. \cite{Born-Huang-DynamicalTheory-1954,Maradudin-harm-appr-1971,Maradudin-DynamicalPropertiesOfSolids-1974,Srivastava-PhysicsOfPhonons-1990}. In the present approach it is assumed that the vibrational Hamiltonian is the sum of kinetic and potential energies $H = T_{n} + \Phi$, where
\begin{equation} 
T_{n} =\frac{1}{2} \sum_{l,\kappa,\alpha} \frac{p^{2}_{\alpha}\left(l \kappa\right)}{M_{\kappa}}, 
\label{eq:HarkThermPerturbKineticEnergy}
\end{equation}
and the potential energy is expressed as a Taylor series
\begin{eqnarray} 
\Phi =&& \sum_{n=0} \frac{1}{n!} \sum_{l_{1},\kappa_{1},\alpha_{1}} \cdots \sum_{l_{n},\kappa_{n},\alpha_{n}} \Phi_{\alpha_{1} \cdots \alpha_{n}}\left(l_{1} \kappa_{1};\ldots;l_{n} \kappa_{n} \right) \nonumber \\
			&&\times u_{\alpha_{1}}\left(l_{1} \kappa_{1} \right) \ldots u_{\alpha_{n}}\left(l_{n} \kappa_{n} \right), 
\label{eq:HarkThermPerturbPotentialEnergy}
\end{eqnarray}
where the interatomic force constants (IFC) are defined as 
\begin{eqnarray} 
\Phi&&_{\alpha_{1} \cdots \alpha_{n}}\left(l_{1} \kappa_{1};\ldots;l_{n} \kappa_{n} \right) \nonumber \\
&&\equiv \left.\frac{\partial^{n}{\Phi}}{\partial{x'_{\alpha_{1}}\left(l_{1} \kappa_{1} \right)} \cdots \partial{x'_{\alpha_{n}}\left(l_{n} \kappa_{n} \right)}  } \right|_{\left\{x'\left(l_{i}\kappa_{i}\right) = x\left(l_{i}\kappa_{i}\right)\right\}}.
\label{eq:HarkThermPerturbAtomicForceConstants}
\end{eqnarray}
In Eqs. \ref{eq:HarkThermPerturbKineticEnergy}-\ref{eq:HarkThermPerturbAtomicForceConstants}, $u_{\alpha_{i}}\left(l_{i} \kappa_{i} \right)$ is the displacement of the atom $\kappa_{i}$ in the unit cell labeled by $l_{i}$ from the equilibrium position $\vec{x}\left(l_{i}\kappa_{i}\right)$ in the direction $\alpha_{i}$, $p_{\alpha_{i}}\left(l_{i} \kappa_{i} \right)$ the corresponding momentum and $M_{\kappa_{i}}$ is the atomic mass of atom $\kappa_{i}$. The equilibrium positions can be written as $\vec{x}\left(l\kappa\right) = \vec{x}\left(l\right) + \vec{x}\left(\kappa\right)$, where $\vec{x}\left(\kappa\right)$ is the equilibrium position vector of atom $\kappa$ within each unit cell and the lattice translational vector can be written as $\vec{x}\left(l\right) \equiv \vec{x}\left(l_{1},l_{2},l_{3}\right)= l_{1}\vec{a}_{1}+l_{2}\vec{a}_{2}+l_{3}\vec{a}_{3}$. Here, $l_{i}$ are integers and the vectors $\vec{a}_{i}$ are the so-called primitive translational vectors of the lattice.

One may use the normal coordinate transformation to diagonalize the harmonic Hamiltonian, this can be done by using the following expansions (quantization is made by introducing the canonical commutation relations)
\begin{eqnarray} 
\hat{u}_{\alpha}\left(l \kappa\right) =&& \left(\frac{\hbar}{2 N M_{\kappa}}\right)^{1/2} \sum_{\vec{q},j} \omega^{-1/2}_{j}\left(\vec{q} \right) e^{i \vec{q}\cdot \vec{x}\left(l\right)} e_{\alpha}\left(\kappa| \vec{q} j\right) \nonumber \\
                                      &&\times \left( \hat{a}_{\vec{q}j} + \hat{a}^{\dagger}_{-\vec{q}j} \right),
\label{eq:HarkThermPerturbDisplacementExpansion}
\end{eqnarray}
\begin{eqnarray} 
\hat{p}_{\alpha}\left(l \kappa\right) = && -i \left(\frac{\hbar M_{\kappa}}{2 N }\right)^{1/2} \sum_{\vec{q},j} \omega^{1/2}_{j}\left(\vec{q} \right) e^{i \vec{q}\cdot \vec{x}\left(l\right)} e_{\alpha}\left(\kappa| \vec{q} j\right) \nonumber \\
                                      &&\times \left( \hat{a}_{\vec{q}j} - \hat{a}^{\dagger}_{-\vec{q}j} \right),
\label{eq:HarkThermPerturbMomentumExpansion}
\end{eqnarray}
where $N$ is the number of permitted $\vec{q}$ values, $\left\{\omega_{j}\left(\vec{q} \right)\right\}$ is the set of eigenvalues of the dynamical matrix and $\hat{a}^{\dagger}_{\vec{q}j}$ and $\hat{a}_{\vec{q}j}$ are the creation and annihilation operators, resprectively. The components of the eigenvector $\vec{e}\left(\kappa|\vec{q}j\right)$ can be chosen to satisfy the orthonormality and closure conditions
\begin{equation}
\sum_{\kappa,\alpha}e_{\alpha}\left(\kappa|\vec{q}j'\right)e^{*}_{\alpha}\left(\kappa|\vec{q}j\right)=\delta_{jj'},
\label{eq:HarkThermPerturbOrthonormality}
\end{equation}
\begin{equation}
\sum_{j}e_{\alpha}\left(\kappa|\vec{q}j\right)e^{*}_{\beta}\left(\kappa'|\vec{q}j\right)=\delta_{\alpha\beta}\delta_{\kappa\kappa'},
\label{eq:HarkThermPerturbClosure}
\end{equation}
where $\delta_{\alpha\beta}$ is the Kronecker delta. From now on, the following notation is used interchangeably
\begin{equation}
\vec{q}j \leftrightarrow \lambda, \quad -\vec{q}j \leftrightarrow -\lambda, \quad \vec{q}_{i}j_{i} \leftrightarrow\lambda_{i}, \quad  \vec{q}'j' \leftrightarrow \lambda'.
\label{eq:HarkThermPerturbAbbreviation}
\end{equation}
After the substitution of Eqs. \ref{eq:HarkThermPerturbDisplacementExpansion} and \ref{eq:HarkThermPerturbMomentumExpansion} to Eqs. \ref{eq:HarkThermPerturbKineticEnergy} and \ref{eq:HarkThermPerturbPotentialEnergy}, one may write
\begin{equation} 
\hat{H} = \hat{H}_{0} + \hat{H}_{a},
\label{eq:HarkThermPerturbLatticeHamiltonian_2}
\end{equation}
where
\begin{eqnarray} 
\hat{H}_{0} &=& \sum_{\lambda}\hbar \omega_{\lambda} \left(\frac{1}{2}+ \hat{a}^{\dagger}_{\lambda} \hat{a}_{\lambda} \right) \nonumber \\
            &=& \frac{1}{4} \sum_{\lambda}\hbar \omega_{\lambda} \left(\hat{A}^{\dagger}_{\lambda} \hat{A}_{\lambda} + \hat{B}^{\dagger}_{\lambda} \hat{B}_{\lambda} \right),
\label{eq:HarkThermPerturbHarmonicCrystalHamiltonianSecondQuant}
\end{eqnarray}
\begin{eqnarray} 
\hat{H}_{a} =&& \sum_{\lambda} V\left(\lambda\right) \hat{A}_{\lambda} +\sum_{n=3} \sum_{\lambda_{1}} \cdots \sum_{\lambda_{n}} \nonumber \\
             &&\times V\left(\lambda_{1};\ldots;\lambda_{n}\right) \hat{A}_{\lambda_{1}}	\cdots \hat{A}_{\lambda_{n}},
\label{eq:HarkThermPerturbLatticeHamiltonianAnharmonicSecondQuantEq_1} 
\end{eqnarray}
and 
\begin{equation} 
\hat{A}_{\lambda} =   \hat{a}_{\lambda} + \hat{a}^{\dagger}_{-\lambda}, \quad \hat{B}_{\lambda} =   \hat{a}_{\lambda} - \hat{a}^{\dagger}_{-\lambda}.
\label{eq:HarkThermPerturbCondensedNotatCreAnnihil}
\end{equation}
Furthermore, it can be shown that \cite{Born-Huang-DynamicalTheory-1954}
\begin{eqnarray} 
V&& \left(\lambda_{1};\ldots;\lambda_{n}\right) \nonumber \\
&&= \frac{1}{n!} \left(\frac{\hbar}{2 N}\right)^{n/2} \frac{N \Delta\left( \vec{q}_{1} +\cdots+\vec{q}_{n} \right) }{ \left[\omega_{\lambda_{1}} \cdots \omega_{\lambda_{n}} \right]^{1/2} } \nonumber \\
&&\times \sum_{\kappa_{1},\alpha_{1}} \sum_{l_{2},\kappa_{2},\alpha_{2}} \cdots \sum_{l_{n},\kappa_{n},\alpha_{n}} \Phi_{\alpha_{1} \cdots \alpha_{n}}\left(0 \kappa_{1};l_{2} \kappa_{2};\ldots;l'_{n} \kappa'_{n} \right) \nonumber \\
&&\times \frac{e_{\alpha_{1}}\left(\kappa_{1}|\lambda_{1}\right)}{M^{1/2}_{\kappa_{1}}} \cdots \frac{e_{\alpha_{n}}\left(\kappa_{n}|\lambda_{n}\right)}{M^{1/2}_{\kappa_{n}}} e^{i \left[\vec{q}_{2}\cdot \vec{x}\left(l_{2}\right)+\cdots+\vec{q}_{n}\cdot \vec{x}\left(l_{n}\right) \right]}, \nonumber \\
\label{eq:HarkThermPerturbLatticeHamiltonianAnharmonicSecondQuantEq_2} 
\end{eqnarray}
where
\begin{equation} 
\Delta\left( \vec{q} \right) =  \frac{1}{N} \sum^{N}_{l} e^{i \vec{q} \cdot \vec{x}\left(l\right)}.
\label{eq:HarkThermPerturbWaveVectorDelta}
\end{equation}
The commutation rules for the operators $\hat{a}_{\lambda},\hat{A}_{\lambda},\hat{B}_{\lambda}$ \textit{et cetera} follow from the commutation relations of displacement and momenta.  The eigenkets of the harmonic Hamiltonian $\hat{H}_{0}$ can be written as
\begin{equation} 
\prod_{\lambda} \left(n_{\lambda}!\right)^{-1/2} \left(\hat{a}^{\dagger}_{\lambda}\right)^{n_{\lambda}}  \ket{0}= \ket{n_{\lambda_{1}}\cdots n_{\lambda_{N,3n}}}\equiv \ket{n}.
\label{eq:HarkThermPerturbCreationAnnihilationGeneralKetNormalized}
\end{equation} 
As already mentioned in Sec. \ref{cha:Introduction}, at the present, the second-order IFCs $\left\{ \Phi_{\alpha\beta}\left(l\kappa;l'\kappa'\right) \right\}$, the eigenvalues $\left\{ \omega_{j}\left(\vec{q}\right) \right\}$ and eigenvectors $\left\{ \vec{e}\left(\kappa|\vec{q}j\right) \right\}$ can be calculated by using the DFPT \cite{Baroni-RevModPhys.73.515-DFTP-2001,QE-2009}.

For a physical interpretation of the harmonic phonon eigenvectors $\left\{\vec{e}\left(\kappa|\vec{q}j\right)\right\}$, consider the harmonic potential energy written as
\begin{equation} 
\hat{\Phi}_{2} = \frac{1}{2}\sum_{l,\kappa,\alpha} \sum_{ l',\kappa',\beta} D_{\alpha \beta}\left(l \kappa;l' \kappa' \right) \hat{w}_{\alpha}\left(l \kappa \right) \hat{w}_{\beta}\left(l' \kappa' \right),
\label{eq:HarkThermPerturbLatticeDynamicsEq_15}
\end{equation}
where $D_{\alpha \beta}\left(l \kappa;l' \kappa' \right) = \Phi_{\alpha \beta}\left(l \kappa;l' \kappa' \right)/  \sqrt{M_{\kappa} M_{\kappa'}}$ and $\hat{w}_{\alpha}\left(l \kappa \right) = \sqrt{M_{\kappa}} \hat{u}_{\alpha}\left(l \kappa \right)$. In obtaining the diagonal form for the potential energy with the expansion for the displacement given by Eq. \ref{eq:HarkThermPerturbDisplacementExpansion}, one can divide the steps need to be made as follows \cite{Born-Huang-DynamicalTheory-1954}
\begin{eqnarray} 
\hat{w}_{\alpha}\left(l \kappa \right) &=& \frac{1}{N } \sum^{N}_{\vec{q}} \hat{w}_{\alpha}\left(\kappa |\vec{q}\right) e^{i \vec{q} \cdot \vec{x}\left(l\right)}, \nonumber \\
\hat{w}_{\alpha}\left(\kappa |\vec{q}\right) &=& \sum_{j} e_{\alpha}\left(\kappa|\lambda\right) \hat{Q}_{\lambda},
\label{eq:HarkThermPerturbLatticeDynamicsEq_16}
\end{eqnarray}
and thus
\begin{equation} 
\hat{w}_{\alpha}\left(l \kappa \right) = \frac{1}{N } \sum_{\lambda} e_{\alpha}\left(\kappa|\lambda\right) e^{i \vec{q} \cdot \vec{x}\left(l\right)} \hat{Q}_{\lambda}.
\label{eq:HarkThermPerturbLatticeDynamicsEq_18}
\end{equation}
After using Eq. \ref{eq:HarkThermPerturbLatticeDynamicsEq_18} in Eq. \ref{eq:HarkThermPerturbLatticeDynamicsEq_15}
\begin{eqnarray} 
\hat{\Phi}_{2}= && \frac{1}{2 N^{2} }\sum_{l,\kappa,\alpha} \sum_{l',\kappa',\beta} \sum_{\lambda} \sum_{\lambda'} e_{\alpha}\left(\kappa|\lambda\right) e^{i \vec{q} \cdot \vec{x}\left(l\right)} \nonumber \\
&&\times  D_{\alpha \beta}\left(l \kappa;l' \kappa' \right) e_{\beta}\left(\kappa'|\lambda'\right) e^{i \vec{q}' \cdot \vec{x}\left(l'\right)} \hat{Q}_{\lambda} \hat{Q}_{\lambda'}.
\label{eq:HarkThermPerturbLatticeDynamicsEq_19_2}
\end{eqnarray}
Now, by using the bra-ket notation
\begin{equation} 
\hat{\Phi}_{2} =  \frac{1}{2} \sum_{l,\kappa,\alpha} \sum_{ l',\kappa',\beta} \Phi_{\alpha \beta,2}\left(l \kappa;l' \kappa' \right)  \ket{l,\kappa,\alpha} \bra{\beta,\kappa',l'}, 
\label{eq:HarkThermPerturbLatticeDynamicsEq_20}
\end{equation}
where
\begin{equation} 
\Phi_{\alpha \beta,2}\left(l \kappa;l' \kappa' \right) \equiv \braket{\alpha,\kappa,l|\hat{\Phi}_{2}|l',\kappa',\beta}.
\label{eq:HarkThermPerturbLatticeDynamicsEq_21}
\end{equation}
and it is assumed that $\left\{\ket{l,\kappa,\alpha}\right\}$ form a complete set (basis). A comparison of Eqs. \ref{eq:HarkThermPerturbLatticeDynamicsEq_20}, \ref{eq:HarkThermPerturbLatticeDynamicsEq_21} and \ref{eq:HarkThermPerturbLatticeDynamicsEq_15} indicates that one may identify $\Phi_{\alpha \beta,2}\left(l \kappa;l' \kappa' \right) = D_{\alpha \beta}\left(l \kappa;l' \kappa' \right)$ and $\ket{l,\kappa,\alpha} \bra{\beta,\kappa',l'} = \hat{w}_{\alpha}\left(l \kappa \right) \hat{w}_{\beta}\left(l' \kappa' \right)$. Furthermore, one may write Eq. \ref{eq:HarkThermPerturbLatticeDynamicsEq_20} as
\begin{eqnarray} 
\hat{\Phi}_{2} =&& \frac{1}{2 N^{2} } \sum_{l,\kappa,\alpha} \sum_{ l',\kappa',\beta} \sum_{\vec{q},j} \sum_{\vec{q}',j'} \braket{j,\vec{q}|l,\kappa,\alpha} D_{\alpha \beta}\left(l \kappa;l' \kappa' \right)  \nonumber \\
&&\times \braket{\beta,\kappa',l'|\vec{q}',j'} \ket{\vec{q},j} \bra{j',\vec{q}'}.
\label{eq:HarkThermPerturbLatticeDynamicsEq_24}
\end{eqnarray}
A comparison of Eqs. \ref{eq:HarkThermPerturbLatticeDynamicsEq_19_2} and \ref{eq:HarkThermPerturbLatticeDynamicsEq_24} indicates that one may identify
\begin{eqnarray} 
\braket{\vec{q},j|l,\kappa,\alpha} &=& e_{\alpha}\left(\kappa|\vec{q}j\right) e^{i \vec{q} \cdot \vec{x}\left(l\right)}, \nonumber \\
 \ket{\vec{q},j} \bra{j',\vec{q}'} &=& \hat{Q}\left(\vec{q}j\right) \hat{Q}\left(\vec{q}'j'\right).
\label{eq:HarkThermPerturbLatticeDynamicsEq_25}
\end{eqnarray}
From the interpretation of quantum mechanics \cite{Dirac-PrinciplesOfQM-1958}, the quantities in Eq. \ref{eq:HarkThermPerturbLatticeDynamicsEq_25} can be considered to be the probability amplitudes and
\begin{equation} 
\left|\braket{j,\vec{q}|l,\kappa,\alpha}\right|^{2} = \left|\braket{j,\vec{q}|\kappa,\alpha}\right|^{2} = \left|e_{\alpha}\left(\kappa|\vec{q}j\right)\right|^{2},
\label{eq:HarkThermPerturbLatticeDynamicsEq_27}
\end{equation}
as the probability of $\left\{j'\right\}$ having the value $j$ (for each $\vec{q}$) when $\left\{\kappa',\alpha'\right\}$ certainly have the values $\kappa,\alpha$ or the probability of $\left\{\kappa',\alpha'\right\}$ having the values $\kappa,\alpha$ when $\left\{j'\right\}$ certainly have the value $j$ (for each $\vec{q}$). The probability is not affected by the phase factor $e^{i \vec{q} \cdot \vec{x}\left(l\right)}$ and thus, it is not affected by the cell index $l$ (only phase difference matters). By using the present notation, the conditions for eigenvectors given by Eqs. \ref{eq:HarkThermPerturbOrthonormality} and \ref{eq:HarkThermPerturbClosure} show that these probabilities are normalized (cell index $l$ neglected)
\begin{equation}
\sum_{\kappa} \left|\braket{j,\vec{q}|\kappa}\right|^{2} = \sum_{j} \left|\braket{j,\vec{q}|\kappa,\alpha}\right|^{2} = 1,
\label{eq:HarkThermPerturbLatticeDynamicsEq_28}
\end{equation}
where $\left|\braket{j,\vec{q}|\kappa}\right|^{2} = \left| \vec{e}\left(\kappa|\vec{q}j\right) \right|^{2}$. With the preceding discussion in mind, one may interpret $\left| \vec{e}\left(\kappa|\vec{q}j\right) \right|^{2}$ as the probability that the atom $\kappa$ vibrates in the phonon mode $\vec{q}j$. An extreme example is such that $\left| \vec{e}\left(\kappa|\vec{q}j\right) \right|^{2} = 1$ for some $\kappa$ and thus by Eq. \ref{eq:HarkThermPerturbLatticeDynamicsEq_28} it is the only atom within a particular unit cell which vibrates in the phonon mode $\vec{q}j$.

\subsection{Hamiltonian for deformed lattice}
\label{cha:HamiltonianForDeformedLattice}
In order to calculate the elastic properties of crystal, for example, it can be convenient to expand the IFCs to a Taylor series in homogenous infinitesimal strain parameters $\left\{u_{\mu\nu}\right\}$ (Ref. \cite{Born-Huang-DynamicalTheory-1954}, p. 306 or Ref. \cite{Barron-DynamicalPropertiesOfSolids-1974}). After the expansion, one may write the Hamiltonian for the deformed lattice as
\begin{equation} 
\hat{H}_{d} = \hat{H}_{0} + \hat{H}_{a} + \hat{H}_{s} = \hat{H}_{0} + \hat{H}_{int},
\label{eq:HarkThermPerturbCrystalHamiltonianSecondQuantDeformedCondensed}
\end{equation}
where $\hat{H}_{0}$ and $\hat{H}_{a}$ are given by Eqs. \ref{eq:HarkThermPerturbHarmonicCrystalHamiltonianSecondQuant} and \ref{eq:HarkThermPerturbLatticeHamiltonianAnharmonicSecondQuantEq_1} and (see Appendix \ref{CondensedNotation} for the notation)
\begin{eqnarray} 
\hat{H}_{s} =&& \sum_{m=1} \frac{1}{m!} \sum_{\bar{\mu}_m} \sum_{\bar{\nu}_{m}} \sum_{n=0} \sum_{\bar{\lambda}_{n}}  V_{\bar{\mu}_{m} \bar{\nu}_{m}} \left(\bar{\lambda}_{n}\right) \bar{u}_{\mu_{m} \nu_{m}} \nonumber \\
&&\times \hat{A}_{\lambda_{1}} \cdots \hat{A}_{\lambda_{n}}.
\label{eq:HarkThermPerturbCrystalHamiltonianSecondQuantStrained}
\end{eqnarray}
In Eq. \ref{eq:HarkThermPerturbCrystalHamiltonianSecondQuantStrained}
\begin{eqnarray} 
V&&_{\bar{\mu}_{m} \bar{\nu}_{m}} \left(\bar{\lambda}_{n}\right) \nonumber \\
&&= \frac{1}{n!} \sum_{ \kappa_{1},\alpha_{1}} \sum_{l_{2} \kappa_{2},\alpha_{2}} \cdots \sum_{l_{n} \kappa_{n},\alpha_{n}} \sum_{l'_{1} \kappa'_{1}} \sum_{l'_{2} \kappa'_{2}} \cdots \sum_{l'_{m} \kappa'_{m}}  \nonumber \\
&&\times \Delta\left(\vec{q}_{1} + \vec{q}_{2} + \cdots + \vec{q}_{n}\right) \nonumber \\
&&\times \Phi_{ \alpha_{1} \cdots \alpha_{n} \mu_{1} \cdots \mu_{m}}\left(0 \kappa_{1};l_{2} \kappa_{2};\cdots;l_{n} \kappa_{n};l'_{1} \kappa'_{1};\cdots;l'_{m} \kappa'_{m} \right) \nonumber \\
&&\times T_{\alpha_{1}}\left(\lambda_{1}|0 \kappa_{1} \right) T_{\alpha_{2}}\left(\lambda_{2}|l_{2} \kappa_{2} \right) \cdots T_{\alpha_{n}}\left(\lambda_{n}|l_{n} \kappa_{n} \right) \nonumber \\
&&\times x_{\nu_{1}}\left(l'_{1} \kappa'_{1}\right) x_{\nu_{2}}\left(l'_{2} \kappa'_{2}\right) \cdots x_{\nu_{m}}\left(l'_{m} \kappa'_{m}\right).
\label{eq:HarkThermPerturbCrystalHamiltonianSecondQuantStrainedCoefficientsEq_1}
\end{eqnarray}
and
\begin{equation} 
T_{\alpha}\left(\lambda|l \kappa \right) = \left(\frac{\hbar}{2 M_{\kappa}}\right)^{1/2} \omega^{-1/2}_{\lambda} e^{i \vec{q}\cdot \vec{x}\left(l\right)} e_{\alpha}\left(\kappa|\lambda \right).
\label{eq:HarkThermPerturbCrystalHamiltonianSecondQuantStrainedCoefficientsEq_2}
\end{equation}
Eventually, the Hamiltonian given by Eq. \ref{eq:HarkThermPerturbCrystalHamiltonianSecondQuantDeformedCondensed} is normalized to the unit cell volume. For the sake of convenience, the factors $1/N$ are not explicitly shown in Eq. \ref{eq:HarkThermPerturbCrystalHamiltonianSecondQuantStrainedCoefficientsEq_1}. In all results of this work, one must add the factor $1/N^{n}$ when there is $n$ different sums over the quantities $\left\{\lambda_{i}\right\}$. In the present work, the Hamiltonian $\hat{H}_{d}$ is used to derive the thermal and elastic properties of crystals (Sec. \ref{cha:PerturbationExpansionForThermodynamicalQuantities}). In order to obtain the coefficients $V\left(\bar{\lambda}_{n}\right)$ (Eq. \ref{eq:HarkThermPerturbLatticeHamiltonianAnharmonicSecondQuantEq_2}), one have to obtain the quantities within harmonic approximation and $n$th-order IFCs. Also, in order to obtain the coefficients $V_{\bar{\mu}_{m} \bar{\nu}_{m}} \left(\bar{\lambda}_{n}\right)$ (Eq. \ref{eq:HarkThermPerturbCrystalHamiltonianSecondQuantStrainedCoefficientsEq_1}), one have to obtain the quantities within harmonic approximation and $\left(n + m\right)$th-order IFCs.

The expansion of the Hamiltonian can be made in more general manner to include a wider class of parameters as in Ref. \cite{Born-Huang-DynamicalTheory-1954}, that is, one may write 
\begin{eqnarray} 
\hat{H}_{p} =&&  \sum_{m=1} \sum_{n_{1}=0} \cdots \sum_{n_{m}=0} \sum_{ \alpha^{\left(1,1\right)}_{1} } \cdots \sum_{\alpha^{\left(m,h_{m}\right)}_{ n_{m} } }\nonumber \\
&&\times \sum_{k_{1}=0}  \sum_{\lambda_{1_{1}}} \cdots \sum_{ \lambda_{k_{1}}} \cdots \sum_{k_{m}=0} \sum_{\lambda_{1_{m}} } \cdots \sum_{ \lambda_{k_{m}}} \nonumber \\
&&\times  g^{ \alpha^{\left(1,1\right)}_{1} \cdots  \alpha^{\left(m,h_{m}\right)}_{ n_{m} } }_{ \lambda_{1_{1}} \cdots \lambda_{k_{m}} } f_{ \alpha^{\left(1,1\right)}_{1} \cdots \alpha^{\left(1,h_{1}\right)}_{1} } \cdots f_{ \alpha^{\left(1,1\right)}_{ n_{1} } \cdots  \alpha^{\left(1,h_{1}\right)}_{ n_{1} } } \cdots \nonumber \\
&&\times  f_{ \alpha^{\left(2,1\right)}_{1} \cdots \alpha^{\left(2,h_{2}\right)}_{1} } \cdots f_{ \alpha^{\left(2,1\right)}_{ n_{2} } \cdots  \alpha^{\left(2,h_{2}\right)}_{ n_{2} } } \cdots  f_{ \alpha^{\left(m,1\right)}_{1} \cdots \alpha^{\left(m,h_{m}\right)}_{1} }  \nonumber \\
&&\times \cdots f_{ \alpha^{\left(m,1\right)}_{ n_{m} } \cdots  \alpha^{\left(m,h_{m}\right)}_{ n_{m} } } \hat{A}_{\lambda_{1}} \cdots \hat{A}_{\lambda_{k_{1}}} \cdots \hat{A}_{\lambda_{1_{m}}} \cdots \hat{A}_{\lambda_{k_{m}}}, \nonumber \\
\label{eq:HarkThermPerturbCrystalHamiltonianSecondQuantStrainedCoefficientsEq_3}
\end{eqnarray}
where $\left\{f_{ \alpha^{\left(m,1\right)}_{ n_{m} } \cdots  \alpha^{\left(m,h_{m}\right)}_{ n_{m} } }\right\}$ is the set of $m$th macroscopic parameters with $h_{m}$ indices. For example, let the macroscopic parameters be the electric field components $f_{ \alpha_{1} } = E_{\alpha_{1}}$ \textit{et cetera} ($m=h_{1}=1$), for this case Eq. \ref{eq:HarkThermPerturbCrystalHamiltonianSecondQuantStrainedCoefficientsEq_3} can be written as 
\begin{eqnarray} 
\hat{H}^{E}_{p} \equiv && \sum_{n=1} \sum_{\alpha_{1}} \cdots \sum_{\alpha_{n}} \sum_{k=0} \sum_{\lambda_{1}}\cdots \sum_{\lambda_{k}} \nonumber \\
&&\times g^{\alpha_{1} \cdots \alpha_{n} }_{ \lambda_{1} \cdots \lambda_{k} } E_{\alpha_{1}} \cdots E_{\alpha_{n}} \hat{A}_{\lambda_{1}} \cdots \hat{A}_{\lambda_{k}}.
\label{eq:HarkThermPerturbCrystalHamiltonianSecondQuantStrainedCoefficientsEq_4}
\end{eqnarray}
Now, let the macroscopic parameters be the strain parameters $f_{\alpha^{\left(1,1\right)}_{1} \alpha^{\left(1,2\right)}_{1} } = u_{\alpha^{\left(1,1\right)}_{1} \alpha^{\left(1,2\right)}_{1}} = u_{\alpha_{1} \beta_{1}}$ ($m=1,h_{1}=2$), for this case Eq. \ref{eq:HarkThermPerturbCrystalHamiltonianSecondQuantStrainedCoefficientsEq_3} can be written as
\begin{eqnarray} 
\hat{H}_{p} = && \sum_{n=1}  \sum_{\alpha_{1}} \cdots \sum_{\alpha_{n}} \sum_{\beta_{1}} \cdots \sum_{\beta_{n}} \sum_{k=0}  \sum_{\lambda_{1}}\cdots \sum_{\lambda_{k}} \nonumber \\
&&\times g^{\alpha_{1} \cdots \alpha_{n}  \beta_{1} \cdots \beta_{n} }_{ \lambda_{1} \cdots \lambda_{k} } u_{\alpha_{1} \beta_{1}} \cdots u_{\alpha_{n}\beta_{n}} \hat{A}_{\lambda_{1}} \cdots \hat{A}_{\lambda_{k}}.
\label{eq:HarkThermPerturbCrystalHamiltonianSecondQuantStrainedCoefficientsEq_5}
\end{eqnarray}
A comparison of Eq. \ref{eq:HarkThermPerturbCrystalHamiltonianSecondQuantStrainedCoefficientsEq_5} to Eqs. \ref{eq:HarkThermPerturbCrystalHamiltonianSecondQuantStrainedCoefficientsEq_2} and \ref{eq:HarkThermPerturbCrystalHamiltonianSecondQuantStrainedCoefficientsEq_3} shows that $\hat{H}_{p} = \hat{H}_{s}$ and one may identify the coefficients
\begin{eqnarray} 
g^{\alpha_{1} \cdots \alpha_{n}  \beta_{1} \cdots \beta_{n} }_{ \lambda_{1} \cdots \lambda_{k} } = \frac{1}{n!} V_{ \alpha_{1} \cdots \alpha_{n}  \beta_{1} \cdots \beta_{n} } \left(\lambda_{1};\ldots;\lambda_{k}\right).
\label{eq:HarkThermPerturbCrystalHamiltonianSecondQuantStrainedCoefficientsEq_6}
\end{eqnarray}
One may write in a similar way for the mixed coefficient ($m > 1$) as it has been done in Ref. \cite{Born-Huang-DynamicalTheory-1954}. The reason why Eq. \ref{eq:HarkThermPerturbCrystalHamiltonianSecondQuantStrainedCoefficientsEq_5} is introduced is that the same form of perturbation theory (Sec. \ref{cha:PerturbationExpansionForThermodynamicalQuantities}) applies for any set of parameters when $m$ is the same for different cases, the difference is in the coefficients $g$ (indices neglected). The matter is how many and how many different operators $\left\{\hat{A}_{\lambda_{k}}\right\}$ belonging to different parameters there is in the expression of a particular order provided the coefficients $g$ have the same symmetry properties with respect to the phonon labels.

\section{Thermodynamical relations}
\label{cha:ThermodynamicalRelations}
The thermodynamics and statistical mechanics of crystals is described, for example, in Refs. \cite{Born-Huang-DynamicalTheory-1954,Leibfried-SolidStatePhysics-1961,Wallace-ThermodynamicsOfCrystals-1972,Fultz-2010247-VibrationalThermodynOfMat-2010}. In this section, the results used to describe the thermal and elastic properties of crystals are only listed.

\subsection{Expansion of free energy and internal energy}
\label{cha:ExpansionOfFreeEnergyAndInternalEnergy}
The definition of isothermal elastic constants can be made by expanding the Helmholtz free energy to Taylor series in strains $\eta_{ij}$ and temperature, namely \cite{Leibfried-SolidStatePhysics-1961,Brugger-ElastConstDef-PhysRev.133.A1611-1964,Wallace-ThermodynamicsOfCrystals-1972}
\begin{eqnarray} 
F =&& F_{\eta_{0}} + \frac{\partial{F}}{\partial{T'}} T + \sum^{3}_{\mu,\nu=1} \frac{\partial{F}}{\partial{\eta'_{\mu\nu}}} \eta_{\mu\nu} \nonumber \\
  &&+  \frac{1}{2} \sum^{3}_{\mu_{1},\nu_{1},\mu_{2},\nu_{2},l=1}\frac{\partial^{2}{F}}{\partial{\eta'_{\mu_{1}\nu_{1}}} \partial{\eta'_{\mu_{2}\nu_{2}}} } \eta_{\mu_{1}\nu_{1}} \eta_{\mu_{2}\nu_{2}} \nonumber \\
  &&+  \sum^{3}_{\mu,\nu=1}\frac{\partial^{2}{F}}{\partial{\eta'_{\mu\nu}} \partial{T'} } \eta_{\mu\nu} T + \cdots \: ,
\label{eq:HarkThermPerturbHelmholtzEnergyExpansionStrainTemperatureEq_1}
\end{eqnarray}
where the finite strain parameters can be written as
\begin{equation} 
\eta_{\mu\nu} = \frac{1}{2} \left(u_{\mu\nu} + u_{\nu\mu} + \sum_{\epsilon} u_{\epsilon\mu} u_{\epsilon \nu} \right).
\label{eq:HarkThermPerturbHelmholtzEnergyExpansionStrainTemperatureEq_2}
\end{equation}
The coefficients of the expansion in Eq. \ref{eq:HarkThermPerturbHelmholtzEnergyExpansionStrainTemperatureEq_1} are entropy and isothermal stress
\begin{equation} 
\frac{\partial{F}}{\partial{T}} = -S, \quad \frac{\partial{F}}{\partial{\eta_{\mu\nu}}}=\sigma^{T}_{\mu\nu},
\label{eq:HarkThermPerturbHelmholtzEnergyExpansionStrainTemperatureEq_3}
\end{equation}
second-order isothermal elastic constants
\begin{equation} 
\frac{\partial^{2}{F}}{\partial{\eta_{\mu_{1}\nu_{1}}} \partial{\eta_{\mu_{2}\nu_{2}}} } =\frac{\partial{\sigma^{T}_{\mu_{1}\nu_{1}}}}{\partial{\eta_{\mu_{2}\nu_{2}}}} =  c^{T}_{\mu_{1}\nu_{1} \mu_{2}\nu_{2}},
\label{eq:HarkThermPerturbHelmholtzEnergyExpansionStrainTemperatureEq_5}
\end{equation}
$k$th-order isothermal elastic constants
\begin{equation} 
\frac{\partial^{k}{F}}{\partial{\eta_{\mu_{1}\nu_{1}}} \cdots \partial{\eta_{\mu_{k}\nu_{k}}} } =  c^{T}_{\mu_{1}\nu_{1} \cdots \mu_{k}\nu_{k}},
\label{eq:HarkThermPerturbHelmholtzEnergyExpansionStrainTemperatureEq_6}
\end{equation}
and so forth.

The adiabatic elastic constants can be defined in a similar way in terms of the coefficients which appear in the expansion of internal energy 
\begin{eqnarray} 
U =&& U_{\eta_{0}} + \frac{\partial{U}}{\partial{T'}} T + \sum^{3}_{\mu,\nu=1} \frac{\partial{U}}{\partial{\eta'_{\mu\nu}}} \eta_{\mu\nu} \nonumber \\
  &&+  \frac{1}{2} \sum^{3}_{\mu_{1},\nu_{1},\mu_{2},\nu_{2},l=1}\frac{\partial^{2}{U}}{\partial{\eta'_{\mu_{1}\nu_{1}}} \partial{\eta'_{\mu_{2}\nu_{2}}} } \eta_{\mu_{1}\nu_{1}} \eta_{\mu_{2}\nu_{2}} \nonumber \\
  &&+  \sum^{3}_{\mu,\nu=1}\frac{\partial^{2}{U}}{\partial{\eta'_{\mu\nu}} \partial{T'} } \eta_{\mu\nu} T + \cdots \: ,
\label{eq:HarkThermPerturbInternalEnergyExpansionStrainTemperatureEq_1}
\end{eqnarray}
and one may identify the heat capacity at constant strain, adiabatic stress and $k$th-order adiabatic elastic constants by writing
\begin{equation} 
\frac{\partial{U}}{\partial{T}} = C_{\eta}, \quad \frac{\partial{U}}{\partial{\eta_{\mu\nu}}}=\sigma^{A}_{\mu\nu},
\label{eq:HarkThermPerturbInternalEnergyExpansionStrainTemperatureEq_3}
\end{equation}
and
\begin{equation} 
\frac{\partial^{k}{U}}{\partial{\eta_{\mu_{1}\nu_{1}}} \cdots \partial{\eta_{\mu_{k}\nu_{k}}} } =  c^{A}_{\mu_{1}\nu_{1} \cdots \mu_{k}\nu_{k}}.
\label{eq:HarkThermPerturbInternalEnergyExpansionStrainTemperatureEq_6}
\end{equation}

As in Ref. \cite{Born-Huang-DynamicalTheory-1954}, the perturbation expansion is made for the Hamiltonian $\hat{H}_{int}$, which includes the expansion in terms of the infinitesimal strain parameters $\left\{u_{\mu\nu}\right\}$ instead of the parameters $\left\{\eta_{\mu\nu}\right\}$. One may write the expansion of free and internal energies as in Eqs. \ref{eq:HarkThermPerturbHelmholtzEnergyExpansionStrainTemperatureEq_1} and \ref{eq:HarkThermPerturbInternalEnergyExpansionStrainTemperatureEq_1} in terms of the infinitesimal strain parameters and define the elastic constants (and stress) as the derivatives with respect to infinitesimal strain parameters. In both cases, the stress is the same, but the elastic constants have different form. The elastic constant defined in terms of the finite strain parameters can be expressed in terms of the elastic constants and stress which in turn are defined in terms of the infinitesimal strain parameters. In addition to the preceding, further transformations are needed to write the elastic constants defined by Eqs. \ref{eq:HarkThermPerturbHelmholtzEnergyExpansionStrainTemperatureEq_5} and \ref{eq:HarkThermPerturbInternalEnergyExpansionStrainTemperatureEq_6} in terms of the elastic constants calculated from the Hamiltonian $\hat{H}_{s}$ (Eq. \ref{eq:HarkThermPerturbCrystalHamiltonianSecondQuantStrained}). If one considers the stressed and unstressed crystal lattice, different relations are obtained. These relations are considered, for example, in Refs. \cite{Born-Huang-DynamicalTheory-1954,Leibfried-SolidStatePhysics-1961,Wallace-ThermodynamicsOfCrystals-1972}. For instance, if the crystal lattice is unstressed, one may write for the second-order elastic constants \cite{Huang-OntheAtomicTheoryofElasticity-1950,Born-Huang-DynamicalTheory-1954}
\begin{eqnarray} 
c_{\mu_{1} \nu_{1} \mu_{2} \nu_{2}} = && \left[\mu_{1} \mu_{2} , \nu_{1} \nu_{2}\right] + \left[\mu_{2} \nu_{1} ,\mu_{1} \nu_{2}\right] \nonumber \\
&&- \left[\mu_{2} \nu_{2} ,\mu_{1} \nu_{1} \right] + \left(\mu_{1} \nu_{1}, \mu_{2} \nu_{2}\right),
\label{eq:HarkThermPerturbInternalEnergyExpansionStrainTemperatureEq_7}
\end{eqnarray}
where 
\begin{equation} 
\left[\mu_{1} \mu_{2} , \nu_{1} \nu_{2}\right] = \frac{1}{2} \left(\tilde{c}_{\mu_{1} \nu_{2} \mu_{2}  \nu_{1} } + \tilde{c}_{\mu_{1} \mu_{2} \nu_{1} \nu_{2} }\right),
\label{eq:HarkThermPerturbInternalEnergyExpansionStrainTemperatureEq_8}
\end{equation}
and terms $\left\{\tilde{c}_{\mu_{1} \nu_{2} \mu_{2}  \nu_{1} }\right\}$ are given by the relations with the coefficients $V_{\mu_{1} \nu_{2} \mu_{2}  \nu_{1}}\left(\bar{\lambda}_{n}\right), \, n=0,1,2,\ldots$ (Eq. \ref{eq:HarkThermPerturbCrystalHamiltonianSecondQuantStrainedCoefficientsEq_1}). These relations are given in Secs. \ref{cha:FirstOrderResults}-\ref{cha:ThirdOrderResults}. Furthermore, in Eq. \ref{eq:HarkThermPerturbInternalEnergyExpansionStrainTemperatureEq_7}, the brackets $\left(\mu_{1} \nu_{1}, \mu_{2} \nu_{2}\right)$ are given by the relations with terms $\left\{\tilde{c}_{\mu_{1} \nu_{2} \mu_{2}  \nu_{1} }\right\}$ of the form $V_{\mu_{1} \nu_{1}}\left(\bar{\lambda}_{n}\right) V_{\mu_{2} \nu_{2}}\left(\bar{\lambda}'_{n'}\right)$, for example, relations as given by Eq. \ref{eq:HarkThermPerturbSecondOrderResultsIsothermalElasticConstantsAndStressEq_2}. It has been shown \cite{Leibfried-SolidStatePhysics-1961} that the square brackets given in Ref. \cite{Born-Huang-DynamicalTheory-1954} (p. 235) can be written as (see Eq. \ref{eq:HarkThermPerturbCrystalHamiltonianSecondQuantStrainedCoefficientsEq_1})
\begin{equation} 
\left[\mu_{1} \mu_{2} , \nu_{1} \nu_{2}\right] = \frac{1}{2} \left(V_{\mu_{1} \nu_{2} \mu_{2}  \nu_{1} } + V_{\mu_{1} \mu_{2} \nu_{1} \nu_{2} }\right).
\label{eq:HarkThermPerturbInternalEnergyExpansionStrainTemperatureEq_9}
\end{equation}
Similar transformation equations for the third-order elastic constants have been considered, for example, in Ref. \cite{Wallace-ThermodynamicsOfCrystals-1972} (p. 89).

\section{Perturbation expansion for thermodynamical quantities}
\label{cha:PerturbationExpansionForThermodynamicalQuantities}

The technique used here is the same as in Refs. \cite{Matsubara-NewApproachToQuantumStatisticalMechanics-1955,Cowley-TheLatticeDynamicsOfAnAnharmonicCrystal-1963,Barron-DynamicalPropertiesOfSolids-1974}. One may write the partition function as in Eq. \ref{eq:HarkThermPerturbCrystalCanonicalPartitionFunction}, but this time for the Hamiltonian $\hat{H}_{d}$ which is given by Eq. \ref{eq:HarkThermPerturbCrystalHamiltonianSecondQuantDeformedCondensed}. Thus
\begin{equation} 
e^{-\beta \hat{H}_{d}} = e^{-\beta \left(\hat{H}_{0} + \hat{H}_{int} \right)} = e^{-\beta \hat{H}_{0}} \hat{S}\left(\beta\right), 
\label{eq:HarkThermPerturbCrystalCanonicalPartitionFunctionPertExpEq_2}
\end{equation} 
since $\hat{H}_{0}$ and $\hat{H}_{int}$ commute. With this notation
\begin{equation} 
Z =   Z_{0}   \left\langle \hat{S}\left(\beta\right) \right\rangle_{0},
\label{eq:HarkThermPerturbCrystalCanonicalPartitionFunctionPertExpEq_2_2}
\end{equation} 
where
\begin{equation} 
Z_{0} = \sum_{n} \braket{n|e^{-\beta \hat{H}_{0}}|n}, 
\label{eq:HarkThermPerturbCrystalCanonicalPartitionFunctionPertExpEq_2_3}
\end{equation}
and
\begin{equation} 
\left\langle \hat{S}\left(\beta\right) \right\rangle_{0} =  \sum_{n} \braket{n|e^{-\beta \hat{H}_{0}} \hat{S}\left(\beta\right)|n}.
\label{eq:HarkThermPerturbCrystalCanonicalPartitionFunctionPertExpEq_2_4}
\end{equation}
One may differentiate Eq. \ref{eq:HarkThermPerturbCrystalCanonicalPartitionFunctionPertExpEq_2} with respect to $\beta$ and write
\begin{equation} 
\frac{\partial}{\partial{\beta}} \hat{S}\left(\beta\right)   = - \hat{H}_{int}\left(\beta\right) \hat{S}\left(\beta\right). 
\label{eq:HarkThermPerturbCrystalCanonicalPartitionFunctionPertExpEq_3}
\end{equation} 
From Eq. \ref{eq:HarkThermPerturbCrystalCanonicalPartitionFunctionPertExpEq_2} it follows that $\hat{S}\left(0\right) = 1$, then by integration and iteration of Eq. \ref{eq:HarkThermPerturbCrystalCanonicalPartitionFunctionPertExpEq_3} and by writing the result as a time-ordered product
\begin{eqnarray} 
\hat{S}\left(\beta\right) =&&1 - \sum^{\infty}_{h=1} \frac{\left(-1\right)^{h}}{h!} \int^{\beta}_{0}d\tau_{1} \cdots \int^{\beta}_{0}d\tau_{h}  \nonumber \\
&&\times  \mathcal{T}\left\{ \hat{H}_{int}\left(\tau_{1}\right) \cdots \hat{H}_{int}\left(\tau_{h}\right)  \right\}, \nonumber \\
\label{eq:HarkThermPerturbCrystalCanonicalPartitionFunctionPertExpEq_5}
\end{eqnarray} 
where the Hamiltonians
\begin{equation} 
\hat{H}_{int}\left(\tau_{i}\right)  = e^{\tau_{i} \hat{H}_{0}} \hat{H}_{int} e^{-\tau_{i} \hat{H}_{0}},
\label{eq:HarkThermPerturbCrystalCanonicalPartitionFunctionPertExpEq_6}
\end{equation} 
are operators in the interaction picture.

From Eqs. \ref{eq:HarkThermPerturbCrystalCanonicalPartitionFunctionPertExpEq_2_2} and \ref{eq:HarkThermPerturbCrystalCanonicalPartitionFunctionPertExpEq_5}
\begin{eqnarray} 
Z = && Z_{0}  \sum^{\infty}_{h=0} \frac{\left(-1\right)^{h}}{h!} \int^{\beta}_{0}d\tau_{1} \cdots \int^{\beta}_{0}d\tau_{h}  \nonumber \\
&&\times \left\langle  \mathcal{T}\left\{ \hat{H}_{int}\left(\tau_{1}\right) \cdots \hat{H}_{int}\left(\tau_{h}\right)  \right\}  \right\rangle_{0},
\label{eq:HarkThermPerturbCrystalCanonicalPartitionFunctionPertExpEq_7}
\end{eqnarray} 
and thus
\begin{equation} 
F = \Phi_{0} - \frac{1}{\beta} \ln Z_{0} - \frac{1}{\beta} \ln \left\langle \hat{S}\left(\beta\right) \right\rangle_{0} = F_{0} + \tilde{F}_{A}.
\label{eq:HarkThermPerturbCrystalCanonicalPartitionFunctionPertExpEq_8}
\end{equation}
where $F_{0}$ is given by Eq. \ref{eq:HarkThermPerturbCrystalHarmonicHelmholtzFreeEnergy} and $\tilde{F}_{A}$ is the last term after the first equality in Eq. \ref{eq:HarkThermPerturbCrystalCanonicalPartitionFunctionPertExpEq_8}. It can be shown by using combinatorial arguments that $\tilde{F}_{A}$ can be written as \cite{Cowley-TheLatticeDynamicsOfAnAnharmonicCrystal-1963,Barron-DynamicalPropertiesOfSolids-1974}
\begin{equation} 
\tilde{F}_{A} =  - \frac{1}{\beta} \left\langle \hat{S}\left(\beta\right) \right\rangle_{0,c},
\label{eq:HarkThermPerturbCrystalCanonicalPartitionFunctionPertExpEq_10}
\end{equation} 
where the subscript $c$ indicates that only terms corresponding to the connected diagrams are considered. By using the result given by Eq. \ref{eq:HarkThermPerturbCrystalCanonicalPartitionFunctionPertExpEq_10}, one may write all the quantities considered in Sec. \ref{cha:ThermodynamicalRelationsForHarmonicAndQuasiHarmonicPhonons} in terms of the ensemble average on the right hand side of Eq. \ref{eq:HarkThermPerturbCrystalCanonicalPartitionFunctionPertExpEq_10}, that is, the anharmonic part of internal energy 
\begin{equation} 
\tilde{U}_{A} = -  \frac{\partial}{\partial{\beta}} \left\langle \hat{S}\left(\beta\right) \right\rangle_{0,c},
\label{eq:HarkThermPerturbCrystalCanonicalPartitionFunctionPertExpEq_11}
\end{equation} 
entropy 
\begin{eqnarray} 
\tilde{S}_{A} &=&  \frac{1}{T \beta} \left\langle \hat{S}\left(\beta\right) \right\rangle_{0,c} - \frac{1}{T } \frac{\partial}{\partial{\beta}} \left\langle \hat{S}\left(\beta\right) \right\rangle_{0,c} \nonumber \\
      &=&\frac{1}{T}\left(\tilde{U}_{A}-\tilde{F}_{A}\right),
\label{eq:HarkThermPerturbCrystalCanonicalPartitionFunctionPertExpEq_12}
\end{eqnarray} 
heat capacity at constant strain
\begin{equation} 
\tilde{C}_{A,\eta} =  \frac{\beta}{T }  \frac{\partial^{2}}{\partial{\beta^{2}}} \left\langle \hat{S}\left(\beta\right) \right\rangle_{0,c},
\label{eq:HarkThermPerturbCrystalCanonicalPartitionFunctionPertExpEq_13}
\end{equation} 
isothermal $k$th-order elastic constants
\begin{eqnarray} 
\tilde{c}&&^{T}_{\mu_{1} \nu_{1} \cdots \mu_{k} \nu_{k} } \nonumber \\
&&= - \frac{1}{\beta} \left. \frac{\partial^{k}{ \left\langle  \hat{S}\left(\beta\right) \right\rangle_{0,c} }}{\partial{u_{\mu_{1} \nu_{1}}} \cdots \partial{u_{\mu_{k} \nu_{k}}} } \right|_{u_{\mu_{1} \nu_{1}} = 0, \ldots, u_{\mu_{k} \nu_{k}} = 0},
\label{eq:HarkThermPerturbCrystalCanonicalPartitionFunctionPertExpEq_14}
\end{eqnarray}
and adiabatic $k$th-order elastic constants
\begin{eqnarray} 
\tilde{c}&&^{A}_{\mu_{1} \nu_{1} \cdots \mu_{k} \nu_{k} } \nonumber \\
&&=   -\frac{\partial}{\partial{\beta}} \left. \frac{\partial^{k}{  \left\langle \hat{S}\left(\beta\right) \right\rangle_{0,c} }}{\partial{u_{\mu_{1} \nu_{1}}} \cdots \partial{u_{\mu_{k} \nu_{k}}} } \right|_{u_{\mu_{1} \nu_{1}} = 0, \ldots, u_{\mu_{k} \nu_{k}} = 0}. 
\label{eq:HarkThermPerturbCrystalCanonicalPartitionFunctionPertExpEq_15}
\end{eqnarray} 
The isothermal and adiabatic stress can be obtained from Eqs. \ref{eq:HarkThermPerturbCrystalCanonicalPartitionFunctionPertExpEq_14} and \ref{eq:HarkThermPerturbCrystalCanonicalPartitionFunctionPertExpEq_15} with $j=1$. It should be noticed that the elastic constants given by Eqs. \ref{eq:HarkThermPerturbCrystalCanonicalPartitionFunctionPertExpEq_14} and \ref{eq:HarkThermPerturbCrystalCanonicalPartitionFunctionPertExpEq_15} are defined in terms of infinitesimal strain parameters $\left\{u_{\mu_{i} \nu_{i}} \right\}$. The transformation to elastic constants defined in terms of finite strain parameters $\left\{\eta_{\mu_{i} \nu_{i}} \right\}$ is considered, for example in Refs. \cite{Born-Huang-DynamicalTheory-1954,Wallace-ThermodynamicsOfCrystals-1972} and the lowest-order transformation for the second-order elastic constants is given in Sec. \ref{cha:ExpansionOfFreeEnergyAndInternalEnergy} (Eqs. \ref{eq:HarkThermPerturbInternalEnergyExpansionStrainTemperatureEq_7}-\ref{eq:HarkThermPerturbInternalEnergyExpansionStrainTemperatureEq_9}).

In Secs. \ref{cha:FirstOrderResults}-\ref{cha:ThirdOrderResults}, the ensemble average on the right hand side of Eq. \ref{eq:HarkThermPerturbCrystalCanonicalPartitionFunctionPertExpEq_10} is written up to third-order. In the actual calculations, the quantities  $\tilde{F}_{A},\tilde{U}_{A},\tilde{S}_{A}$ and $\tilde{C}_{A,\eta}$ given in Secs. \ref{cha:FirstOrderResults}-\ref{cha:ThirdOrderResults}, are to be normalized, for instance, to a unit volume.

\subsection{Evaluation of perturbation expansion}
\label{cha:EvaluationOfPerturbationExpansion}
In the evaluation of the perturbation expansion, the following quantities, for example, appear (Eq. \ref{eq:HarkThermPerturbLatticeHamiltonianAnharmonicSecondQuantEq_1}, see also Appendix \ref{CondensedNotation} for the notation)
\begin{eqnarray} 
\left\langle  \mathcal{T}\left\{ \hat{H}_{a}\left(\tau\right) \right\}  \right\rangle_{0,c} =&&  \sum_{n=3} \sum_{\bar{\lambda}_{n}} V\left(\bar{\lambda}_{n}\right)  \nonumber \\
&&\times \left\langle  \mathcal{T}\left\{ \hat{A}_{\lambda_{1}}\left(\tau\right)	\cdots \hat{A}_{\lambda_{n}}\left(\tau\right) \right\}  \right\rangle_{0,c}. \nonumber \\
\label{eq:HarkThermPerturbEvaluationOfPerturbationExpansionEq_1}
\end{eqnarray} 
The ensemble averages as in Eq. \ref{eq:HarkThermPerturbEvaluationOfPerturbationExpansionEq_1} can be simplified by using the Wick's theorem of the form \cite{Choquard-TheAnharmonicCrystal-1967,Barron-DynamicalPropertiesOfSolids-1974,Rammer-QuantumFieldTheoryOfNonEquilibriumStates-2007}
\begin{eqnarray} 
\left\langle  \mathcal{T}\left\{ \hat{A}_{\lambda_{1}}	\cdots \hat{A}_{\lambda_{2n}} \right\}  \right\rangle_{0,c} =&& \sum^{2n}_{k=2} \left\langle \mathcal{T}\left\{   \hat{A}_{\lambda_{1}}  \hat{A}_{\lambda_{k}}  \right\} \right\rangle_{0} \nonumber \\
&&\times \left\langle \mathcal{T}\left\{  \prod^{2n}_{l \neq 1,k} \hat{A}_{\lambda_{l}}  \right\} \right\rangle_{0,c},
\label{eq:HarkThermPerturbEvaluationOfPerturbationExpansionEq_2}
\end{eqnarray}
which is valid for all time arguments (real or imaginary times). Furthermore, one may use the following symmetry properties of the coefficients
\begin{eqnarray} 
V\left(\lambda_{1};\lambda_{2}\ldots;\lambda_{n}\right) &=& V\left(\lambda_{2};\lambda_{1}\ldots;\lambda_{n}\right) = \cdots, \nonumber \\
V_{\bar{\mu}_{m} \bar{\nu}_{m}} \left(\lambda_{1};\lambda_{2}\ldots;\lambda_{n}\right) &=& V_{\bar{\mu}_{m} \bar{\nu}_{m}} \left(\lambda_{2};\lambda_{1}\ldots;\lambda_{n}\right) = \cdots,\nonumber \\
\label{eq:HarkThermPerturbEvaluationOfPerturbationExpansionEq_3}
\end{eqnarray}
to simplify the expressions.

The following results from the theory of the many-body Green's functions \cite{Zubarev-GreensFunc0038-5670-3-3-R02-1960,Kadanoff-Baym-Quant.Stat.Mech-1962,Maradudin-Fein-ScatteringOfNeutrons-1962,Fetter-Walencka-q-theory-of-many-particle-1971,Doniach-Sondheimer-Green-1974,Barron-DynamicalPropertiesOfSolids-1974,Rickayzen-GreensFunctions-1980,Mahan-many-particle-1990,Rammer-QuantumFieldTheoryOfNonEquilibriumStates-2007,Stefanucci-Leeuwen-many-body-book-2013} are used to evaluate the ensemble averages which result from Eq. \ref{eq:HarkThermPerturbEvaluationOfPerturbationExpansionEq_2} 
\begin{eqnarray} 
G&&_{0}\left(\lambda\tau|\lambda'\tau'\right) \nonumber \\
&&\equiv \left\langle  \mathcal{T}\left\{ \hat{A}_{\lambda}\left(\tau\right)	\hat{A}_{\lambda'}\left(\tau'\right) \right\}  \right\rangle_{0} = G_{0}\left(\lambda\tau|\lambda' 0\right) \nonumber \\
&&= \delta_{\lambda\left(-\lambda'\right)}\theta\left(\tau\right) \left[ e^{\hbar \omega_{\lambda} \tau} \bar{n}_{\lambda} + e^{- \hbar \omega_{\lambda} \tau}  \left(\bar{n}_{\lambda}+1\right) \right] \nonumber \\
 &&~~+ \delta_{\lambda\left(-\lambda'\right)} \theta\left(-\tau\right) \left[ e^{- \hbar \omega_{\lambda} \tau} \bar{n}_{\lambda} +  e^{ \hbar \omega_{\lambda} \tau}  \left(\bar{n}_{\lambda}+1\right)\right],
\label{eq:HarkThermPerturbEvaluationOfPerturbationExpansionEq_4}
\end{eqnarray}
\begin{equation} 
G_{0}\left(\lambda\tau|\lambda'\tau\right) = \delta_{\lambda\left(-\lambda'\right)} \left(2 \bar{n}_{\lambda} +1 \right),
\label{eq:HarkThermPerturbEvaluationOfPerturbationExpansionEq_5}
\end{equation}
\begin{equation} 
G_{0}\left(\lambda\tau|\lambda0\right) \equiv G_{0}\left(\lambda\tau\right) = \sum^{\infty}_{n=-\infty} G_{0}\left(\lambda|\omega_{n}\right)e^{i \omega_{n} \tau},  
\label{eq:HarkThermPerturbEvaluationOfPerturbationExpansionEq_6}
\end{equation} 
where
\begin{equation} 
G_{0}\left(\lambda|\omega_{n}\right) = \frac{2 \omega_{\lambda} }{\beta \hbar \left[\omega^{2}_{\lambda}+ \omega^{2}_{n} \right]}, \quad \omega_{n} = \frac{2 n \pi }{\beta \hbar}.
\label{eq:HarkThermPerturbEvaluationOfPerturbationExpansionEq_7}
\end{equation}

The integrals over the ensemble averages as in Eq. \ref{eq:HarkThermPerturbCrystalCanonicalPartitionFunctionPertExpEq_7} are evaluated by using the Wick's theorem given by Eq. \ref{eq:HarkThermPerturbEvaluationOfPerturbationExpansionEq_2}, by writing the resulting Green's functions in terms of their Fourier series given in Eq. \ref{eq:HarkThermPerturbEvaluationOfPerturbationExpansionEq_6}, by using the result
\begin{equation} 
\int^{\beta}_{0}d\tau e^{i \omega_{n} \tau} = \beta \delta_{n0},
\label{eq:HarkThermPerturbEvaluationOfPerturbationExpansionEq_9}
\end{equation} 
and then, the resulting summations are simplified by applying the residue theorem.

All the results of this work are written in terms of the coefficients $V\left(\bar{\lambda}_{n}\right)$ (Eq. \ref{eq:HarkThermPerturbLatticeHamiltonianAnharmonicSecondQuantEq_2}, quantities within harmonic approximation and $n$th-order IFCs are needed), $V_{\bar{\mu}_{m} \bar{\nu}_{m}} \left(\bar{\lambda}_{n}\right)$ (Eq. \ref{eq:HarkThermPerturbCrystalHamiltonianSecondQuantStrainedCoefficientsEq_1}, quantities within harmonic approximation and $\left(n + m\right)$th-order IFCs are needed) and Bose-Einstein distribution functions $\bar{n}_{\lambda}$ (Eq. \ref{eq:HarkThermPerturbCrystalHarmonicHelmholtzFreeEnergy_2}). Some terms to be represented may vanish for crystals of some particular kind, for example, due to the absence of an internal strain. However, in the present work all terms are just listed and symmetry considerations are neglected. In the actual calculations, these terms should vanish for some crystal structures if the calculations are based on IFCs with correct symmetry imposed on them.

\section{First-order results}
\label{cha:FirstOrderResults}
To first-order ($h=1$), one may write by using Eq. \ref{eq:HarkThermPerturbCrystalCanonicalPartitionFunctionPertExpEq_7}
\begin{equation} 
\left\langle   \hat{S}\left(\beta\right)  \right\rangle_{0,c,h=1}  = - \int^{\beta}_{0}d\tau_{1} \left\langle  \mathcal{T}\left\{ \hat{H}_{a}\left(\tau_{1}\right) + \hat{H}_{s}\left(\tau_{1}\right) \right\}  \right\rangle_{0,c}. 
\label{eq:HarkThermPerturbFirstOrderResultsEq_1}
\end{equation} 
By using the procedure described in Sec. \ref{cha:EvaluationOfPerturbationExpansion}, one may write (see Appendix \ref{CondensedNotation} for the notation)
\begin{eqnarray} 
\left\langle \hat{S}\left(\beta\right) \right\rangle_{0,c,h=1}  =&& -  \beta \sum_{n= 2 l } \sum_{\bar{\lambda}_{n/2}} V\left(\bar{\lambda}_{n/2};-\bar{\lambda}_{n/2}\right)  \xi^{\left(1\right)}_{n/2}  \nonumber \\
&&-  \beta  \sum_{n = 2 m' } \sum_{\bar{\lambda}_{n/2}}   \sum_{m=1} \frac{1}{m!} \sum_{\bar{\mu}_{m}} \sum_{\bar{\nu}_{m}} \bar{u}_{\mu_{m} \nu_{m}}   \nonumber \\
&&\times V_{\bar{\mu}_{m}\bar{\nu}_{m}}\left(\bar{\lambda}_{n/2};-\bar{\lambda}_{n/2}\right)  \xi^{\left(1\right)}_{n/2}  \nonumber \\
&&-  \beta \sum_{m=1} \frac{1}{m!} \sum_{\bar{\mu}_{m}} \sum_{\bar{\nu}_{m}} V_{\bar{\mu}_{m}\bar{\nu}_{m}} \bar{u}_{\mu_{m} \nu_{m}}, \nonumber \\
l =&& 2,3,\ldots, \quad m'=1,2,\ldots,
\label{eq:HarkThermPerturbFirstOrderResultsEq_2}
\end{eqnarray}
where the notation for $\xi^{\left(1\right)}_{n/2}$ is given by Eq. \ref{eq:HarkThermPerturbFirstOrderResultsEq_3} and the short hand notation for the summation is shown in Eq. \ref{eq:HarkThermPerturbFirstOrderResultsEq_3_1}.

\begin{figure}
\includegraphics[]{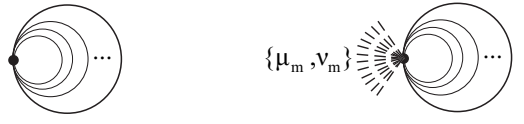}
\caption{The diagrams corresponding to Eq. \ref{eq:HarkThermPerturbFirstOrderResultsEq_2}.} 
\label{fig:FirstOrderStrainedSEnsembleDiagrams}
\end{figure}

Diagrams corresponding to Eq. \ref{eq:HarkThermPerturbFirstOrderResultsEq_2} are shown in Fig. \ref{fig:FirstOrderStrainedSEnsembleDiagrams}. In these diagrams, a dot encountered with $n$ lines refers to the interaction coefficient $V\left(\bar{\lambda}_{n}\right)$ (Eq. \ref{eq:HarkThermPerturbLatticeHamiltonianAnharmonicSecondQuantEq_2}). Also, a dot encountered with $n$ lines and a fan of dashed lines in conjunction with the symbol $\left\{\mu_{m},\nu_{m}\right\}$ (for instance, second diagram in Fig. \ref{fig:FirstOrderStrainedSEnsembleDiagrams}), refers to the strained interaction coefficient $V_{\bar{\mu}_{m} \bar{\nu}_{m}} \left(\bar{\lambda}_{n}\right)$ (Eq. \ref{eq:HarkThermPerturbCrystalHamiltonianSecondQuantStrainedCoefficientsEq_1}).
As discussed in Sec. \ref{cha:HamiltonianForDeformedLattice}, the perturbation expansion for a wider class of macroscopical parameters has the same form. The second diagram from the left in Fig. \ref{fig:FirstOrderStrainedSEnsembleDiagrams} could be written for any set of parameters given in Eq. \ref{eq:HarkThermPerturbCrystalHamiltonianSecondQuantStrainedCoefficientsEq_3}, by just using a different notation for the vertex (some other symbol than a fan of dashed lines). Similar argument can be used for the second and third-order terms discussed in Secs. \ref{cha:SecondOrderResults} and \ref{cha:ThirdOrderResults}. Such an expansion in the electric field components have been considered in Ref. \cite{Cowley-TheLatticeDynamicsOfAnAnharmonicCrystal-1963}, where some of the lowest-order terms were taken into account. For instance, consider the Hamiltonian $\hat{H}^{E}_{p}$ given by Eq. \ref{eq:HarkThermPerturbCrystalHamiltonianSecondQuantStrainedCoefficientsEq_4}. In this case ($\hat{H}_{s} \rightarrow \hat{H}^{E}_{p}$), the first term on the right hand side of Eq. \ref{eq:HarkThermPerturbFirstOrderResultsEq_2} is the same in both cases and the remaining terms can be written as
\begin{equation} 
- \beta  \sum_{n = 2 m' } \sum_{\bar{\lambda}_{n/2}}   \sum_{m=1} \frac{1}{m!} \sum_{\bar{\mu}_{m}} \bar{E}_{\mu_{m}}  g^{\bar{\mu}_{m}}_{\bar{\lambda}_{n/2};-\bar{\lambda}_{n/2}}  \xi^{\left(1\right)}_{n/2}, 
\label{eq:HarkThermPerturbFirstOrderResultsEq_2_2}
\end{equation}
where $m'=0,1,2,\ldots$ and $g^{\bar{\mu}_{m}}_{\bar{\lambda}_{0/2};-\bar{\lambda}_{0/2}} \equiv g^{\bar{\mu}_{m}}$. The diagram for Eq. \ref{eq:HarkThermPerturbFirstOrderResultsEq_2_2} is the same than depicted in Fig. \ref{fig:FirstOrderStrainedSEnsembleDiagrams}, except that one should use a different symbol for the coefficients $g^{\bar{\mu}_{m}}_{\bar{\lambda}_{n/2};-\bar{\lambda}_{n/2}}$.

The first term on the right hand side of Eq. \ref{eq:HarkThermPerturbFirstOrderResultsEq_2} for $n=4$ is the same than have been obtained earlier, for example, in Refs. \cite{Cowley-TheLatticeDynamicsOfAnAnharmonicCrystal-1963,Shukla-AnharmonicFreeEnergy-PhysRevB.3.4055-1971,Barron-DynamicalPropertiesOfSolids-1974} (in Ref. \cite{Shukla-AnharmonicFreeEnergy-PhysRevB.3.4055-1971}, also the case $n=6$ is obtained). The general diagrammatic (but not algebraic) form of this term (for any $n$) have been obtained in Ref. \cite{Horner-DynamicalPropertiesOfSolids-1974}. Terms of Eq. \ref{eq:HarkThermPerturbFirstOrderResultsEq_2} with the strained coefficients are considered in Ref. \cite{Barron-DynamicalPropertiesOfSolids-1974} up to $n=6$ with $m=1$ (stress) and up to $n=4$ with $m = 2$ (second-order elastic constants). The algebraic expressions for the lowest-order case ($n=2$) with $m=1,2$ were shown. The present result provides the diagram and corresponding algebraic expression for arbitrary $n$ and $m$. By using Eqs. \ref{eq:HarkThermPerturbCrystalCanonicalPartitionFunctionPertExpEq_15} and \ref{eq:HarkThermPerturbFirstOrderResultsEq_2}, an explicit algebraic expression is given for the adiabatic elastic constants (Sec. \ref{cha:FirstOrderResultsAdiabaticElasticConstantsAndStress}) such that the first-order contribution to the internal energy and entropy can be calculated for arbitrary $n$ and $m$. Moreover, these results can be used to calculate the first-order contribution to the CTE for arbitrary $n$ (Sec. \ref{cha:FirstOrderResultsThermalExpansion}).

\subsection{First-order results: free energy, internal energy, heat capacity and entropy}
\label{cha:FirstOrderResultsFreeEnergyInternalEnergyHeatCapacityAndEntropy}
By using Eqs. \ref{eq:HarkThermPerturbCrystalCanonicalPartitionFunctionPertExpEq_10} and \ref{eq:HarkThermPerturbFirstOrderResultsEq_2}
and the elastic constants given in Sec. \ref{cha:FirstOrderResultsIsothermalElasticConstantsAndStress}, one may write for the free energy (Eq. \ref{eq:HarkThermPerturbFirstOrderResultsIsotElConstStressEq_1})
\begin{eqnarray}  
\tilde{F}^{\left(1\right)}_{A}  =&&  \sum_{n= 2 m' } \sum_{\bar{\lambda}_{n/2}} V\left(\bar{\lambda}_{n/2};-\bar{\lambda}_{n/2}\right) \xi^{\left(1\right)}_{n/2} \nonumber \\
&&+ \sum_{m=1} \sum_{\bar{\mu}_{m}} \sum_{\bar{\nu}_{m}} \tilde{c}^{T\left(1\right)}_{ \bar{\mu}_{m} \bar{\nu}_{m} } \bar{u}_{\mu_{m} \nu_{m}}  \nonumber \\
m' =&& 2,3,\ldots \,.
\label{eq:HarkThermPerturbFirstOrderResultsFreeEnergyEq_2}
\end{eqnarray} 
One may transform $\tilde{F}^{\left(1\right)}_{A} \rightarrow F^{\left(1\right)}_{A}$ in Eq. \ref{eq:HarkThermPerturbFirstOrderResultsFreeEnergyEq_2} for terms with $m=2$, by using Eqs. \ref{eq:HarkThermPerturbInternalEnergyExpansionStrainTemperatureEq_7}-\ref{eq:HarkThermPerturbInternalEnergyExpansionStrainTemperatureEq_9}. The same procedure can be made for the other quantities considered below.

By using Eqs. \ref{eq:HarkThermPerturbCrystalCanonicalPartitionFunctionPertExpEq_11} and \ref{eq:HarkThermPerturbFirstOrderResultsEq_2}
and the adiabatic elastic constants given in Sec. \ref{cha:FirstOrderResultsAdiabaticElasticConstantsAndStress}, the first-order contribution to the internal energy can be written as (Eq. \ref{eq:HarkThermPerturbFirstOrderResultsAdiabaticElConstStressEq_1})
\begin{eqnarray} 
\tilde{U}^{\left(1\right)}_{A} =&& \sum_{n= 2 m' } \sum_{\bar{\lambda}_{n/2}} V\left(\bar{\lambda}_{n/2};-\bar{\lambda}_{n/2}\right) \left(  \xi^{\left(1\right)}_{n/2} -  \beta \hbar \xi^{\left(2\right)}_{n/2} \right) \nonumber \\
&&+ \sum_{m=1} \sum_{\bar{\mu}_{m}} \sum_{\bar{\nu}_{m}} \tilde{c}^{A\left(1\right)}_{\bar{\mu}_{m} \bar{\nu}_{m} }  \bar{u}_{\mu_{m} \nu_{m}}, \nonumber \\
 m' =&& 2,3,\ldots,
\label{eq:HarkThermPerturbFirstOrderResultsInternalEnergyEq_2}
\end{eqnarray} 
where $\xi^{\left(2\right)}_{n/2}$ is given by Eq. \ref{eq:HarkThermPerturbFirstOrderResultsEq_3}.

From Eqs. \ref{eq:HarkThermPerturbCrystalCanonicalPartitionFunctionPertExpEq_13} and \ref{eq:HarkThermPerturbFirstOrderResultsEq_2}, it follows that the first-order heat capacity at constant strain may be written as
\begin{eqnarray} 
\tilde{C}^{\left(1\right)}_{A,\mu}  =&& 2  \frac{\beta \hbar }{T}   \sum_{n= 2 m } \sum_{\bar{\lambda}_{n/2}} \left( \xi^{\left(2\right)}_{n/2} -\beta \hbar \xi^{\left(3\right)}_{n/2} \right) V\left(\bar{\lambda}_{n/2};-\bar{\lambda}_{n/2}\right)  \nonumber \\
&&+ 2  \frac{\beta \hbar }{T}   \sum_{n = 2 m' } \sum_{\bar{\lambda}_{n/2}}  \sum_{m=1} \frac{1}{m!} \sum_{\bar{\mu}_{m}} \sum_{\bar{\nu}_{m}} \nonumber \\
&&\times V_{\bar{\mu}_{m} \bar{\nu}_{m}}\left(\bar{\lambda}_{n/2};-\bar{\lambda}_{n/2}\right) \nonumber \\
&&\times\left( \xi^{\left(2\right)}_{n/2} -\beta \hbar \xi^{\left(3\right)}_{n/2} \right) \bar{u}_{\mu_{m} \nu_{m}},
\label{eq:HarkThermPerturbFirstOrderResultsHeatCapacityEq_1}
\end{eqnarray} 
where $\xi^{\left(3\right)}_{n/2}$ is given by Eq. \ref{eq:HarkThermPerturbFirstOrderResultsHeatCapacityEq_2}.

By using Eqs. \ref{eq:HarkThermPerturbCrystalCanonicalPartitionFunctionPertExpEq_12}, \ref{eq:HarkThermPerturbFirstOrderResultsEq_2}
and Eqs. \ref{eq:HarkThermPerturbFirstOrderResultsIsotElConstStressEq_1} and \ref{eq:HarkThermPerturbFirstOrderResultsAdiabaticElConstStressEq_1}, the first-order contribution to the entropy can be written as
\begin{eqnarray} 
\tilde{S}^{\left(1\right)}_{A} =&& - \frac{   \beta \hbar }{T}  \sum_{n= 2 m' } \sum_{\bar{\lambda}_{n/2}} V\left(\bar{\lambda}_{n/2};-\bar{\lambda}_{n/2}\right) \xi^{\left(2\right)}_{n/2} \nonumber \\
&&- \frac{1}{T}  \sum_{m=1} \sum_{\bar{\mu}_{m}} \sum_{\bar{\nu}_{m}} \left(\tilde{c}^{T\left(1\right)}_{ \bar{\mu}_{m} \bar{\nu}_{m} }- \tilde{c}^{A\left(1\right)}_{\bar{\mu}_{m} \bar{\nu}_{m} } \right) \bar{u}_{\mu_{m} \nu_{m}}, \nonumber \\
m' =&& 2,3,\ldots \,.
\label{eq:HarkThermPerturbFirstOrderResultsEntropyEq_2}
\end{eqnarray}
The strain derivative of the second term on the right hand side of Eq. \ref{eq:HarkThermPerturbFirstOrderResultsEntropyEq_2} can be used to calculate the first-order contribution to the CTE (see Sec. \ref{cha:FirstOrderResultsThermalExpansion}).

\subsection{First-order results: isothermal elastic constants and stress}
\label{cha:FirstOrderResultsIsothermalElasticConstantsAndStress}
By using Eqs. \ref{eq:HarkThermPerturbCrystalCanonicalPartitionFunctionPertExpEq_14} and \ref{eq:HarkThermPerturbFirstOrderResultsEq_2}
\begin{eqnarray} 
\tilde{c}^{T\left(1\right)}_{\bar{\mu}_{k} \bar{\nu}_{k} } =&& \frac{1}{k!} \sum_{n = 2 m'' } \sum_{\bar{\lambda}_{n/2}} V_{\bar{\mu}_{k} \bar{\nu}_{k}}\left(\bar{\lambda}_{n/2};-\bar{\lambda}_{n/2}\right) \xi^{\left(1\right)}_{n/2} \nonumber \\
&&+ \frac{1}{k!} V_{\bar{\mu}_{k} \bar{\nu}_{k}}, \quad m''=1,2,\ldots \,.
\label{eq:HarkThermPerturbFirstOrderResultsIsotElConstStressEq_1}
\end{eqnarray} 
The isothermal stress can be obtained from Eq. \ref{eq:HarkThermPerturbFirstOrderResultsIsotElConstStressEq_1}, that is
\begin{eqnarray} 
\tilde{\sigma}^{T\left(1\right)}_{\mu_{1} \nu_{1} } =&& \sum_{n = 2 m'' } \sum_{\bar{\lambda}_{n/2}}  V_{\mu_{1}\nu_{1} }\left(\bar{\lambda}_{n/2};-\bar{\lambda}_{n/2}\right) \xi^{\left(1\right)}_{n/2} + V_{\mu_{1}\nu_{1}}, \nonumber \\
&&m''=1,2,\ldots \,,
\label{eq:HarkThermPerturbFirstOrderResultsIsotElConstStressEq_2}
\end{eqnarray} 
and for $m''=1$ (Eq. \ref{eq:HarkThermPerturbFirstOrderResultsEq_3})
\begin{equation} 
\tilde{\sigma}^{T\left(1\right)}_{\mu \nu, n=2 }  =2 \sum_{\lambda} V_{\mu \nu } \left(\lambda;-\lambda\right)\left( \bar{n}_{\lambda} + \frac{1}{2}\right). 
\label{eq:HarkThermPerturbFirstOrderResultsIsotElConstStressEq_3}
\end{equation} 
A comparison of Eqs. \ref{eq:HarkThermPerturbStressHelmholtzEnergyCrystalHarmonicGruneisen} and \ref{eq:HarkThermPerturbFirstOrderResultsIsotElConstStressEq_3} shows that
\begin{equation} 
\gamma_{\mu \nu }\left(\lambda\right) =- \frac{2 V_{\mu \nu }\left(\lambda;-\lambda\right)}{\hbar \omega_{\lambda}}.
\label{eq:HarkThermPerturbFirstOrderResultsIsotElConstStressEq_4}
\end{equation} 
Further corrections to $\gamma_{\mu \nu }\left(\lambda\right)$ can be obtained from higher-order perturbation terms as in Ref. \cite{Barron-DynamicalPropertiesOfSolids-1974}, in which Eq. \ref{eq:HarkThermPerturbFirstOrderResultsIsotElConstStressEq_4} is obtained as well. In a similar way for the elastic constants, for example
\begin{eqnarray} 
\tilde{c}^{T\left(1\right)}_{\mu_{1} \nu_{1} \mu_{2} \nu_{2} ,n=0 } &=&  \frac{1}{2} V_{\mu_{1}\mu_{2} \nu_{1} \nu_{2}}, \nonumber \\
\tilde{c}^{T\left(1\right)}_{\mu_{1} \nu_{1} \mu_{2} \nu_{2} ,n=1 } &=& \sum_{\lambda} V_{\mu_{1}\mu_{2} \nu_{1} \nu_{2}}\left(\lambda;-\lambda\right) \left( \bar{n}_{\lambda} + \frac{1}{2}\right), \nonumber \\
\label{eq:HarkThermPerturbFirstOrderResultsIsotElConstStressEq_5}
\end{eqnarray} 
\begin{eqnarray} 
\tilde{c}^{T\left(1\right)}_{\mu_{1} \nu_{1} \mu_{2} \nu_{2} ,n=2 } =&& 6 \sum_{\lambda_{1}} \sum_{\lambda_{2}} V_{\mu_{1}\mu_{2} \nu_{1}\nu_{2}} \left(\lambda_{1};-\lambda_{1};\lambda_{2};-\lambda_{2}\right)   \nonumber \\
              &&\times   \left( \bar{n}_{\lambda_{1}} + \frac{1}{2}\right) \left( \bar{n}_{\lambda_{2}} + \frac{1}{2}\right),
\label{eq:HarkThermPerturbFirstOrderResultsIsotElConstStressEq_6}
\end{eqnarray} 
and
\begin{eqnarray} 
\tilde{c}^{T\left(1\right)}_{  \bar{\mu}_{3} \bar{\nu}_{3} ,n=0 } &=& \frac{1}{6} V_{  \bar{\mu}_{3} \bar{\nu}_{3} }, \nonumber \\
\tilde{c}^{T\left(1\right)}_{  \bar{\mu}_{3} \bar{\nu}_{3} ,n=1 } &=& \frac{1}{3} \sum_{\lambda}  V_{ \bar{\mu}_{3} \bar{\nu}_{3}} \left(\lambda;-\lambda\right) \left( \bar{n}_{\lambda} + \frac{1}{2}\right).  \nonumber \\
\label{eq:HarkThermPerturbFirstOrderResultsIsotElConstStressEq_7}
\end{eqnarray}
The static term contributions to elastic constants like $V_{\mu_{1}\mu_{2} \nu_{1} \nu_{2}}$ and $V_{ \bar{\mu}_{3} \bar{\nu}_{3}}$ can also be obtained, for instance, by establishing \textit{ab initio} single point total energy calculations for crystals with different strains imposed and then by calculating the elastic constants from the derivatives of the energy (or stress) \cite{Nielsen-AbInitionStress-PhysRevLett.50.697-1983,Golesorkhtabar-ElaStic-20131861-2013}. A comparison of Eqs. \ref{eq:HarkThermPerturbIsothermaElasticConstGruneisen} and \ref{eq:HarkThermPerturbFirstOrderResultsIsotElConstStressEq_5} to Eqs. \ref{eq:HarkThermPerturb3RdOrderIsothermaElasticConstGruneisen} and \ref{eq:HarkThermPerturbFirstOrderResultsIsotElConstStressEq_7} indicates that
\begin{equation} 
\gamma_{ \mu_{1}\nu_{1} \mu_{2}\nu_{2} }\left(\lambda\right) \propto - \frac{ V_{\mu_{1}\mu_{2} \nu_{1} \nu_{2}}\left(\lambda;-\lambda\right) }{\hbar \omega_{\lambda}},
\label{eq:HarkThermPerturbFirstOrderResultsIsotElConstStressEq_8}
\end{equation} 
\begin{equation} 
\gamma_{\mu_{1}\nu_{1} \cdots \mu_{3}\nu_{3}}\left(\lambda\right) \propto - \frac{ V_{\mu_{1}\cdots \mu_{3} \nu_{1}\cdots \nu_{3}} \left(\lambda;-\lambda\right) }{3 \hbar \omega_{\lambda}}.
\label{eq:HarkThermPerturbFirstOrderResultsIsotElConstStressEq_9}
\end{equation} 
Differentiation of Eq. \ref{eq:HarkThermPerturb3RdOrderIsothermaElasticConstGruneisen} with respect to strains furthermore shows that there exists terms of the form
\begin{equation} 
-  \sum_{\lambda} U_{0}\left( \lambda \right)   \gamma_{\bar{\mu}_{k} \bar{\nu}_{k}}\left(\lambda\right),
\label{eq:HarkThermPerturbFirstOrderResultsIsotElConstStressEq_10}
\end{equation} 
which shows that $k$th-order Gr\"{u}neisen parameters $\gamma_{\bar{\mu}_{k} \bar{\nu}_{k} }\left(\lambda\right)$ are proportional to
\begin{equation} 
\gamma_{\bar{\mu}_{k} \bar{\nu}_{k} }\left(\lambda\right) \propto - \frac{2}{k!} \frac{ V_{\bar{\mu}_{k} \bar{\nu}_{k}} \left(\lambda;-\lambda\right) }{\hbar \omega_{\lambda}},
\label{eq:HarkThermPerturbFirstOrderResultsIsotElConstStressEq_11}
\end{equation} 
where the coefficients $V_{ \bar{\mu}_{k} \bar{\nu}_{k} } \left(\lambda;-\lambda\right)$ can be calculated from IFCs of order $k + 2$ (Eq. \ref{eq:HarkThermPerturbCrystalHamiltonianSecondQuantStrainedCoefficientsEq_1}).

The Gr\"{u}neisen parameters are defined in terms of finite strain parameters $\left\{\eta_{\mu_{m} \nu_{m}}\right\}$, while the coefficients $V_{ \bar{\mu}_{k} \bar{\nu}_{k} }\left(\lambda;-\lambda\right)$ are defined in terms of the infinitesimal parameters $\left\{u_{\mu_{m} \nu_{m}}\right\}$, thus, one must use the transformation equations to obtain equality in Eqs. \ref{eq:HarkThermPerturbFirstOrderResultsIsotElConstStressEq_8}, \ref{eq:HarkThermPerturbFirstOrderResultsIsotElConstStressEq_9} and \ref{eq:HarkThermPerturbFirstOrderResultsIsotElConstStressEq_11}. For instance
\begin{equation} 
\gamma_{ \mu_{1}\nu_{1} \mu_{2}\nu_{2} }\left(\lambda\right) = \frac{2 \delta_{\mu_{1} \nu_{1}} V_{\mu_{2} \nu_{2} }\left(\lambda;-\lambda\right) }{\hbar \omega_{\lambda}} - \frac{V_{\mu_{1}\mu_{2} \nu_{1} \nu_{2}}\left(\lambda;-\lambda\right)}{\hbar \omega_{\lambda}}.
\label{eq:HarkThermPerturbFirstOrderResultsIsotElConstStressEq_12}
\end{equation} 
In order to calculate $\gamma_{ \mu_{1}\nu_{1} \mu_{2}\nu_{2} }\left(\lambda\right)$ by using Eq. \ref{eq:HarkThermPerturbFirstOrderResultsIsotElConstStressEq_12}, one needs the third and fourth-order IFCs.

\subsection{First-order results: adiabatic elastic constants and stress}
\label{cha:FirstOrderResultsAdiabaticElasticConstantsAndStress}
By using Eqs. \ref{eq:HarkThermPerturbCrystalCanonicalPartitionFunctionPertExpEq_15} and \ref{eq:HarkThermPerturbFirstOrderResultsEq_2}
\begin{eqnarray} 
\tilde{c}^{A\left(1\right)}_{ \bar{\mu}_{k} \bar{\nu}_{k} } =&& -  \frac{ \beta \hbar }{k!} \sum_{n = 2 m'' } \sum_{\bar{\lambda}_{n/2}} V_{\bar{\mu}_{k} \bar{\nu}_{k}}\left(\bar{\lambda}_{n/2};-\bar{\lambda}_{n/2}\right) \xi^{\left(2\right)}_{n/2}  \nonumber \\
&&+  \frac{1}{k!} \sum_{n = 2 m'' } \sum_{\bar{\lambda}_{n/2}} V_{\bar{\mu}_{k} \bar{\nu}_{k}}\left(\bar{\lambda}_{n/2};-\bar{\lambda}_{n/2}\right) \xi^{\left(1\right)}_{n/2}  \nonumber \\
&&+  \frac{1}{k!} V_{\bar{\mu}_{k} \bar{\nu}_{k}}, \quad m''=1,2,\ldots \,.
\label{eq:HarkThermPerturbFirstOrderResultsAdiabaticElConstStressEq_1}
\end{eqnarray} 
The adiabatic stress is the special case of Eq. \ref{eq:HarkThermPerturbFirstOrderResultsAdiabaticElConstStressEq_1} with $k=1$, that is
\begin{eqnarray} 
\tilde{\sigma}^{A\left(1\right)}_{\mu_{1} \nu_{1} } =&& -   \beta \hbar \sum_{n = 2 m'' } \sum_{\bar{\lambda}_{n/2}} V_{\mu_{1}\nu_{1}}\left(\bar{\lambda}_{n/2};-\bar{\lambda}_{n/2}\right) \xi^{\left(2\right)}_{n/2}  \nonumber \\
&&+ \sum_{n = 2 m'' } \sum_{\bar{\lambda}_{n/2}} V_{\mu_{1} \nu_{1}}\left(\bar{\lambda}_{n/2};-\bar{\lambda}_{n/2}\right) \xi^{\left(1\right)}_{n/2}  \nonumber \\
&&+  V_{\mu_{1} \nu_{1} }, \quad m''=1,2,\ldots,
\label{eq:HarkThermPerturbFirstOrderResultsAdiabaticElConstStressEq_2}
\end{eqnarray} 
and for $m''=1$ (Eq. \ref{eq:HarkThermPerturbFirstOrderResultsEq_3})
\begin{eqnarray} 
\tilde{\sigma}^{A\left(1\right)}_{\mu \nu, m''=1} =&& - 2 \beta \hbar \sum_{\lambda} V_{\mu\nu} \left(\lambda;-\lambda\right) \omega_{\lambda}  \bar{n}_{\lambda} \left(\bar{n}_{\lambda} + 1\right)  \nonumber \\
&&+   2 \sum_{\lambda} V_{\mu \nu} \left(\lambda;-\lambda\right)\left( \bar{n}_{\lambda} + \frac{1}{2}\right). 
\label{eq:HarkThermPerturbFirstOrderResultsAdiabaticElConstStressEq_3}
\end{eqnarray} 
By using Eq. \ref{eq:HarkThermPerturbFirstOrderResultsIsotElConstStressEq_4} and the results of Sec. \ref{cha:ThermodynamicalRelationsForHarmonicAndQuasiHarmonicPhonons}, Eqs. \ref{eq:HarkThermPerturbStressInternalEnergyCrystalHarmonicGruneisen} and \ref{eq:HarkThermPerturbFirstOrderResultsAdiabaticElConstStressEq_3} become equivalent (in Eq. \ref{eq:HarkThermPerturbStressInternalEnergyCrystalHarmonicGruneisen}, the static term $V_{\mu \nu }$ is neglected). These results show that the QHA includes the first-order perturbation in the calculation of stress (Eqs. \ref{eq:HarkThermPerturbFirstOrderResultsIsotElConstStressEq_3} and \ref{eq:HarkThermPerturbFirstOrderResultsAdiabaticElConstStressEq_3}) and the lowest-order processes in $\left\{\lambda_{i}\right\}$.

\subsection{First-order results, thermal expansion}
\label{cha:FirstOrderResultsThermalExpansion}
One may write the CTE as in Eq. \ref{eq:HarkThermPerturbThermalExpansionCoefficientGruneisen_3} and in the present case
\begin{equation} 
\alpha^{\left(1\right)}_{\mu_{1} \nu_{1}} = \frac{1}{T} \sum^{3}_{ \mu_{2}, \nu_{2}=1} s^{T\left(1\right)}_{\mu_{1} \nu_{1} \mu_{2} \nu_{2}}  \left( \sigma^{A\left(1\right)}_{\mu_{2} \nu_{2}} - \sigma^{T\left(1\right)}_{\mu_{2} \nu_{2}} \right),
\label{eq:HarkThermPerturbFirstOrderResultsThermalExpansionEq_1}
\end{equation}
As an example, in the case of cubic crystals, Eq. \ref{eq:HarkThermPerturbFirstOrderResultsThermalExpansionEq_1} can be written as
\begin{equation} 
\alpha_{\mu \mu} = \frac{1}{T} \frac{ \sigma^{A\left(1\right)}_{\mu \mu } - \sigma^{T\left(1\right)}_{ \mu \mu } }{c^{T\left(1\right)}_{1111} + 2 c^{T\left(1\right)}_{1122}}.
\label{eq:HarkThermPerturbFirstOrderResultsThermalExpansionEq_4}
\end{equation} 
The inversion of the second-order compliance tensor $s_{\mu_{1} \nu_{1} \mu_{2} \nu_{2}}$, established here, is considered in Ref. \cite{Nye-PhysicalPropertiesOfCrystals-1985}. Since it can be shown that in the case of cubic crystals \cite{Nye-PhysicalPropertiesOfCrystals-1985} $c_{1111} + 2 c_{1122} > 0$, the NTE occurs only if $\sigma^{A}_{\mu \mu } - \sigma^{T}_{ \mu \mu } < 0$. In Eq. \ref{eq:HarkThermPerturbFirstOrderResultsThermalExpansionEq_1}, the results for $\sigma^{T\left(1\right)}_{\mu_{2} \nu_{2}}, \sigma^{A\left(1\right)}_{\mu_{2} \nu_{2}}$ and $\tilde{c}^{T\left(1\right)}_{\mu_{1} \nu_{1} \mu_{2} \nu_{2}}$ (in order to calculate $\tilde{s}^{T\left(1\right)}_{\mu_{1} \nu_{1} \mu_{2} \nu_{2}}$) are given by Eqs. \ref{eq:HarkThermPerturbFirstOrderResultsIsotElConstStressEq_1}, \ref{eq:HarkThermPerturbFirstOrderResultsIsotElConstStressEq_2} and \ref{eq:HarkThermPerturbFirstOrderResultsAdiabaticElConstStressEq_2}, thus (the notation is given by Eq. \ref{eq:HarkThermPerturbFirstOrderResultsEq_3})
\begin{eqnarray} 
\tilde{c}^{T\left(1\right)}_{\mu_{1} \nu_{1} \mu_{2} \nu_{2} } =&& \frac{1}{2} \sum_{n = 2 m } \sum_{\bar{\lambda}_{n/2}}  V_{\mu_{1}\mu_{2} \nu_{1} \nu_{2}}\left(\bar{\lambda}_{n/2};-\bar{\lambda}_{n/2}\right) \xi^{\left(1\right)}_{n/2} \nonumber \\
&&+ \frac{1}{2} V_{\mu_{1} \mu_{2} \nu_{1} \nu_{2}}, \quad m=1,2,\ldots,
\label{eq:HarkThermPerturbFirstOrderResultsThermalExpansionEq_2}
\end{eqnarray} 
\begin{eqnarray} 
\sigma^{A\left(1\right)}_{\mu_{2} \nu_{2}} - \sigma^{T\left(1\right)}_{\mu_{2} \nu_{2}} =&&  -\beta \hbar \sum_{n = 2 m } \sum_{\bar{\lambda}_{n/2}} V_{\mu_{2} \nu_{2}}\left(\bar{\lambda}_{n/2};-\bar{\lambda}_{n/2}\right)  \xi^{\left(2\right)}_{n/2}, \nonumber \\
&&m=1,2,\ldots,
\label{eq:HarkThermPerturbFirstOrderResultsThermalExpansionEq_3}
\end{eqnarray}
and the transformation $\tilde{c}^{T\left(1\right)}_{\mu_{1} \nu_{1} \mu_{2} \nu_{2}} \rightarrow c^{T\left(1\right)}_{\mu_{1} \nu_{1} \mu_{2} \nu_{2}}$ is given by Eqs. \ref{eq:HarkThermPerturbInternalEnergyExpansionStrainTemperatureEq_7}-\ref{eq:HarkThermPerturbInternalEnergyExpansionStrainTemperatureEq_9}. A comparison of Eqs. \ref{eq:HarkThermPerturbStressDerivativeHelmholtzEnergyCrystalHarmonicGruneisen} and \ref{eq:HarkThermPerturbIsothermaElasticConstGruneisen} to Eqs. \ref{eq:HarkThermPerturbFirstOrderResultsThermalExpansionEq_2} and \ref{eq:HarkThermPerturbFirstOrderResultsThermalExpansionEq_3} shows that in calculating the CTE, QHA takes into account the stress contributions up to first-order, elastic constants up to first-order (both with $n = 2$) and one second-order term, which is also included in the present approach when some of the second-order terms $c^{T\left(2\right)}_{\mu_{1} \nu_{1} \mu_{2} \nu_{2}}$ are taken into account. In Eq. \ref{eq:HarkThermPerturbFirstOrderResultsThermalExpansionEq_3} there is no static term, which indicates that the NTE occurs due to vibrational contribution.

\section{Second-order results}
\label{cha:SecondOrderResults}
To second-order ($h=2$), one may write by using Eq. \ref{eq:HarkThermPerturbCrystalCanonicalPartitionFunctionPertExpEq_7}
\begin{eqnarray} 
\left\langle \hat{S}\left(\beta\right) \right\rangle_{0,c,h=2} = && \left\langle \hat{S}\left(\beta\right) \right\rangle_{0,c,aa} + \left\langle \hat{S}\left(\beta\right) \right\rangle_{0,c,ss} \nonumber \\
&&+ \left\langle \hat{S}\left(\beta\right) \right\rangle_{0,c,sa}, 
\label{eq:HarkThermPerturbSecondOrderResultsEq_1}
\end{eqnarray} 
where
\begin{eqnarray} 
&&\left\langle \hat{S}\left(\beta\right) \right\rangle_{0,c,aa} \nonumber \\
&&= \frac{1}{2} \int^{\beta}_{0}d\tau_{1} \int^{\beta}_{0}d\tau_{2}   \left\langle  \mathcal{T}\left\{ \hat{H}_{a}\left(\tau_{1}\right) \hat{H}_{a}\left(\tau_{2}\right) \right\}  \right\rangle_{0,c},  \nonumber \\
&&\left\langle \hat{S}\left(\beta\right) \right\rangle_{0,c,ss} \nonumber \\
&&= \frac{1}{2} \int^{\beta}_{0}d\tau_{1} \int^{\beta}_{0}d\tau_{2}  \left\langle  \mathcal{T}\left\{ \hat{H}_{s}\left(\tau_{1}\right)  \hat{H}_{s}\left(\tau_{2}\right)  \right\}  \right\rangle_{0,c}, \nonumber \\
&&\left\langle \hat{S}\left(\beta\right) \right\rangle_{0,c,sa} \nonumber \\
&&= \int^{\beta}_{0}d\tau_{1} \int^{\beta}_{0}d\tau_{2}  \left\langle  \mathcal{T}\left\{  \hat{H}_{s}\left(\tau_{2}\right)  \hat{H}_{a}\left(\tau_{1}\right)  \right\}  \right\rangle_{0,c}. 
\label{eq:HarkThermPerturbSecondOrderResultsEq_2}
\end{eqnarray}

\begin{figure*}
\includegraphics[]{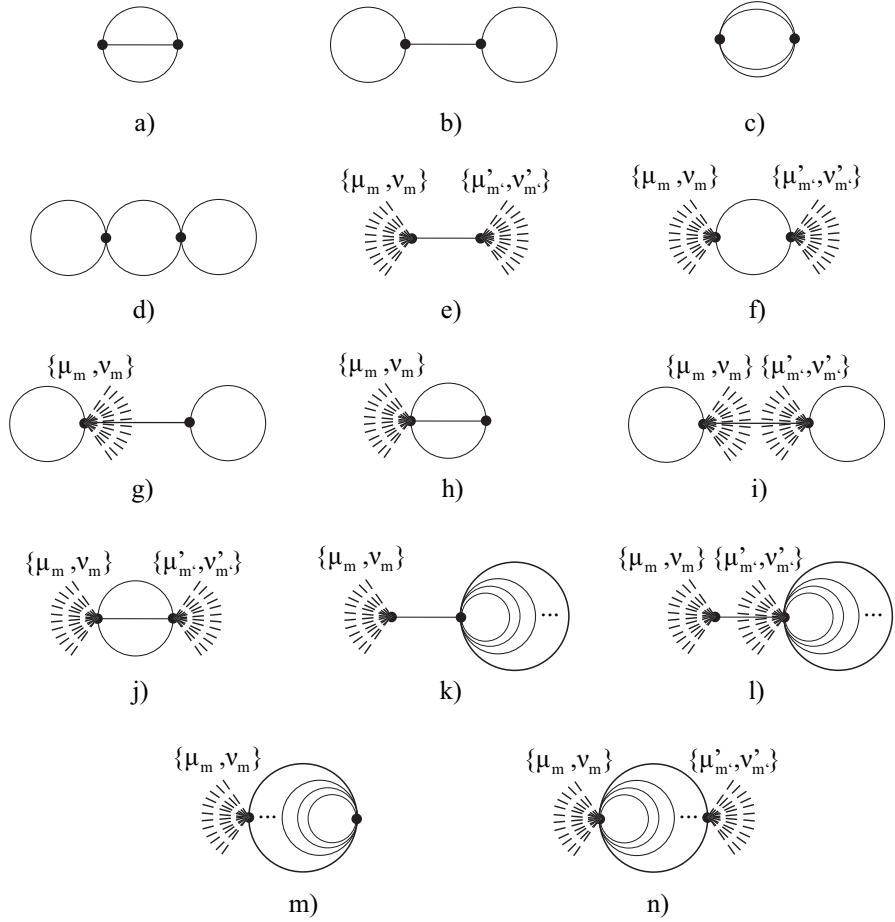}
\caption{Some diagrams corresponding to the equations of this section: a), b) Eq. \ref{eq:HarkThermPerturbSecondOrderResultsFreeEnergyEq_2}, c), d) second-order diagrams for the fourth-order coefficients $V\left(\bar{\lambda}_{4}\right)$ given for example in Ref. \cite{Shukla-AnharmonicFreeEnergy-PhysRevB.3.4055-1971}, e) Eq. \ref{eq:HarkThermPerturbSecondOrderResultsIsothermalElasticConstantsAndStressEq_2}, f) Eq. \ref{eq:HarkThermPerturbSecondOrderResultsIsothermalElasticConstantsAndStressEq_3}, g), h) Eq. \ref{eq:HarkThermPerturbSecondOrderResultsIsothermalElasticConstantsAndStressEq_9}, i), j) Eq. \ref{eq:HarkThermPerturbSecondOrderResultsIsothermalElasticConstantsAndStressEq_4}, k) Eq. \ref{eq:HarkThermPerturbSecondOrderResultsIsothermalElasticConstantsAndStressEq_8} is a special case of this diagram, l) Eqs. \ref{eq:HarkThermPerturbSecondOrderResultsIsothermalElasticConstantsAndStressEq_5} and \ref{eq:HarkThermPerturbSecondOrderResultsIsothermalElasticConstantsAndStressEq_7}, m) Eq. \ref{eq:HarkThermPerturbSecondOrderResultsIsothermalElasticConstantsAndStressEq_10} is a special case of this diagram and n) Eq. \ref{eq:HarkThermPerturbSecondOrderResultsIsothermalElasticConstantsAndStressEq_6}.} 
\label{fig:SecondOrderStrainedSEnsembleDiagrams}
\end{figure*}

Diagrams corresponding to the equations of this section are shown in Fig. \ref{fig:SecondOrderStrainedSEnsembleDiagrams}. 
Several authors have considered, for example, the diagrams a)-d) of Fig. \ref{fig:SecondOrderStrainedSEnsembleDiagrams} (see for instance Ref. \cite{Shukla-AnharmonicFreeEnergy-PhysRevB.3.4055-1971}). Some of the strained diagrams in Fig. \ref{fig:SecondOrderStrainedSEnsembleDiagrams} have been considered in Refs. \cite{Cowley-TheLatticeDynamicsOfAnAnharmonicCrystal-1963} and \cite{Barron-DynamicalPropertiesOfSolids-1974}. In the case of stress, special cases ($m=1$) of the diagrams h), k) ($n=1,n'=3$) and m) ($n=1, n'=4$) of Fig. \ref{fig:SecondOrderStrainedSEnsembleDiagrams} were considered in Ref. \cite{Barron-DynamicalPropertiesOfSolids-1974}, while the algebraic expression was given for the diagram k) with $n=1, n'=3$. For the second-order isothermal elastic constants ($m+m'=2$), the lowest-order special cases of the diagrams e), f), h), i), j), l), m) and n) of Fig. \ref{fig:SecondOrderStrainedSEnsembleDiagrams} were shown in Ref. \cite{Barron-DynamicalPropertiesOfSolids-1974}, while the algebraic expressions were given for diagrams e) and f) with $n=n'=1$ and $n=n'=2$, respectively. To the author's knowledge, the diagrams and corresponding algebraic expressions for the higher-order elastic constants ($m+m' \geq 3$) have not been given before. The present second-order results are valid for arbitrary $m$ and for various $n$ and $n'$ extending the previous results. The corresponding expressions for the adiabatic quantities are also represented.

By using Eq. \ref{eq:HarkThermPerturbCrystalCanonicalPartitionFunctionPertExpEq_14} and the result for Eq. \ref{eq:HarkThermPerturbSecondOrderResultsEq_2}, one may approximate
\begin{equation} 
\tilde{c}^{T\left(2\right)}_{ \bar{\mu}_{k} \bar{\nu}_{k} }  \approx \sum^{9}_{i=1} \tilde{c}^{T\left(2\right),i}_{ \bar{\mu}_{k} \bar{\nu}_{k} }, \quad c^{A\left(2\right)}_{ \bar{\mu}_{k} \bar{\nu}_{k} }  \approx \sum^{7}_{i=1} c^{A\left(2\right),i}_{ \bar{\mu}_{k} \bar{\nu}_{k} }.
\label{eq:HarkThermPerturbSecondOrderResultsIsothermalElasticConstantsAndStressEq_1} 
\end{equation}
The different contributions to the isothermal elastic constants subsumed into Eq. \ref{eq:HarkThermPerturbSecondOrderResultsIsothermalElasticConstantsAndStressEq_1} are given in Appendix \ref{ExpressionsForSecondOrderElasticConstants} by Eqs. \ref{eq:HarkThermPerturbSecondOrderResultsIsothermalElasticConstantsAndStressEq_2}-\ref{eq:HarkThermPerturbSecondOrderResultsIsothermalElasticConstantsAndStressEq_10} and these terms represented in diagrammatic form are depicted in Fig. \ref{fig:SecondOrderStrainedSEnsembleDiagrams}. The adiabatic elastic constants in Eq. \ref{eq:HarkThermPerturbSecondOrderResultsIsothermalElasticConstantsAndStressEq_1}, are given by Eqs. \ref{eq:HarkThermPerturbSecondOrderResultsAdiabaticElasticConstantsAndStressEq_2}-\ref{eq:HarkThermPerturbSecondOrderResultsAdiabaticElasticConstantsAndStressEq_8}. Further, the second-order contribution to the adiabatic stress $\sigma^{A\left(2\right)}_{\mu_{1} \nu_{1} }$ can be obtained from Eqs. \ref{eq:HarkThermPerturbSecondOrderResultsAdiabaticElasticConstantsAndStressEq_6}-\ref{eq:HarkThermPerturbSecondOrderResultsAdiabaticElasticConstantsAndStressEq_8}.

After the calculation, the second-order contribution to the Helmholtz free energy can be approximated as
\begin{eqnarray} 
\tilde{F}^{\left(2\right)}_{A} \approx&& F^{\left(2\right)}_{aa,n=n'=3} \nonumber \\
&&+ \sum_{m=1} \sum_{\bar{\mu}_{m}} \sum_{\bar{\nu}_{m}} \sum^{9}_{i=7} \tilde{c}^{T\left(2\right),i}_{\bar{\mu}_{m} \bar{\nu}_{m} } \bar{u}_{\mu_{m} \nu_{m}}   \nonumber \\
&&+ \sum_{m=1} \sum_{\bar{\mu}_{m}} \sum_{\bar{\nu}_{m}} \sum_{m'=1} \sum_{\bar{\mu}'_{m'}} \sum_{\bar{\nu}'_{m'}} \sum^{6}_{i=1}  \nonumber \\
&&\times \tilde{c}^{T\left(2\right),i}_{\bar{\mu}_{m} \bar{\nu}_{m} \bar{\mu}'_{m'} \bar{\nu}'_{m'}} \bar{u}_{\mu_{m} \nu_{m}}  \bar{u}'_{\mu'_{m'} \nu'_{m'}}, 
\label{eq:HarkThermPerturbSecondOrderResultsFreeEnergyEq_1} 
\end{eqnarray}
where 
\begin{eqnarray} 
F^{\left(2\right)}_{aa,n=n'=3} =&& -  \frac{ 36 }{ \hbar } \sum_{\lambda,\lambda',\lambda''} V\left(\lambda;-\lambda;\lambda'\right) \nonumber \\
&&\times V\left(-\lambda';\lambda'';-\lambda''\right)  \frac{ \left( \bar{n}_{\lambda} + \frac{1}{2}\right) \left( \bar{n}_{\lambda''} + \frac{1}{2}\right) }{ \omega_{\lambda'}}  \nonumber \\
&&- \frac{6 }{  \hbar }  \sum_{\lambda,\lambda',\lambda''} \left|V\left(\lambda;\lambda';\lambda''\right)\right|^{2}  G^{\left(3\right)}\left(\lambda;\lambda';\lambda''\right). \nonumber \\
\label{eq:HarkThermPerturbSecondOrderResultsFreeEnergyEq_2} 
\end{eqnarray} 
The isothermal elastic constants in Eq. \ref{eq:HarkThermPerturbSecondOrderResultsFreeEnergyEq_1} are given by Eqs. \ref{eq:HarkThermPerturbSecondOrderResultsIsothermalElasticConstantsAndStressEq_2}-\ref{eq:HarkThermPerturbSecondOrderResultsIsothermalElasticConstantsAndStressEq_10} and $G^{\left(3\right)}\left(\lambda;\lambda';\lambda''\right)$ in Eq. \ref{eq:HarkThermPerturbSecondOrderResultsFreeEnergyEq_2} is given by Eq. \ref{eq:HarkThermPerturbSecondOrderResultsFreeEnergyEq_3}. The result given by Eq. \ref{eq:HarkThermPerturbSecondOrderResultsFreeEnergyEq_2} is the same than have been obtained earlier, for example, in Refs. \cite{Cowley-TheLatticeDynamicsOfAnAnharmonicCrystal-1963,Cowley-AnharmonicCrystals-1968,Shukla-AnharmonicFreeEnergy-PhysRevB.3.4055-1971,Wallace-ThermodynamicsOfCrystals-1972,Barron-DynamicalPropertiesOfSolids-1974}.

The second-order contribution to the internal energy can be approximated as $\tilde{U}^{\left(2\right)}_{A} \approx U^{\left(2\right)}_{aa,n=n'=3} + U^{\left(2\right)}_{s}$, where the strained contribution $U^{\left(2\right)}_{s}$ is the same than in Eq. \ref{eq:HarkThermPerturbSecondOrderResultsFreeEnergyEq_2}, but the isothermal elastic constants are replaced by the adiabatic ones and the static contribution is given by Eq. \ref{eq:ExpressionsForThermalQuantitiesEq_1}. The adiabatic elastic constants, included in the expression of $U^{\left(2\right)}_{s}$, are given by Eqs. \ref{eq:HarkThermPerturbSecondOrderResultsAdiabaticElasticConstantsAndStressEq_2}-\ref{eq:HarkThermPerturbSecondOrderResultsAdiabaticElasticConstantsAndStressEq_8}. The second-order results for heat capacity are not explicitly shown, but can be obtained from the results of internal energy by differentiating with respect to $T$. By using Eqs. \ref{eq:HarkThermPerturbCrystalCanonicalPartitionFunctionPertExpEq_12}, \ref{eq:HarkThermPerturbSecondOrderResultsEq_2}
and Eq. \ref{eq:HarkThermPerturbSecondOrderResultsIsothermalElasticConstantsAndStressEq_1}, the second-order contribution to the entropy can be written as
\begin{eqnarray} 
\tilde{S}&&^{\left(2\right)}_{A} \approx S^{\left(2\right)}_{aa,n=n'=3}   \nonumber \\
&&- \frac{1}{T}  \sum_{m=1} \sum_{\bar{\mu}_{m}} \sum_{\bar{\nu}_{m}} \sum^{9}_{i=7} \left(\tilde{c}^{T\left(2\right),i}_{  \bar{\mu}_{m} \bar{\nu}_{m} }- \tilde{c}^{A\left(2\right),i}_{  \bar{\mu}_{m} \bar{\nu}_{m} } \right) \bar{u}_{\mu_{m} \nu_{m}}\nonumber \\
&&- \frac{1}{T}  \sum_{m=1} \sum_{\bar{\mu}_{m}} \sum_{\bar{\nu}_{m}} \sum_{m'=1} \sum_{\bar{\mu}'_{m'}} \sum_{\bar{\nu}'_{m'}} \sum^{6}_{i=1} \nonumber \\
&&\times \left( \tilde{c}^{T\left(2\right),i}_{ \bar{\mu}_{m} \bar{\nu}_{m}  \bar{\mu}'_{m'} \bar{\nu}'_{m'} } - \tilde{c}^{A\left(2\right),i}_{ \bar{\mu}_{m} \bar{\nu}_{m}  \bar{\mu}'_{m'} \bar{\nu}'_{m'} } \right) \bar{u}_{\mu_{m} \nu_{m}}  \bar{u}'_{\mu'_{m'} \nu'_{m'}}, \nonumber \\
\label{eq:HarkThermPerturbSecondOrderResultsEntropyEq_1}
\end{eqnarray}
where the static contribution is given by Eq. \ref{eq:ExpressionsForThermalQuantitiesEq_2}

The second-order contribution to the CTE can be obtained by using Eq. \ref{eq:HarkThermPerturbFirstOrderResultsThermalExpansionEq_1}. The results for $\sigma^{T\left(2\right)}_{\gamma \delta}, \sigma^{A\left(2\right)}_{\gamma \delta}$ and $c^{T\left(2\right)}_{\mu \nu \gamma \delta}$ (in order to calculate $s^{T\left(2\right)}_{\mu \nu \gamma \delta}$) are given by Eq. \ref{eq:HarkThermPerturbSecondOrderResultsIsothermalElasticConstantsAndStressEq_1} and the contributions are listed in Appendix \ref{ExpressionsForSecondOrderElasticConstants}. The contribution to the second-order isothermal elastic constants can be obtained from Eqs. \ref{eq:HarkThermPerturbSecondOrderResultsIsothermalElasticConstantsAndStressEq_2}-\ref{eq:HarkThermPerturbSecondOrderResultsIsothermalElasticConstantsAndStressEq_10}, while the contribution to the isothermal and adiabatic stress can be obtained from Eqs. \ref{eq:HarkThermPerturbSecondOrderResultsIsothermalElasticConstantsAndStressEq_8}-\ref{eq:HarkThermPerturbSecondOrderResultsIsothermalElasticConstantsAndStressEq_10} and \ref{eq:HarkThermPerturbSecondOrderResultsAdiabaticElasticConstantsAndStressEq_6}-\ref{eq:HarkThermPerturbSecondOrderResultsAdiabaticElasticConstantsAndStressEq_8}, respectively.

\section{Third-order results}
\label{cha:ThirdOrderResults}
\begin{figure*}
\includegraphics[]{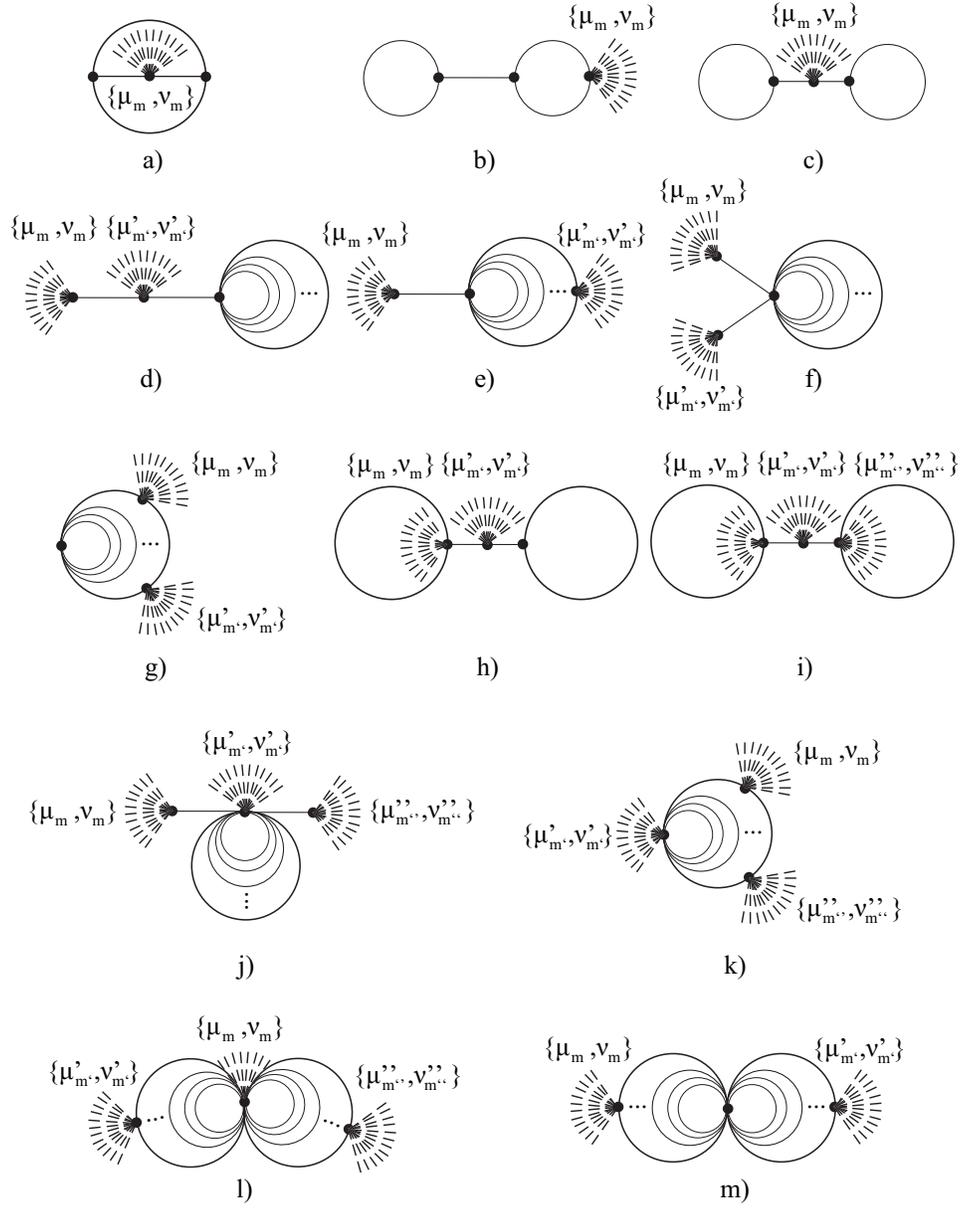}
\caption{Some diagrams corresponding to equations of this section: a), b), c) Eq. \ref{eq:HarkThermPerturbThirdOrderResultsIsothermalElasticConstantsAndStressAppendixEq_1}, d), e) Eqs. \ref{eq:HarkThermPerturbThirdOrderResultsIsothermalElasticConstantsAndStressAppendixEq_4} and \ref{eq:HarkThermPerturbThirdOrderResultsIsothermalElasticConstantsAndStressAppendixEq_7}, f) Eqs. \ref{eq:HarkThermPerturbThirdOrderResultsIsothermalElasticConstantsAndStressAppendixEq_2} and \ref{eq:HarkThermPerturbThirdOrderResultsIsothermalElasticConstantsAndStressAppendixEq_5}, g) Eqs. \ref{eq:HarkThermPerturbThirdOrderResultsIsothermalElasticConstantsAndStressAppendixEq_3} and \ref{eq:HarkThermPerturbThirdOrderResultsIsothermalElasticConstantsAndStressAppendixEq_6} , h), i) Eq. \ref{eq:HarkThermPerturbThirdOrderResultsIsothermalElasticConstantsAndStressAppendixEq_1} by replacing one or two of the coefficients $V\left(\lambda;\lambda';\lambda''\right)$ with the corresponding strained coefficients, j) Eqs. \ref{eq:HarkThermPerturbThirdOrderResultsIsothermalElasticConstantsAndStressAppendixEq_8} and \ref{eq:HarkThermPerturbThirdOrderResultsIsothermalElasticConstantsAndStressAppendixEq_11}, k) Eqs. \ref{eq:HarkThermPerturbThirdOrderResultsIsothermalElasticConstantsAndStressAppendixEq_9} and \ref{eq:HarkThermPerturbThirdOrderResultsIsothermalElasticConstantsAndStressAppendixEq_10}, l) Eq. \ref{eq:HarkThermPerturbThirdOrderResultsIsothermalElasticConstantsAndStressAppendixEq_14} and m) Eq. \ref{eq:HarkThermPerturbThirdOrderResultsIsothermalElasticConstantsAndStressAppendixEq_14} by replacing one of the coefficients with an unstrained one.} 
\label{fig:ThirdOrderStrainedSEnsembleDiagrams}
\end{figure*}

By using Eq. \ref{eq:HarkThermPerturbCrystalCanonicalPartitionFunctionPertExpEq_7}, the third-order terms ($h=3$) can be written as 
\begin{eqnarray} 
&&\left\langle \hat{S}\left(\beta\right) \right\rangle_{0,c,h=3} = -\frac{1}{6} \int^{\beta}_{0}d\tau_{1} \int^{\beta}_{0}d\tau_{2} \int^{\beta}_{0}d\tau_{3} \nonumber \\
&&\times \left\langle  \mathcal{T}\left\{ \hat{H}_{a}\left(\tau_{1}\right) \hat{H}_{a}\left(\tau_{2}\right)  \hat{H}_{a}\left(\tau_{3}\right) \right\}  \right\rangle_{0,c}  \nonumber \\
&&- \frac{1}{2} \int^{\beta}_{0}d\tau_{1} \int^{\beta}_{0}d\tau_{2} \int^{\beta}_{0}d\tau_{3}  \nonumber \\
&&\times \left\langle  \mathcal{T}\left\{ \hat{H}_{a}\left(\tau_{1}\right) \hat{H}_{a}\left(\tau_{2}\right)  \hat{H}_{s}\left(\tau_{3}\right)  \right\}  \right\rangle_{0,c} \nonumber \\
&&- \frac{1}{2} \int^{\beta}_{0}d\tau_{1} \int^{\beta}_{0}d\tau_{2} \int^{\beta}_{0}d\tau_{3} \nonumber \\
&&\times  \left\langle  \mathcal{T}\left\{  \hat{H}_{s}\left(\tau_{1}\right)  \hat{H}_{s}\left(\tau_{2}\right) \hat{H}_{a}\left(\tau_{3}\right)  \right\}  \right\rangle_{0,c} \nonumber \\
&&- \frac{1}{6} \int^{\beta}_{0}d\tau_{1} \int^{\beta}_{0}d\tau_{2} \int^{\beta}_{0}d\tau_{3}  \nonumber \\
&&\times \left\langle  \mathcal{T}\left\{ \hat{H}_{s}\left(\tau_{1}\right) \hat{H}_{s}\left(\tau_{2}\right)  \hat{H}_{s}\left(\tau_{3}\right)  \right\}  \right\rangle_{0,c}.
\label{eq:HarkThermPerturbThirdOrderResultsEq_1}
\end{eqnarray} 
Diagrams corresponding to the equations of this section are shown in Fig. \ref{fig:ThirdOrderStrainedSEnsembleDiagrams}. Some special cases of the third-order diagrams shown in Fig. \ref{fig:ThirdOrderStrainedSEnsembleDiagrams} (to lowest-order in $n,n',n''$) for the stress ($m=1$) and second-order elastic constants ($m=2$) were considered in Ref. \cite{Barron-DynamicalPropertiesOfSolids-1974}, but no algebraic expressions were given. In particular, the highest-order in IFCs considered by Barron and Klein \cite{Barron-DynamicalPropertiesOfSolids-1974} is $n=4$. Also, one third-order diagram contributing to the third-order elastic constants was shown, namely, the diagram k) of Fig. \ref{fig:ThirdOrderStrainedSEnsembleDiagrams} with $n=n'=n''=2$. In Ref. \cite{Cowley-TheLatticeDynamicsOfAnAnharmonicCrystal-1963}, the special case $m=m'=1, n=n'=1, n''=3$, of the diagram e) of Fig. \ref{fig:ThirdOrderStrainedSEnsembleDiagrams} and the corresponding algebraic expression was considered.
These previous results are extended in the present work by providing diagrams and expressions for arbitrary $m,m',m''$ and to various orders in $n,n',n''$.

By using Eqs. \ref{eq:HarkThermPerturbCrystalCanonicalPartitionFunctionPertExpEq_14} and \ref{eq:HarkThermPerturbThirdOrderResultsEq_1}, one may approximate
\begin{eqnarray} 
\tilde{c}^{T\left(3\right)}_{ \bar{\mu}_{j} \bar{\nu}_{j} }  &\approx& \sum^{14}_{i=1} \tilde{c}^{T\left(3\right),i}_{ \bar{\mu}_{j} \bar{\nu}_{j} }, \nonumber \\ \tilde{c}^{A\left(3\right)}_{ \bar{\mu}_{j} \bar{\nu}_{j} } &\approx&  \sum^{14}_{i=1} \left[\tilde{c}^{T\left(3\right),i}_{ \bar{\mu}_{j} \bar{\nu}_{j} } + \beta \frac{\partial}{\partial{\beta}}  \tilde{c}^{T\left(3\right),i}_{ \bar{\mu}_{j} \bar{\nu}_{j} }\right].
\label{eq:HarkThermPerturbThirdOrderResultsIsothermalElasticConstantsAndStressEq_1} 
\end{eqnarray}
The different terms of Eq. \ref{eq:HarkThermPerturbThirdOrderResultsIsothermalElasticConstantsAndStressEq_1} are given in Appendix \ref{ExpressionsForThirdOrderElasticConstants} by Eqs. \ref{eq:HarkThermPerturbThirdOrderResultsIsothermalElasticConstantsAndStressAppendixEq_1}-\ref{eq:HarkThermPerturbThirdOrderResultsIsothermalElasticConstantsAndStressAppendixEq_14} and these terms represented in diagrammatic form are shown in Fig. \ref{fig:ThirdOrderStrainedSEnsembleDiagrams}. The third-order approximation of the free energy can be written as
\begin{eqnarray} 
\tilde{F}^{\left(3\right)}_{A}  \approx&&  \sum_{m=1} \sum_{\bar{\mu}_{m}} \sum_{\bar{\nu}_{m}} \tilde{c}^{T\left(3\right),1}_{ \bar{\mu}_{m} \bar{\nu}_{m} } \bar{u}_{\mu_{m} \nu_{m}} \nonumber \\
&&+ \sum_{m,m'=1} \sum_{\bar{\mu}_{m}} \sum_{\bar{\nu}_{m}}  \sum_{\bar{\mu}'_{m'}} \sum_{\bar{\nu}'_{m'}} \sum^{7}_{i=2}  \nonumber \\
&&\times \tilde{c}^{T\left(3\right),i}_{ \bar{\mu}_{m} \bar{\nu}_{m} \bar{\mu}'_{m'} \bar{\nu}'_{m'} } \bar{u}_{\mu_{m} \nu_{m}}  \bar{u}'_{\mu'_{m'} \nu'_{m'}} \nonumber \\
&&+ \sum_{m,m',m''=1} \sum_{\bar{\mu}_{m}} \sum_{\bar{\nu}_{m}}  \sum_{\bar{\mu}'_{m'}} \sum_{\bar{\nu}'_{m'}} \sum_{\bar{\mu}''_{m''}} \sum_{\bar{\nu}''_{m''}} \sum^{14}_{i=8}  \nonumber \\
&&\times \tilde{c}^{T\left(3\right),i}_{\bar{\mu}_{m} \bar{\nu}_{m} \bar{\mu}'_{m'} \bar{\nu}'_{m'} \bar{\mu}''_{m''} \bar{\nu}''_{m''} } \bar{u}_{\mu_{m} \nu_{m}} \bar{u}'_{\mu'_{m'} \nu'_{m'}} \bar{u}''_{\mu''_{m''} \nu''_{m''}}. \nonumber \\
\label{eq:HarkThermPerturbThirdOrderResultsFreeEnergyEq_1} 
\end{eqnarray}
The isothermal elastic constants $\tilde{c}^{T\left(3\right),i}_{\mu_{1} \nu_{1} \mu_{2} \nu_{2} \cdots \mu_{j} \nu_{j} }$ in Eq. \ref{eq:HarkThermPerturbThirdOrderResultsFreeEnergyEq_1} are given by Eqs. \ref{eq:HarkThermPerturbThirdOrderResultsIsothermalElasticConstantsAndStressAppendixEq_1}-\ref{eq:HarkThermPerturbThirdOrderResultsIsothermalElasticConstantsAndStressAppendixEq_14}.

The third-order contribution to the internal energy can be approximated by using the same expression as given by Eq. \ref{eq:HarkThermPerturbThirdOrderResultsFreeEnergyEq_1}, except by replacing the isothermal elastic constants with the corresponding adiabatic ones. The adiabatic elastic constants can be obtained by using Eqs. \ref{eq:HarkThermPerturbThirdOrderResultsIsothermalElasticConstantsAndStressAppendixEq_1}-\ref{eq:HarkThermPerturbThirdOrderResultsIsothermalElasticConstantsAndStressAppendixEq_14} with Eq. \ref{eq:HarkThermPerturbThirdOrderResultsIsothermalElasticConstantsAndStressEq_1}. The third-order contribution to heat capacity is not explicitly shown but can be obtained by the differentiation of internal energy contribution with respect to $T$. The third-order contribution to the entropy can be written in a similar way than in the second-order case (Eq. \ref{eq:HarkThermPerturbSecondOrderResultsEntropyEq_1}), but in addition the third-order term in strains, as in Eq. \ref{eq:HarkThermPerturbThirdOrderResultsFreeEnergyEq_1}, must be included.  Lastly, the contribution $\alpha^{\left(3\right)}_{\mu_{1} \nu_{1}}$ to the CTE can be obtained by using Eq. \ref{eq:HarkThermPerturbFirstOrderResultsThermalExpansionEq_1}. The results for $\sigma^{T\left(3\right)}_{\gamma \delta}, \sigma^{A\left(3\right)}_{\gamma \delta}$ and $c^{T\left(3\right)}_{\mu \nu \gamma \delta}$ are given by Eq. \ref{eq:HarkThermPerturbThirdOrderResultsIsothermalElasticConstantsAndStressEq_1} and the expressions are listed in Appendix \ref{ExpressionsForThirdOrderElasticConstants}. The contribution to the second-order isothermal elastic constants can be obtained from Eqs. \ref{eq:HarkThermPerturbThirdOrderResultsIsothermalElasticConstantsAndStressAppendixEq_1}-\ref{eq:HarkThermPerturbThirdOrderResultsIsothermalElasticConstantsAndStressAppendixEq_7}, while the contribution to isothermal stress can be obtained from Eq. \ref{eq:HarkThermPerturbThirdOrderResultsIsothermalElasticConstantsAndStressAppendixEq_1} and the adiabatic stress by using Eq. \ref{eq:HarkThermPerturbThirdOrderResultsIsothermalElasticConstantsAndStressAppendixEq_1} with Eq. \ref{eq:HarkThermPerturbThirdOrderResultsIsothermalElasticConstantsAndStressEq_1}.

\section{Conclusions}
\label{cha:Conclusions}
Expressions for different thermodynamical quantities were derived up to third-order in perturbation and in some cases, up to $n$th-order in IFCs. The results allow, for instance, the perturbative calculation of the following quantities for a crystal lattice: Helmholtz free energy, internal energy, entropy, heat capacity, isothermal and adiabatic elastic constants and CTE. The present method can be used to study, for example, the NTE beyond the QHA whenever the IFCs needed are available for the system under consideration.  In each order of perturbation, the results of the present work were compared with those obtained earlier by other authors and the original contribution of the present work was emphasized. In particular, expressions for the elastic constants of arbitrary order were considered and in the case of the lowest-order elastic constants, relations which are higher-order in the IFCs than have been obtained earlier were given. Similarity of the perturbation theory for the different macroscopic parameters was emphasized. A quantum mechanical physical interpretation for the harmonic phonon eigenvectors and phase factors was given.

\appendix

\section{Condensed notation}
\label{CondensedNotation}
In this section, condensed notations for different quantities are given. The following notations are used
\begin{equation} 
\sum_{\bar{\mu}_{m}} \equiv \sum_{\mu_{1}} \cdots \sum_{\mu_{m}}, \quad \sum_{\bar{\nu}_{m}} \equiv  \sum_{\nu_{1}} \cdots \sum_{\nu_{m}},
\label{eq:HarkThermPerturbFirstOrderResultsEq_3_1}
\end{equation}
\begin{equation} 
\bar{\mu}_{m} \equiv \mu_{1} \cdots \mu_{m}, \quad \bar{\nu}_{m} \equiv \nu_{1} \cdots \nu_{m},
\label{eq:HarkThermPerturbFirstOrderResultsEq_3_2}
\end{equation}
\begin{equation} 
\bar{u}_{\mu_{m} \nu_{m}} \equiv u_{\mu_{1} \nu_{1}} \cdots u_{\mu_{m} \nu_{m}},
\label{eq:HarkThermPerturbFirstOrderResultsEq_3_3}
\end{equation}
\begin{equation} 
c_{\bar{\mu}_{m} \bar{\nu}_{m}} \equiv  c_{\mu_{1} \nu_{1} \cdots \mu_{m} \nu_{m} },
\label{eq:HarkThermPerturbFirstOrderResultsEq_3_4}
\end{equation}
\begin{equation} 
c_{\bar{\mu}_{m} \bar{\nu}_{m} \bar{\mu}'_{m'} \bar{\nu}'_{m'}} \equiv  c_{\mu_{1} \nu_{1} \cdots \mu_{m} \nu_{m} \mu'_{1} \nu'_{1} \cdots \mu'_{m'} \nu'_{m'} },
\label{eq:HarkThermPerturbFirstOrderResultsEq_3_5}
\end{equation}
\begin{eqnarray} 
c&&_{\bar{\mu}_{m} \bar{\nu}_{m} \bar{\mu}'_{m'} \bar{\nu}'_{m'} \bar{\mu}''_{m''} \bar{\nu}''_{m''}} \nonumber \\
&&\equiv  c_{\mu_{1} \nu_{1} \cdots \mu_{m} \nu_{m} \mu'_{1} \nu'_{1} \cdots \mu'_{m'} \nu'_{m'} \mu''_{1} \nu''_{1} \cdots \mu''_{m''} \nu'_{m''}},
\label{eq:HarkThermPerturbFirstOrderResultsEq_3_6}
\end{eqnarray}
\begin{equation} 
\bar{\lambda}_{n} \equiv  \lambda_{1};\lambda_{2};\cdots;\lambda_{n}, \quad \sum_{\bar{\lambda}_{n}} \equiv \sum_{\lambda_{1}} \cdots \sum_{\lambda_{n}},
\label{eq:HarkThermPerturbFirstOrderResultsEq_3_7}
\end{equation}
\begin{equation} 
V_{\bar{\mu}_{m} \bar{\nu}_{m}} \left(\bar{\lambda}_{n}\right) \equiv  V_{\mu_{1} \nu_{1} \cdots \mu_{m} \nu_{m}} \left(\lambda_{1};\lambda_{2};\cdots;\lambda_{n}\right),
\label{eq:HarkThermPerturbFirstOrderResultsEq_3_8}
\end{equation}
\begin{equation} 
V_{\bar{\mu}_{m} \bar{\nu}_{m}} \left(\bar{\lambda}_{n};-\bar{\lambda}_{n}\right) \equiv  V_{ \bar{\mu}_{m} \bar{\nu}_{m} } \left(\lambda_{1};-\lambda_{1};\cdots;\lambda_{n};-\lambda_{n}\right).
\label{eq:HarkThermPerturbFirstOrderResultsEq_3_9}
\end{equation}
The following notations are sometimes used
\begin{eqnarray} 
\xi^{\left(1\right)}_{n/2} \equiv && \prod^{n-1}_{k=1 \wedge k= \, odd } k \, 2^{n/2} \nonumber \\
&&\times \left( \bar{n}_{\lambda_{1}} + \frac{1}{2}\right) \left( \bar{n}_{\lambda_{2}} + \frac{1}{2}\right) \cdots \left( \bar{n}_{\lambda_{n/2}} + \frac{1}{2}\right),  \nonumber \\
\xi^{\left(2\right)}_{n/2} \equiv&& \prod^{n-1}_{k=1 \wedge k= \, odd } k \, 2^{n/2} \nonumber \\
&&\times \sum^{n/2}_{i=1} \omega_{\lambda_{i}}  \bar{n}_{\lambda_{i}} \left(\bar{n}_{\lambda_{i}} + 1\right) \prod^{n/2}_{l=1\wedge l\neq i} \left(\bar{n}_{\lambda_{l}} +\frac{1}{2}\right), \nonumber \\
\label{eq:HarkThermPerturbFirstOrderResultsEq_3}
\end{eqnarray}
\begin{eqnarray} 
\xi&&^{\left(3\right)}_{n/2} \nonumber \\
&&\equiv \prod^{n-1}_{k=1 \wedge k= \, odd } k \left[ 2^{n/2} \sum^{n/2}_{i=1} \right.  \nonumber \\
&&\times  \omega^{2}_{\lambda_{i}} \bar{n}_{\lambda_{i}}  \left(\bar{n}_{\lambda_{i}} + 1\right) \left( \bar{n}_{\lambda_{i}} + 1 \right)  \prod^{n/2}_{l=1\wedge l\neq i} \left( \bar{n}_{\lambda_{l}} +\frac{1}{2}\right)  \nonumber \\
&&+ \left. 2 \sum^{n/2}_{i=1} \omega_{\lambda_{i}} \bar{n}_{\lambda_{i}} \left(\bar{n}_{\lambda_{i}} + 1\right)  \prod^{n/2}_{l=1\wedge l\neq i}  \omega_{\lambda_{l}}  \bar{n}_{\lambda_{l}} \left( \bar{n}_{\lambda_{l}} + 1 \right) \right], \nonumber \\
\label{eq:HarkThermPerturbFirstOrderResultsHeatCapacityEq_2}
\end{eqnarray}
\begin{eqnarray} 
G&&^{\left(3\right)}\left(\lambda;\lambda';\lambda''\right) \nonumber \\
&&\equiv \left[ 3 \frac{ \bar{n}_{\lambda} \left( \bar{n}_{\lambda'}+\bar{n}_{\lambda''}+1 \right) - \bar{n}_{\lambda'} \bar{n}_{\lambda''} }{-\omega_{\lambda}+\omega_{\lambda'}+\omega_{\lambda''}} \right. \nonumber \\
&& +  \left. \frac{ \left(\bar{n}_{\lambda} + 1 \right)  \left(\bar{n}_{\lambda'} + 1 \right) \left(\bar{n}_{\lambda''} + 1 \right) - \bar{n}_{\lambda}\bar{n}_{\lambda'} \bar{n}_{\lambda''}  }{\omega_{\lambda}+\omega_{\lambda'}+\omega_{\lambda''}} \right], \nonumber \\
\label{eq:HarkThermPerturbSecondOrderResultsFreeEnergyEq_3} 
\end{eqnarray}  
\begin{eqnarray} 
\xi^{\left(1\right)}_{n} \equiv && \prod^{n-1}_{k=1 \wedge k= \, odd }\left(k\right)  2^{n} \nonumber \\
&&\times \left( \bar{n}_{\lambda_{1}} + \frac{1}{2}\right) \left( \bar{n}_{\lambda_{2}} + \frac{1}{2}\right) \cdots \left( \bar{n}_{\lambda_{n}} + \frac{1}{2}\right), \nonumber \\ 
\label{eq:HarkThermPerturbSecondOrderResultsIsothermalElasticConstantsAndStressEq_12}
\end{eqnarray}
\begin{eqnarray} 
\xi^{\left(2\right)}_{n} \equiv && \prod^{n-1}_{k=1 \wedge k= \, odd }\left(k\right) 2^{n} \nonumber \\
&&\times \sum^{n}_{l=1}  \omega_{\lambda_{l}}  \bar{n}_{\lambda_{l}} \left(\bar{n}_{\lambda_{l}} + 1\right) \prod^{n}_{k'=1\wedge k'\neq l} \left(\bar{n}_{\lambda_{k'}} +\frac{1}{2}\right), \nonumber \\
\label{eq:HarkThermPerturbSecondOrderResultsAdiabaticElasticConstantsAndStressEq_12} 
\end{eqnarray}  
\begin{eqnarray} 
G&&^{\left(3,\beta\right)}\left(\lambda;\lambda';\lambda''\right) \nonumber \\
&&\equiv \left[ \frac{ 2 \omega_{\lambda'}  \bar{n}_{\lambda'} \left(\bar{n}_{\lambda'} + 1\right)\left( \bar{n}_{\lambda''} - \bar{n}_{\lambda} \right)  }{-\omega_{\lambda}+\omega_{\lambda'}+\omega_{\lambda''}} \right. \nonumber \\
&&- \frac{  \omega_{\lambda}  \bar{n}_{\lambda} \left(\bar{n}_{\lambda} + 1\right) \left( \bar{n}_{\lambda'}+\bar{n}_{\lambda''}+1 \right) }{-\omega_{\lambda}+\omega_{\lambda'}+\omega_{\lambda''}}  \nonumber \\
&&+\frac{  \omega_{\lambda} \bar{n}_{\lambda} \left(\bar{n}_{\lambda} + 1\right) \bar{n}_{\lambda'} \bar{n}_{\lambda''} }{\omega_{\lambda}+\omega_{\lambda'}+\omega_{\lambda''}} \nonumber \\
&&-  \left. \frac{   \omega_{\lambda} \bar{n}_{\lambda} \left(\bar{n}_{\lambda} + 1\right) \left(\bar{n}_{\lambda'} + 1\right) \left( \bar{n}_{\lambda''} +1 \right)  }{\omega_{\lambda}+\omega_{\lambda'}+\omega_{\lambda''}} \right].
\label{eq:HarkThermPerturbSecondOrderResultsAdiabaticElasticConstantsAndStressEq_13} 
\end{eqnarray}

\section{Thermodynamical relations for harmonic and quasi-harmonic phonons}
\label{cha:ThermodynamicalRelationsForHarmonicAndQuasiHarmonicPhonons}
In this section, some thermodynamical relations within the QHA are listed so that these results can be compared with those obtained by using the perturbation theory. The results listed in this section have been obtained, for example, in Refs. \cite{Leibfried-SolidStatePhysics-1961,Wallace-ThermodynamicsOfCrystals-1972}. The Helmholtz free energy for a crystal lattice can be written as \cite{Born-Huang-DynamicalTheory-1954}
\begin{equation} 
F =\Phi_{0} - \beta^{-1} \ln Z,
\label{eq:HarkThermPerturbCrystalHelmholtzFreeEnergy}
\end{equation}
where $\Phi_{0}$ is the part arising from the electronic state of the crystal and $Z$ is the canonical partition function
\begin{equation} 
Z = tr\left[e^{-\beta \hat{H}}\right] = \sum_{n} \braket{n|e^{-\beta \hat{H}}|n}.
\label{eq:HarkThermPerturbCrystalCanonicalPartitionFunction}
\end{equation}
Within the harmonic approximation $\hat{H} \rightarrow \hat{H}_{0}$ and
\begin{equation} 
Z_{0} = \sum_{n} \braket{n|e^{-\beta \hat{H}_{0}}|n} = \prod_{\lambda} \frac{e^{-\frac{1}{2}\beta \hbar \omega_{\lambda} }}{1-e^{-\beta \hbar \omega_{\lambda}}},
\label{eq:HarkThermPerturbCrystalHarmonicCanonicalPartitionFunction}
\end{equation}
and thus
\begin{equation} 
F_{0} =\Phi_{0} + \beta^{-1} \sum_{\lambda}\left[\frac{1}{2}\beta \hbar \omega_{\lambda} + \ln \left(1- \frac{\bar{n}_{\lambda}}{\bar{n}_{\lambda} + 1} \right)\right].
\label{eq:HarkThermPerturbCrystalHarmonicHelmholtzFreeEnergy}
\end{equation}
with $\bar{n}_{\lambda}$ being the Bose-Einstein distribution function
\begin{equation} 
\bar{n}_{\lambda} = \frac{1}{e^{\beta \hbar \omega_{\lambda}} - 1} = \sum_{n} \braket{n|e^{-\beta \hat{H}_{0}} \hat{a}^{\dagger}_{\lambda} \hat{a}_{\lambda} |n} Z^{-1}_{0}.
\label{eq:HarkThermPerturbCrystalHarmonicHelmholtzFreeEnergy_2}
\end{equation}
Furthermore, the entropy, internal energy and heat capacity at constant strain can be written as
\begin{eqnarray} 
S_{0}=&&  \frac{\beta}{T} \frac{\partial{F_{0}}}{\partial{\beta}} = \sum_{\lambda} \frac{\hbar \omega_{\lambda} }{T}\left(\bar{n}_{\lambda} + \frac{1}{2}\right) \nonumber \\
      &&- k_{B} \sum_{\lambda} \ln \left(\left[\bar{n}_{\lambda}\left(\bar{n}_{\lambda}+1\right)\right]^{-1/2} \right),
\label{eq:HarkThermPerturbCrystalHarmonicEntropy}
\end{eqnarray}
\begin{equation} 
U_{0} =  \Phi_{0} + \sum_{\lambda}  \hbar \omega_{\lambda} \left(\bar{n}_{\lambda} + \frac{1}{2}\right) = \Phi_{0} + \sum_{\lambda}  U_{0}\left(\lambda\right),
\label{eq:HarkThermPerturbVibrationalInternalEnergyOfCrystal}
\end{equation}
\begin{equation} 
C_{\eta} = \frac{\partial{U_{0}}}{\partial{T}} = k_{B} \sum_{\lambda} \left[ \hbar \omega_{\lambda} \beta \right]^{2}  \bar{n}_{\lambda}\left(\bar{n}_{\lambda}+1\right)= \sum_{\lambda} c_{\eta}\left(\lambda\right).
\label{eq:HarkThermPerturbVibrationalHeatCapConstStrain}
\end{equation}

The elastic constants within the QHA can be obtained from Eqs. \ref{eq:HarkThermPerturbCrystalHarmonicHelmholtzFreeEnergy} and \ref{eq:HarkThermPerturbVibrationalInternalEnergyOfCrystal} by differentiation with respect to strains as in Sec. \ref{cha:ExpansionOfFreeEnergyAndInternalEnergy} [it is thus assumed that $\omega_{\lambda} = \omega_{\lambda}\left(\eta_{\alpha \beta}\right)$]. After the differentiation and by using the definition of the generalized Gr\"{u}neisen parameters
\begin{equation} 
\gamma_{\mu_{1}\nu_{1} \cdots \mu_{n}\nu_{n}}\left(\lambda\right) \equiv - \frac{1}{\omega_{\lambda}}  \frac{\partial^{n}{ \omega_{\lambda} }}{ \partial{\eta_{\mu_{1}\nu_{1}}} \partial{\eta_{\mu_{2}\nu_{2}}} \cdots \partial{\eta_{\mu_{n}\nu_{n}}}  },
\label{eq:HarkThermPerturbGruneisenParameterDefinition2}
\end{equation}
one may write (derivatives of $\Phi_{0}$ are neglected)
\begin{equation} 
\sigma^{T}_{\mu\nu,0} = -\sum_{\lambda} U_{0}\left(\lambda\right)  \gamma_{\mu\nu}\left(\lambda\right),
\label{eq:HarkThermPerturbStressHelmholtzEnergyCrystalHarmonicGruneisen}
\end{equation}
\begin{equation} 
\sigma^{A}_{\mu\nu,0} = \sigma^{T}_{\mu\nu,0} + T \sum_{\lambda} c_{v}\left(\lambda\right)  \gamma_{\mu\nu}\left(\lambda\right),
\label{eq:HarkThermPerturbStressInternalEnergyCrystalHarmonicGruneisen}
\end{equation}
\begin{equation} 
\frac{\partial{\sigma^{T}_{\mu\nu,0}}}{\partial{T}} =- \sum_{\lambda} c_{v}\left(\lambda\right)  \gamma_{\mu\nu}\left(\lambda\right),
\label{eq:HarkThermPerturbStressDerivativeHelmholtzEnergyCrystalHarmonicGruneisen}
\end{equation}  
\begin{eqnarray} 
c^{T}_{\mu_{1}\nu_{1} \mu_{2}\nu_{2},0} =&& - \sum_{\lambda} U_{0}\left(\lambda\right) \gamma_{ \mu_{1}\nu_{1} \mu_{2}\nu_{2} }\left(\lambda\right) \nonumber \\
&&- T \sum_{\lambda}  c_{\eta}\left(\lambda\right) \gamma_{\mu_{1}\nu_{1} }\left(\lambda\right) \gamma_{\mu_{2}\nu_{2}}\left(\lambda\right),
\label{eq:HarkThermPerturbIsothermaElasticConstGruneisen}
\end{eqnarray}       
\begin{equation} 
c^{A}_{\mu_{1}\nu_{1} \mu_{2}\nu_{2},0} = c^{T}_{\mu_{1}\nu_{1} \mu_{2}\nu_{2},0}  - T \frac{\partial{ c^{T}_{\mu_{1}\nu_{1} \mu_{2}\nu_{2},0} }}{\partial{T} },
\label{eq:HarkThermPerturbAdiabaticElasticConstGruneisen}
\end{equation} 
\begin{eqnarray} 
c&&^{T}_{\mu_{1}\nu_{1} \cdots \mu_{3}\nu_{3} ,0} \nonumber \\
&&=   -\frac{ 2 \hbar^{2} }{k_{B}} \sum_{\lambda} U_{0}\left( \lambda \right) c_{\eta}\left(\lambda\right) \nonumber \\
&&\times   \gamma_{\mu_{1}\nu_{1}}\left(\lambda\right) \gamma_{\mu_{2}\nu_{2}}\left(\lambda\right) \gamma_{\mu_{3}\nu_{3}}\left(\lambda\right)  - T \sum_{\lambda} c_{\eta}\left(\lambda\right)  \nonumber \\
  &&\times \left[\gamma_{\mu_{1}\nu_{1}}\left(\lambda\right)   \gamma_{\mu_{2}\nu_{2} \mu_{3}\nu_{3} }\left(\lambda\right) + \gamma_{\mu_{2}\nu_{2}}\left(\lambda\right)   \gamma_{\mu_{1}\nu_{1} \mu_{3}\nu_{3} }\left(\lambda\right) \right. \nonumber \\
&&+	\left. \gamma_{\mu_{3}\nu_{3}}\left(\lambda\right)   \gamma_{\mu_{1}\nu_{1} \mu_{2}\nu_{2} }\left(\lambda\right)  \right] -  \sum_{\lambda} U_{0}\left( \lambda \right)   \gamma_{\mu_{1}\nu_{1} \cdots \mu_{3}\nu_{3}}\left(\lambda\right), \nonumber \\
\label{eq:HarkThermPerturb3RdOrderIsothermaElasticConstGruneisen}
\end{eqnarray} 
\begin{equation} 
c^{A}_{\mu_{1}\nu_{1} \cdots \mu_{3}\nu_{3},0} = c^{T}_{\mu_{1}\nu_{1} \cdots \mu_{3}\nu_{3},0} - T \frac{\partial{ c^{T}_{\mu_{1}\nu_{1} \cdots \mu_{3}\nu_{3},0} }}{\partial{T} }.
\label{eq:HarkThermPerturb3RdOrderAdiabaticElasticConstGruneisen}
\end{equation} 
The CTE within the QHA can be written as \cite{Wallace-ThermodynamicsOfCrystals-1972}
\begin{equation} 
\alpha_{\mu_{1}\nu_{1},0} = - \sum^{3}_{\mu_{2},\nu_{2}=1} s^{T}_{\mu_{1}\nu_{1} \mu_{2}\nu_{2},0}  \frac{\partial{\sigma^{T}_{\mu_{2}\nu_{2},0}}}{\partial{T}},
\label{eq:HarkThermPerturbThermalExpansionCoefficientGruneisen}
\end{equation}
or
\begin{equation} 
\alpha_{\mu_{1}\nu_{1},0} = \frac{1}{T} \sum^{3}_{\mu_{2},\nu_{2}=1} s^{T}_{\mu_{1}\nu_{1} \mu_{2}\nu_{2},0} \left(\sigma^{A}_{\mu_{2}\nu_{2},0}- \sigma^{T}_{\mu_{2}\nu_{2},0}\right),
\label{eq:HarkThermPerturbThermalExpansionCoefficientGruneisen_3}
\end{equation} 
where $s^{T}_{\mu_{1}\nu_{1} \mu_{2}\nu_{2},0}$ is the second-order elastic compliance tensor defined through $\sum_{\mu',\nu'} c^{T}_{\mu\nu \mu'\nu',0} s^{T}_{\mu'\nu' \mu''\nu'',0} = \delta_{\mu\mu''} \delta_{\nu\nu''}$ \cite{Nye-PhysicalPropertiesOfCrystals-1985}. In the case of cubic crystals, the volumetric and generalized Gr\"{u}neisen parameters are related as
\begin{equation} 
\frac{1}{3}\gamma_{\mu\mu}\left(\lambda\right) = -\frac{V}{\omega_{\lambda}} \frac{\partial{\omega_{\lambda}}}{\partial{V}} \equiv \gamma\left(\lambda\right).
\label{eq:HarkThermPerturbThermalExpansionCoefficientGruneisenConnection_1}
\end{equation} 
In Sec. \ref{cha:PerturbationExpansionForThermodynamicalQuantities}, more general approach to calculate the quantities given in this section is considered and it turns out that some results have a similar form in both approaches (Secs. \ref{cha:FirstOrderResultsIsothermalElasticConstantsAndStress}, \ref{cha:FirstOrderResultsAdiabaticElasticConstantsAndStress} and \ref{cha:FirstOrderResultsThermalExpansion}).

\section{Expressions for second-order elastic constants}
\label{ExpressionsForSecondOrderElasticConstants}

The list of second-order contributions to isothermal elastic constants is given below (in Eqs. \ref{eq:HarkThermPerturbSecondOrderResultsIsothermalElasticConstantsAndStressEq_2}-\ref{eq:HarkThermPerturbSecondOrderResultsIsothermalElasticConstantsAndStressEq_7}, $m+m'=k$)
\begin{eqnarray} 
\tilde{c}^{T\left(2\right),1}_{ \bar{\mu}_{k} \bar{\nu}_{k} } =&& -  \frac{ 1 }{ \hbar } \sum_{m=1} \frac{1}{m!} \sum_{m'=1}  \frac{1}{m'!} \sum_{\lambda_{1}} \nonumber \\
&&\times \frac{ V_{ \bar{\mu}_{m} \bar{ \nu}_{m}}\left(\lambda_{1}\right) V_{ \bar{\mu}'_{m'}  \bar{ \nu}'_{m'}}\left(-\lambda_{1}\right) }{ \omega_{\lambda_{1}} }, 
\label{eq:HarkThermPerturbSecondOrderResultsIsothermalElasticConstantsAndStressEq_2} 
\end{eqnarray}
\begin{eqnarray} 
\tilde{c}&&^{T\left(2\right),2}_{ \bar{\mu}_{k} \bar{\nu}_{k} }  \nonumber \\
&&=  - \frac{ 4 }{ \hbar}   \sum_{m=1} \frac{1}{m!} \sum_{m'=1}  \frac{1}{m'!} \sum_{\lambda} \sum_{\lambda'} V_{ \bar{\mu}_{m} \bar{ \nu}_{m}}\left(\lambda;\lambda'\right)  \nonumber \\
&&\times  V_{ \bar{\mu}'_{m'}  \bar{ \nu}'_{m'}}\left(-\lambda;-\lambda'\right) \frac{ \omega_{\lambda} \left( \bar{n}_{\lambda'} +\frac{1}{2}\right) - \omega_{\lambda'} \left( \bar{n}_{\lambda} +\frac{1}{2}\right) }{ \omega^{2}_{\lambda} - \omega^{2}_{\lambda'} }, \nonumber \\
\label{eq:HarkThermPerturbSecondOrderResultsIsothermalElasticConstantsAndStressEq_3} 
\end{eqnarray}
\begin{eqnarray} 
\tilde{c}&&^{T\left(2\right),3}_{ \bar{\mu}_{k} \bar{\nu}_{k} } \nonumber \\
&&=  - \frac{ 6 }{ \hbar } \sum_{m=1} \frac{1}{m!} \sum_{m'=1} \frac{1}{m'!} \sum_{\lambda} \sum_{\lambda'} \sum_{\lambda''} \nonumber \\
&&\left\{  6  V_{ \bar{\mu}_{m} \bar{ \nu}_{m}}\left(\lambda;-\lambda;\lambda'\right) V_{ \bar{\mu}'_{m'}  \bar{ \nu}'_{m'}}\left(-\lambda';\lambda'';-\lambda''\right) \right. \nonumber \\
&&\times  \frac{ \left( \bar{n}_{\lambda} + \frac{1}{2}\right) \left( \bar{n}_{\lambda''} + \frac{1}{2}\right) }{ \omega_{\lambda'}} \nonumber \\
&&+  V_{ \bar{\mu}_{m} \bar{ \nu}_{m}}\left(\lambda;\lambda';\lambda''\right) V_{ \bar{\mu}'_{m'}  \bar{ \nu}'_{m'}}\left(-\lambda;-\lambda';-\lambda''\right) \nonumber \\
&&\times \left.  G^{\left(3\right)}\left(\lambda;\lambda';\lambda''\right) \right\}, 
\label{eq:HarkThermPerturbSecondOrderResultsIsothermalElasticConstantsAndStressEq_4} 
\end{eqnarray}
\begin{eqnarray} 
\tilde{c}^{T\left(2\right),4}_{ \bar{\mu}_{k} \bar{\nu}_{k} } =&& - \frac{ 12 }{ \hbar }   \sum_{m=1} \frac{1}{m!} \sum_{m'=1} \frac{1}{m'!} \sum_{\lambda_{1}} \sum_{\lambda_{2}} \frac{ \left( \bar{n}_{\lambda_{2}} + \frac{1}{2}\right) }{\omega_{\lambda_{1}}} \nonumber \\
&&\times  V_{ \bar{\mu}_{m} \bar{ \nu}_{m}} \left(\lambda_{1}\right) V_{ \bar{\mu}'_{m'}  \bar{ \nu}'_{m'}} \left(-\lambda_{1};\lambda_{2};-\lambda_{2}\right),
\label{eq:HarkThermPerturbSecondOrderResultsIsothermalElasticConstantsAndStressEq_5} 
\end{eqnarray}
\begin{eqnarray} 
\tilde{c}&&^{T\left(2\right),5}_{ \bar{\mu}_{k} \bar{\nu}_{k} } \nonumber \\
&&= -\frac{4}{ \hbar}  \sum_{\lambda_{1}}  \sum_{\lambda_{2}} \sum_{m=1} \frac{1}{m!} V_{ \bar{\mu}_{m} \bar{ \nu}_{m}} \left(\lambda_{1};\lambda_{2}\right) \nonumber \\
&&\times \sum_{n'=3} \sum_{\lambda'_{1}} \sum_{\lambda'_{2}} \cdots \sum_{\lambda'_{n'-1}}  \sum_{m'=1} \frac{1}{m'!}  \nonumber \\
&&\times V_{ \bar{\mu}'_{m'}  \bar{ \nu}'_{m'}} \left(-\lambda_{1};-\lambda_{2};\left\{\lambda'_{n'-1}\right\}\right) \nonumber \\
&&\times  \frac{ \omega_{\lambda_{1}} \left( \bar{n}_{\lambda_{2}} +\frac{1}{2}\right) - \omega_{\lambda_{2}} \left( \bar{n}_{\lambda_{1}} +\frac{1}{2}\right) }{ \omega^{2}_{\lambda_{1}} - \omega^{2}_{\lambda_{2}} } n' \prod^{2n'-3}_{k=1 \wedge k=odd} k \nonumber \\
&&\times 2^{n'} \left( \bar{n}_{\lambda'_{1}} + \frac{1}{2}\right)  \left( \bar{n}_{\lambda'_{2}} + \frac{1}{2}\right) \cdots  \left( \bar{n}_{\lambda'_{n'-1}} + \frac{1}{2}\right), \nonumber \\
&&\left\{\lambda'_{n'-1}\right\} = \lambda'_{1};-\lambda'_{1};\ldots;\lambda'_{n'-1};-\lambda'_{n'-1},
\label{eq:HarkThermPerturbSecondOrderResultsIsothermalElasticConstantsAndStressEq_6} 
\end{eqnarray}
\begin{eqnarray} 
\tilde{c}&&^{T\left(2\right),6}_{ \bar{\mu}_{k} \bar{\nu}_{k} } \nonumber \\
&&= -\frac{ 1 }{ \hbar }   \sum_{\lambda_{1}}  \sum_{m=1} \frac{1}{m!} V_{ \bar{\mu}_{m} \bar{ \nu}_{m}} \left(\lambda_{1}\right) \nonumber \\
&&\times \sum_{n'=2} \sum_{\lambda'_{1}} \sum_{\lambda'_{2}} \cdots \sum_{\lambda'_{n'-1}}  \sum_{m'=1} \frac{1}{m'!} \prod^{2n'-1}_{k=1 \wedge k=odd} k \nonumber \\
&&\times \frac{V_{ \bar{\mu}'_{m'}  \bar{ \nu}'_{m'}} \left(-\lambda_{1};\lambda'_{1};-\lambda'_{1}\ldots;\lambda'_{n'-1};-\lambda'_{n'-1}\right)}{ \omega_{\lambda_{1}} } \nonumber \\
&&\times	2^{n'} \left( \bar{n}_{\lambda'_{1}} + \frac{1}{2}\right) \left( \bar{n}_{\lambda'_{2}} + \frac{1}{2}\right) \cdots \left( \bar{n}_{\lambda'_{n'-1}} + \frac{1}{2} \right), \nonumber \\
\label{eq:HarkThermPerturbSecondOrderResultsIsothermalElasticConstantsAndStressEq_7} 
\end{eqnarray}
\begin{eqnarray} 
\tilde{c}^{T\left(2\right),7}_{ \bar{\mu}_{k} \bar{\nu}_{k} } =&& - \frac{12 }{ \hbar k!}   \sum_{\lambda_{1}} \sum_{\lambda_{2}} V_{ \bar{\mu}_{k} \bar{ \nu}_{k}} \left(\lambda_{1}\right) \nonumber \\
&&\times  V\left(-\lambda_{1};\lambda_{2};-\lambda_{2}\right) \frac{ \bar{n}_{\lambda_{2}} + \frac{1}{2}  }{ \omega_{\lambda_{1}} },
\label{eq:HarkThermPerturbSecondOrderResultsIsothermalElasticConstantsAndStressEq_8} 
\end{eqnarray}
\begin{eqnarray} 
\tilde{c}&&^{T\left(2\right),8}_{ \bar{\mu}_{k} \bar{\nu}_{k} } \nonumber \\
&&= - \frac{ 72 }{ \hbar k! }  \sum_{\lambda} \sum_{\lambda'} \sum_{\lambda''} V_{ \bar{\mu}_{k} \bar{ \nu}_{k}}\left(\lambda;-\lambda;\lambda'\right) \nonumber \\
&&\times  V\left(-\lambda';\lambda'';-\lambda''\right)  \frac{ \left( \bar{n}_{\lambda} + \frac{1}{2}\right) \left( \bar{n}_{\lambda''} + \frac{1}{2}\right) }{ \omega_{\lambda'}}  \nonumber \\
&&- \frac{12 }{  \hbar k!}   \sum_{\lambda} \sum_{\lambda'} \sum_{\lambda''} V_{ \bar{\mu}_{k} \bar{ \nu}_{k}}\left(\lambda;\lambda';\lambda''\right) \nonumber \\
&&\times   V\left(-\lambda;-\lambda';-\lambda''\right) G^{\left(3\right)}\left(\lambda;\lambda';\lambda''\right),
\label{eq:HarkThermPerturbSecondOrderResultsIsothermalElasticConstantsAndStressEq_9} 
\end{eqnarray}
and
\begin{eqnarray} 
\tilde{c}&&^{T\left(2\right),9}_{ \bar{\mu}_{k} \bar{\nu}_{k} } \nonumber \\
&&= - \frac{96 }{ \hbar k!}    \sum_{\lambda} \sum_{\lambda'} \sum_{\lambda''} \nonumber \\
&&\times V_{ \bar{\mu}_{k} \bar{ \nu}_{k}}\left(\lambda;\lambda'\right) V\left(-\lambda;-\lambda',\lambda'';-\lambda''\right) \nonumber \\
&&\times \left( \bar{n}_{\lambda''} + \frac{1}{2}\right) \frac{ \omega_{\lambda} \left( \bar{n}_{\lambda'} +\frac{1}{2}\right) - \omega_{\lambda'} \left( \bar{n}_{\lambda} +\frac{1}{2} \right) }{ \omega^{2}_{\lambda} - \omega^{2}_{\lambda'} }.
\label{eq:HarkThermPerturbSecondOrderResultsIsothermalElasticConstantsAndStressEq_10} 
\end{eqnarray}
In Eqs. \ref{eq:HarkThermPerturbSecondOrderResultsIsothermalElasticConstantsAndStressEq_2}-\ref{eq:HarkThermPerturbSecondOrderResultsIsothermalElasticConstantsAndStressEq_8}, $\xi^{\left(1\right)}_{n'/2}$ is given by Eq. \ref{eq:HarkThermPerturbFirstOrderResultsEq_3}, $G^{\left(3\right)}\left(\lambda;\lambda';\lambda''\right)$ by Eq. \ref{eq:HarkThermPerturbSecondOrderResultsFreeEnergyEq_3} and $\xi^{\left(1\right)}_{n}$ by Eq. \ref{eq:HarkThermPerturbSecondOrderResultsIsothermalElasticConstantsAndStressEq_12}.

The list of second-order contributions to adiabatic elastic constants is given below (in Eqs. \ref{eq:HarkThermPerturbSecondOrderResultsAdiabaticElasticConstantsAndStressEq_2}-\ref{eq:HarkThermPerturbSecondOrderResultsAdiabaticElasticConstantsAndStressEq_5}, $m+m'=k$)
\begin{eqnarray} 
c^{A\left(2\right),1}_{ \bar{\mu}_{k} \bar{\nu}_{k} } =&& - \frac{ 1 }{ \hbar } \sum_{m=1} \frac{1}{m!} \sum_{m'=1} \frac{1}{m'!}   \sum_{\lambda_{1}} \nonumber \\
&&\times \frac{ V_{ \bar{\mu}_{m} \bar{ \nu}_{m}}\left(\lambda_{1}\right) V_{ \bar{\mu}'_{m'}  \bar{ \nu}'_{m'}}\left(-\lambda_{1}\right) }{ \omega_{\lambda_{1}} }, 
\label{eq:HarkThermPerturbSecondOrderResultsAdiabaticElasticConstantsAndStressEq_2} 
\end{eqnarray}
\begin{eqnarray} 
c&&^{A\left(2\right),2}_{ \bar{\mu}_{k} \bar{\nu}_{k} } \nonumber \\
&&=- \frac{ 4 }{ \hbar}   \sum_{m=1}  \frac{1}{m!} \sum_{m'=1} \frac{1}{m'!} \sum_{\lambda} \sum_{\lambda'} V_{ \bar{\mu}_{m} \bar{ \nu}_{m}}\left(\lambda;\lambda'\right)  \nonumber \\
&&\times  V_{ \bar{\mu}'_{m'}  \bar{ \nu}'_{m'}}\left(-\lambda;-\lambda'\right)   \frac{ \omega_{\lambda} \left( \bar{n}_{\lambda'} +\frac{1}{2}\right) - \omega_{\lambda'} \left( \bar{n}_{\lambda} + \frac{1}{2} \right) }{ \omega^{2}_{\lambda} - \omega^{2}_{\lambda'} } \nonumber \\
&&+ 8 \beta  \sum_{m=1}  \frac{1}{m!}\sum_{m'=1} \frac{1}{m'!} \sum_{\lambda} \sum_{\lambda'} V_{ \bar{\mu}_{m} \bar{ \nu}_{m}}\left(\lambda;\lambda'\right)   \nonumber \\
&&\times V_{ \bar{\mu}'_{m'}  \bar{ \nu}'_{m'}}\left(-\lambda;-\lambda'\right)  \frac{ \omega_{\lambda}  \omega_{\lambda'} \bar{n}_{\lambda'} \left(\bar{n}_{\lambda'} + 1\right) }{ \omega^{2}_{\lambda} - \omega^{2}_{\lambda'} },
\label{eq:HarkThermPerturbSecondOrderResultsAdiabaticElasticConstantsAndStressEq_3} 
\end{eqnarray}
\begin{eqnarray} 
c&&^{A\left(2\right),3}_{ \bar{\mu}_{k} \bar{\nu}_{k} } \nonumber \\
&&=- \frac{ 6}{ \hbar } \sum_{m=1}  \frac{1}{m!} \sum_{m'=1}  \frac{1}{m'!} \sum_{\lambda} \sum_{\lambda'} \sum_{\lambda''} \nonumber \\
&&\left\{ 6   V_{ \bar{\mu}_{m} \bar{ \nu}_{m}}\left(\lambda;-\lambda;\lambda'\right) V_{ \bar{\mu}'_{m'}  \bar{ \nu}'_{m'}}\left(-\lambda';\lambda'';-\lambda''\right) \right. \nonumber \\
&&\times  \frac{ \left( \bar{n}_{\lambda} + \frac{1}{2}\right) \left( \bar{n}_{\lambda''} + \frac{1}{2}\right) }{ \omega_{\lambda'}} \nonumber \\
&&+  V_{ \bar{\mu}_{m} \bar{ \nu}_{m}}\left(\lambda;\lambda';\lambda''\right) V_{ \bar{\mu}'_{m'}  \bar{ \nu}'_{m'}}\left(-\lambda;-\lambda';-\lambda''\right) \nonumber \\
&&\times \left.   G^{\left(3\right)}\left(\lambda;\lambda';\lambda''\right) \right\} \nonumber \\
&&- 18 \beta \sum_{m=1}  \frac{1}{m!} \sum_{m'=1} \frac{1}{m'!} \sum_{\lambda} \sum_{\lambda'} \sum_{\lambda''} \nonumber \\
&&\left\{ - 4  V_{ \bar{\mu}_{m} \bar{ \nu}_{m}}\left(\lambda;-\lambda;\lambda'\right) V_{ \bar{\mu}'_{m'}  \bar{ \nu}'_{m'}}\left(-\lambda';\lambda'';-\lambda''\right)  \right. \nonumber \\
&&\times \frac{ \omega_{\lambda}  \bar{n}_{\lambda} \left(\bar{n}_{\lambda} + 1\right) \left(\bar{n}_{\lambda''} +\frac{1}{2}\right) }{ \omega_{\lambda'}} \nonumber \\
&&+ V_{ \bar{\mu}_{m} \bar{ \nu}_{m}}\left(\lambda;\lambda';\lambda''\right) V_{ \bar{\mu}'_{m'}  \bar{ \nu}'_{m'}}\left(-\lambda;-\lambda';-\lambda''\right) \nonumber \\
&&\times \left.  G^{\left(3,\beta\right)}\left(\lambda;\lambda';\lambda''\right) \right\},
\label{eq:HarkThermPerturbSecondOrderResultsAdiabaticElasticConstantsAndStressEq_4} 
\end{eqnarray}
\begin{eqnarray} 
c&&^{A\left(2\right),4}_{ \bar{\mu}_{k} \bar{\nu}_{k} } \nonumber \\
&&= 12 \beta \sum_{m=1} \frac{1}{m!} \sum_{m'=1}  \frac{1}{m'!} \sum_{\lambda_{1}} \sum_{\lambda_{2}} \frac{ \omega_{\lambda_{2}}  \bar{n}_{\lambda_{2}} \left(\bar{n}_{\lambda_{2}} + 1\right) }{\omega_{\lambda_{1}}} \nonumber \\
&&\times  V_{ \bar{\mu}_{m} \bar{ \nu}_{m}} \left(\lambda_{1}\right) V_{ \bar{\mu}'_{m'}  \bar{ \nu}'_{m'}} \left(-\lambda_{1};\lambda_{2};-\lambda_{2}\right)  \nonumber \\
&&- \frac{12}{ \hbar }   \sum_{m=1}  \frac{1}{m!} \sum_{m'=1} \frac{1}{m'!} \sum_{\lambda_{1}} \sum_{\lambda_{2}} \frac{ \bar{n}_{\lambda_{2}} + \frac{1}{2} }{\omega_{\lambda_{1}}} \nonumber \\
&&\times  V_{ \bar{\mu}_{m} \bar{ \nu}_{m}} \left(\lambda_{1}\right) V_{ \bar{\mu}'_{m'}  \bar{ \nu}'_{m'}} \left(-\lambda_{1};\lambda_{2};-\lambda_{2}\right), 
\label{eq:HarkThermPerturbSecondOrderResultsAdiabaticElasticConstantsAndStressEq_5} 
\end{eqnarray}
\begin{eqnarray} 
c^{A\left(2\right),5}_{ \bar{\mu}_{k} \bar{\nu}_{k} } =&& - \frac{ 12 }{  \hbar k!}   \sum_{\lambda_{1}} \sum_{\lambda_{2}} V_{ \bar{\mu}_{k} \bar{ \nu}_{k}} \left(\lambda_{1}\right) \nonumber \\
&&\times V\left(-\lambda_{1};\lambda_{2};-\lambda_{2}\right) \frac{ \bar{n}_{\lambda_{2}} + \frac{1}{2}  }{ \omega_{\lambda_{1}} } \nonumber \\
&&+   \frac{ 12 \beta }{k!} \sum_{\lambda_{1}} \sum_{\lambda_{2}} V_{ \bar{\mu}_{k} \bar{ \nu}_{k}} \left(\lambda_{1}\right) \nonumber \\
&&\times V\left(-\lambda_{1};\lambda_{2};-\lambda_{2}\right) \frac{ \omega_{\lambda_{2}}  \bar{n}_{\lambda_{2}} \left(\bar{n}_{\lambda_{2}} + 1\right)  }{ \omega_{\lambda_{1}} }, 
\label{eq:HarkThermPerturbSecondOrderResultsAdiabaticElasticConstantsAndStressEq_6} 
\end{eqnarray}
\begin{eqnarray} 
c&&^{A\left(2\right),6}_{ \bar{\mu}_{k} \bar{\nu}_{k} } \nonumber \\
&&= - \frac{ 72 }{  \hbar k! } \sum_{\lambda} \sum_{\lambda'} \sum_{\lambda''} V_{ \bar{\mu}_{k} \bar{ \nu}_{k}}\left(\lambda;-\lambda;\lambda'\right) \nonumber \\
&&\times  V\left(-\lambda';\lambda'';-\lambda''\right)   \frac{ \left( \bar{n}_{\lambda} + \frac{1}{2}\right) \left(  \bar{n}_{\lambda''} + \frac{1}{2}\right) }{ \omega_{\lambda'}}  \nonumber \\
&&+   \frac{ 72 \beta }{k!} \sum_{\lambda} \sum_{\lambda'} \sum_{\lambda''} V_{ \bar{\mu}_{k} \bar{ \nu}_{k}}\left(\lambda;-\lambda;\lambda'\right) \nonumber \\
&&\times \frac{V\left(-\lambda';\lambda'';-\lambda''\right)}{\omega_{\lambda'}} \left[ \omega_{\lambda}  \bar{n}_{\lambda} \left(\bar{n}_{\lambda} + 1\right) \left(\bar{n}_{\lambda''} +\frac{1}{2}\right) \right. \nonumber \\
&&+\times \left.   \omega_{\lambda''}  \bar{n}_{\lambda''} \left(\bar{n}_{\lambda''} + 1\right) \left(\bar{n}_{\lambda} +\frac{1}{2}\right) \right]  \nonumber \\
&&- \frac{12 }{ \hbar k! } \sum_{\lambda} \sum_{\lambda'} \sum_{\lambda''} V_{ \bar{\mu}_{k} \bar{ \nu}_{k}}\left(\lambda;\lambda';\lambda''\right)  \nonumber \\
&&\times   V\left(-\lambda;-\lambda';-\lambda''\right)  G^{\left(3\right)}\left(\lambda;\lambda';\lambda''\right) \nonumber \\
&&-  \frac{ 36 \beta }{k!} \sum_{\lambda} \sum_{\lambda'} \sum_{\lambda''}  V_{ \bar{\mu}_{k} \bar{ \nu}_{k}}\left(\lambda;\lambda';\lambda''\right)  \nonumber \\
&&\times V\left(-\lambda;-\lambda';-\lambda''\right) G^{\left(3,\beta\right)}\left(\lambda;\lambda';\lambda''\right),
\label{eq:HarkThermPerturbSecondOrderResultsAdiabaticElasticConstantsAndStressEq_7} 
\end{eqnarray}
\begin{eqnarray} 
c&&^{A\left(2\right),7}_{ \bar{\mu}_{k} \bar{\nu}_{k} } \nonumber \\
&&= - \frac{96}{  \hbar k! } \sum_{\lambda} \sum_{\lambda'} \sum_{\lambda''} \nonumber \\
&&\times V_{ \bar{\mu}_{k} \bar{ \nu}_{k}}\left(\lambda;\lambda'\right) V\left(-\lambda;-\lambda',\lambda'';-\lambda''\right) \nonumber \\
&&\times \left( \bar{n}_{\lambda''} + \frac{1}{2}\right) \frac{ \omega_{\lambda} \left( \bar{n}_{\lambda'} +\frac{1}{2}\right) - \omega_{\lambda'} \left( \bar{n}_{\lambda} +\frac{1}{2}\right) }{ \omega^{2}_{\lambda} - \omega^{2}_{\lambda'} } \nonumber \\
&&+   \frac{ 192 \beta }{k!} \sum_{\lambda} \sum_{\lambda'} \sum_{\lambda''} \nonumber \\
&&\times V_{ \bar{\mu}_{k} \bar{ \nu}_{k}}\left(\lambda;\lambda'\right) V\left(-\lambda;-\lambda',\lambda'';-\lambda''\right) \nonumber \\
&&\times  \left[ \frac{  \omega_{\lambda}  \omega_{\lambda''}  \bar{n}_{\lambda''} \left(\bar{n}_{\lambda''} + 1\right) \left(\bar{n}_{\lambda'} +\frac{1}{2}\right) }{ \omega^{2}_{\lambda} - \omega^{2}_{\lambda'} } \right. \nonumber \\
&&+ \left. \frac{  \omega_{\lambda'}  \bar{n}_{\lambda'} \left(\bar{n}_{\lambda'} + 1\right) \left(\bar{n}_{\lambda''} +\frac{1}{2}\right)   }{ \omega^{2}_{\lambda} - \omega^{2}_{\lambda'} } \right].
\label{eq:HarkThermPerturbSecondOrderResultsAdiabaticElasticConstantsAndStressEq_8} 
\end{eqnarray}
In Eqs. \ref{eq:HarkThermPerturbSecondOrderResultsAdiabaticElasticConstantsAndStressEq_2}-\ref{eq:HarkThermPerturbSecondOrderResultsAdiabaticElasticConstantsAndStressEq_8}, $\xi^{\left(2\right)}_{n}$ and $G^{\left(3,\beta\right)}\left(\lambda;\lambda';\lambda''\right)$ are given by Eqs. \ref{eq:HarkThermPerturbSecondOrderResultsAdiabaticElasticConstantsAndStressEq_12} and \ref{eq:HarkThermPerturbSecondOrderResultsAdiabaticElasticConstantsAndStressEq_13}, respectively.

\section{Expressions for third-order elastic constants}
\label{ExpressionsForThirdOrderElasticConstants}

The list of third-order contributions to isothermal elastic constants is given below
\begin{eqnarray} 
\tilde{c}&&^{T\left(3\right),1}_{ \bar{\mu}_{k} \bar{\nu}_{k} } \nonumber \\
&&=\frac{ 18 }{ \beta } \sum_{\lambda_{1}} \cdots \sum_{\lambda_{3}} \sum_{\lambda'_{3}} V\left(\lambda_{1};\lambda_{2};\lambda_{3}\right)  V\left(-\lambda_{1};-\lambda_{2};\lambda'_{3}\right)    \nonumber \\
&&\times \sum_{m=1} \frac{1}{m!} V_{ \bar{\mu}_{m} \bar{ \nu}_{m}} \left(-\lambda_{3};-\lambda'_{3}\right) \nonumber \\
&&\times \int^{\beta}_{0}d\tau_{1} \int^{\beta}_{0}d\tau_{2} \int^{\beta}_{0}d\tau_{3} G_{0}\left(\lambda_{1}\tau_{1}|\lambda_{1}\tau_{2}\right) \nonumber \\
&&\times   G_{0}\left(\lambda_{2}\tau_{1}|\lambda_{2}\tau_{2}\right) G_{0}\left(\lambda_{3}\tau_{1}|\lambda_{3}\tau_{3}\right) G_{0}\left(\lambda'_{3}\tau_{2}|\lambda'_{3}\tau_{3}\right)  \nonumber \\
&&+ \frac{60 }{ \hbar^{2} }   \sum_{\lambda_{1}} \cdots \sum_{\lambda_{3}} \sum_{\lambda'_{2}} V\left(\lambda_{1};\lambda_{2};\lambda_{3}\right)  V\left(-\lambda_{1};\lambda'_{2};-\lambda'_{2}\right)    \nonumber \\
&&\times \left(2 \bar{n}_{\lambda'_{2}} + 1 \right) \sum_{m=1} \frac{1}{m!} V_{ \bar{\mu}_{m} \bar{ \nu}_{m}} \left(-\lambda_{2};-\lambda_{3}\right)  \nonumber \\ 
&&\times  \frac{1}{ \omega_{\lambda_{1}}}  \frac{ \omega_{\lambda_{2}} \left(2 \bar{n}_{\lambda_{3}} +1\right) - \omega_{\lambda_{3}} \left(2 \bar{n}_{\lambda_{2}} +1\right) }{ \omega^{2}_{\lambda_{2}} - \omega^{2}_{\lambda_{3}} }  \nonumber \\
&&+ \frac{24 }{ \hbar^{2}  }  \sum_{\lambda_{1}} \sum_{\lambda_{3}} \sum_{\lambda'_{1}} \sum_{\lambda'_{3}}  V\left(\lambda_{1};-\lambda_{1};\lambda_{3}\right)  V\left(\lambda'_{1};-\lambda'_{1};\lambda'_{3}\right) \nonumber \\
&&\times \sum_{m=1} \frac{1}{m!} V_{ \bar{\mu}_{m} \bar{ \nu}_{m}} \left(-\lambda_{3};-\lambda'_{3}\right) \frac{\left(2 \bar{n}_{\lambda_{1}} + 1 \right)\left(2 \bar{n}_{\lambda'_{1}} + 1 \right)}{ \omega_{\lambda_{3}} \omega_{\lambda'_{3}} }, \nonumber \\
\label{eq:HarkThermPerturbThirdOrderResultsIsothermalElasticConstantsAndStressAppendixEq_1}
\end{eqnarray}
(in Eqs. \ref{eq:HarkThermPerturbThirdOrderResultsIsothermalElasticConstantsAndStressAppendixEq_2}-\ref{eq:HarkThermPerturbThirdOrderResultsIsothermalElasticConstantsAndStressAppendixEq_6}, $m+m'=k$)
\begin{eqnarray} 
\tilde{c}&&^{T\left(3\right),2}_{ \bar{\mu}_{k} \bar{\nu}_{k} } \nonumber \\
&&= \frac{48 }{ \hbar^{2}} \sum_{\lambda_{1}} \sum_{m=1} \frac{1}{m!} V_{ \bar{\mu}_{m} \bar{ \nu}_{m}} \left(\lambda_{1}\right) \nonumber \\
&&\times \sum_{\lambda'_{1}} \sum_{m'=1} \frac{1}{m'!} V_{ \bar{\mu}'_{m'}  \bar{ \nu}'_{m'}} \left(\lambda'_{1}\right) \nonumber \\
&&\times \sum_{\lambda''_{3}}  V\left(-\lambda_{1};-\lambda'_{1},\lambda''_{3};-\lambda''_{3}\right) \frac{ \bar{n}_{\lambda''_{3}} + \frac{1}{2} }{ \omega_{\lambda_{1}} \omega_{\lambda'_{1}} }, 
\label{eq:HarkThermPerturbThirdOrderResultsIsothermalElasticConstantsAndStressAppendixEq_2}
\end{eqnarray}
\begin{eqnarray} 
\tilde{c}&&^{T\left(3\right),3}_{ \bar{\mu}_{k} \bar{\nu}_{k} } \nonumber \\
&&= \frac{ 192 }{ \hbar^{2}} \sum_{\lambda_{1}} \sum_{\lambda_{2}} \sum_{m=1} \frac{1}{m!} V_{ \bar{\mu}_{m} \bar{ \nu}_{m}} \left(\lambda_{1};\lambda_{2}\right) \nonumber \\
&&\times   \sum_{\lambda'_{2}} \sum_{m'=1} \frac{1}{m'!} V_{ \bar{\mu}'_{m'}  \bar{ \nu}'_{m'}} \left(-\lambda_{1};\lambda'_{2}\right) \nonumber \\
&&\times \sum_{n''=3} \sum_{\lambda''_{3}}  V\left(-\lambda_{2};-\lambda'_{2};\lambda''_{3};-\lambda''_{3}\right) \left( \bar{n}_{\lambda''_{3}} + \frac{1}{2}\right) \nonumber \\
&&\times   \left[ \frac{ \omega_{\lambda_{1}}  \left( \bar{n}_{\lambda_{2}} - \bar{n}_{\lambda'_{2}} \right)  }{\left(\omega_{\lambda_{1}} -\omega_{\lambda_{2}}\right) \left(\omega_{\lambda_{1}} +\omega_{\lambda_{2}}\right) \left(\omega_{\lambda'_{2}}-\omega_{\lambda_{2}}\right)} \right. \nonumber \\
&&+ \frac{ \omega_{\lambda_{1}}  \left( \bar{n}_{\lambda_{2}} +\bar{n}_{\lambda'_{2}}  + 1\right)  }{\left(\omega_{\lambda_{1}}  -\omega_{\lambda_{2}}\right) \left(\omega_{\lambda_{1}} +\omega_{\lambda_{2}}\right) \left(\omega_{\lambda_{2}}+\omega_{\lambda'_{2}}\right)} \nonumber \\
&&+\frac{ \omega_{\lambda_{2}} \left( \bar{n}_{\lambda_{1}} - \bar{n}_{\lambda'_{2}} \right) }{ \left( \omega_{\lambda_{2}}-\omega_{\lambda_{1}} \right) \left( \omega_{\lambda_{1}} +\omega_{\lambda_{2}}  \right)\left( \omega_{\lambda'_{2} }-\omega_{\lambda_{1}} \right)} \nonumber \\
&&+ \left. \frac{ \omega_{\lambda_{2}} \left( \bar{n}_{\lambda_{1}} + \bar{n}_{\lambda'_{2} } + 1  \right) }{\left( \omega_{\lambda_{2}}-\omega_{\lambda_{1}} \right) \left( \omega_{\lambda_{1}} +\omega_{\lambda_{2}} \right) \left( \omega_{\lambda_{1}} +\omega_{\lambda'_{2} } \right)} \right] \nonumber \\
&&+  \frac{192 }{\hbar^{2}} \sum_{\lambda_{1}} \sum_{\lambda_{2}} \sum_{m=1} \frac{1}{m!} V_{ \bar{\mu}_{m} \bar{ \nu}_{m}} \left(\lambda_{1};\lambda_{2}\right) \nonumber \\
&&\times  \sum_{\lambda'_{1}} \sum_{\lambda'_{2}} \sum_{m'=1} \frac{1}{m'!} V_{ \bar{\mu}'_{m'}  \bar{ \nu}'_{m'}} \left(\lambda'_{1};\lambda'_{2}\right) \nonumber \\
&&\times \sum_{n''=3}  V\left(-\lambda_{1};-\lambda_{2};-\lambda'_{1};-\lambda'_{2}\right) \nonumber \\  
&&\times \frac{ \omega_{\lambda_{1}} \left( \bar{n}_{\lambda_{2}} +\frac{1}{2}\right) - \omega_{\lambda_{2}} \left( \bar{n}_{\lambda_{1}} +\frac{1}{2}\right) }{ \omega^{2}_{\lambda_{1}} - \omega^{2}_{\lambda_{2}} } \nonumber \\  
&&\times \frac{ \omega_{\lambda'_{1}} \left( \bar{n}_{\lambda'_{2}} +\frac{1}{2}\right) - \omega_{\lambda'_{2}} \left( \bar{n}_{\lambda'_{1}} +\frac{1}{2} \right) }{ \omega^{2}_{\lambda'_{1}} - \omega^{2}_{\lambda'_{2}} }, 
\label{eq:HarkThermPerturbThirdOrderResultsIsothermalElasticConstantsAndStressAppendixEq_3}
\end{eqnarray}
\begin{eqnarray} 
\tilde{c}&&^{T\left(3\right),4}_{ \bar{\mu}_{k} \bar{\nu}_{k} } \nonumber \\
&&= \frac{48}{ \hbar^{2}  } \sum_{\lambda_{1}} \sum_{\lambda_{2}} \sum_{m=1} \frac{1}{m!} V_{ \bar{\mu}_{m} \bar{ \nu}_{m}} \left(\lambda_{1};\lambda_{2}\right)  \nonumber \\
&&\times \sum_{m'=1} \frac{1}{m'!} V_{ \bar{\mu}'_{m'}  \bar{ \nu}'_{m'}} \left(-\lambda_{1}\right) \nonumber \\
&&\times \sum_{\lambda''_{2}} \sum_{\lambda''_{3}}  V\left(-\lambda_{2};\lambda''_{2};-\lambda''_{2}\right)  \frac{  \bar{n}_{\lambda''_{2}} + \frac{1}{2} }{ \omega_{\lambda_{1}} \omega_{\lambda_{2}} } \nonumber \\
&&+ \frac{48 }{ \hbar^{2}  } \sum_{\lambda_{1}} \sum_{\lambda_{2}} \sum_{m=1} \frac{1}{m!} V_{ \bar{\mu}_{m} \bar{ \nu}_{m}} \left(\lambda_{1};\lambda_{2}\right) \nonumber \\
&&\times \sum_{\lambda'_{1}} \sum_{m'=1} \frac{1}{m'!} V_{ \bar{\mu}'_{m'}  \bar{ \nu}'_{m'}} \left(\lambda'_{1}\right)  \nonumber \\
&&\times \frac{ V\left(-\lambda_{1};-\lambda_{2};-\lambda'_{1}\right) }{ \omega_{\lambda'_{1}} } \frac{ \omega_{\lambda_{2}} \left( \bar{n}_{\lambda_{1}} +\frac{1}{2}\right) - \omega_{\lambda_{1}} \left( \bar{n}_{\lambda_{2}} +\frac{1}{2} \right) }{ \omega^{2}_{\lambda_{2}} - \omega^{2}_{\lambda_{1}} }, \nonumber \\
\label{eq:HarkThermPerturbThirdOrderResultsIsothermalElasticConstantsAndStressAppendixEq_4}
\end{eqnarray}
\begin{eqnarray} 
\tilde{c}&&^{T\left(3\right),5}_{ \bar{\mu}_{k} \bar{\nu}_{k} } \nonumber \\
&&= \frac{4 }{ \hbar^{2}  } \sum_{\lambda_{1}} \sum_{m=1} \frac{1}{m!}  V_{ \bar{\mu}_{m} \bar{ \nu}_{m}} \left(\lambda_{1}\right) \nonumber \\
&&\times  \sum_{\lambda'_{1}} \sum_{m'=1} \frac{1}{m'!} V_{ \bar{\mu}'_{m'}  \bar{ \nu}'_{m'}} \left(\lambda'_{1}\right) \nonumber \\
&&\times \sum_{n''=2} \sum_{\lambda''_{1}} \sum_{\lambda''_{2}} \cdots \sum_{\lambda''_{n''}}  \nonumber \\
&&\times \frac{ V\left(-\lambda_{1};-\lambda'_{1};\lambda''_{1};-\lambda''_{1};\ldots;\lambda''_{n''};-\lambda''_{ n'' }\right) }{ \omega_{\lambda_{1}} \omega_{\lambda'_{1}} } \nonumber \\
&&\times 2^{n''} \left( n'' + 1 \right) \prod^{2n'' + 1}_{k'=1 \wedge k'=odd} k'  \nonumber \\
&&\times \left(\bar{n}_{\lambda''_{1}} + \frac{1}{2}\right) \left(\bar{n}_{\lambda''_{2}} + \frac{1}{2}\right) \cdots  \left(\bar{n}_{\lambda''_{n''}} + \frac{1}{2}\right),
\label{eq:HarkThermPerturbThirdOrderResultsIsothermalElasticConstantsAndStressAppendixEq_5}
\end{eqnarray}
\begin{eqnarray} 
\tilde{c}&&^{T\left(3\right),6}_{ \bar{\mu}_{k} \bar{\nu}_{k} } \nonumber \\
&&=  \frac{8 }{ \hbar^{2}} \sum_{\lambda_{1}} \sum_{\lambda_{2}} \sum_{m=1} \frac{1}{m!} V_{ \bar{\mu}_{m} \bar{ \nu}_{m}} \left(\lambda_{1};\lambda_{2}\right) \nonumber \\
&&\times  \sum_{\lambda'_{2}} \sum_{m'=1} \frac{1}{m'!} V_{ \bar{\mu}'_{m'}  \bar{ \nu}'_{m'}} \left(-\lambda_{1};\lambda'_{2}\right) \nonumber \\
&&\times \sum_{n''=3} \sum_{\bar{\lambda}''_{n''-1}} V\left(-\lambda_{2};-\lambda'_{2};\bar{\lambda}''_{n''-1};-\bar{\lambda}''_{n''-1}\right)    \nonumber \\
&&\times n'' \prod^{2n'' - 1}_{k'=1 \wedge k' = odd} k'   \nonumber \\
&&\times 2^{n''}  \left(\bar{n}_{\lambda''_{1}} + \frac{1}{2}\right) \left(\bar{n}_{\lambda''_{2}} + \frac{1}{2}\right) \cdots \left(\bar{n}_{\lambda''_{n''-1}} + \frac{1}{2} \right)   \nonumber \\
&&\times \left[\frac{ \omega_{\lambda_{1}}  \left( \bar{n}_{\lambda_{2}} - \bar{n}_{\lambda'_{2}} \right)  }{\left(\omega_{\lambda_{1}} -\omega_{\lambda_{2}}\right) \left(\omega_{\lambda_{1}} +\omega_{\lambda_{2}}\right) \left(\omega_{\lambda'_{2}}-\omega_{\lambda_{2}}\right)} \right. \nonumber \\
&&+ \frac{ \omega_{\lambda_{1}}  \left( \bar{n}_{\lambda_{2}} +\bar{n}_{\lambda'_{2}}  + 1\right)  }{\left(\omega_{\lambda_{1}}  -\omega_{\lambda_{2}}\right) \left(\omega_{\lambda_{1}} +\omega_{\lambda_{2}}\right) \left(\omega_{\lambda_{2}}+\omega_{\lambda'_{2}}\right)} \nonumber \\
&&+\frac{ \omega_{\lambda_{2}} \left( \bar{n}_{\lambda_{1}} - \bar{n}_{\lambda'_{2}} \right) }{ \left( \omega_{\lambda_{2}}-\omega_{\lambda_{1}} \right) \left( \omega_{\lambda_{1}} +\omega_{\lambda_{2}}  \right)\left( \omega_{\lambda'_{2} }-\omega_{\lambda_{1}} \right)} \nonumber \\
&&+ \left. \frac{ \omega_{\lambda_{2}} \left( \bar{n}_{\lambda_{1}} + \bar{n}_{\lambda'_{2} } + 1  \right) }{\left( \omega_{\lambda_{2}}-\omega_{\lambda_{1}} \right) \left( \omega_{\lambda_{1}} +\omega_{\lambda_{2}} \right) \left( \omega_{\lambda_{1}} +\omega_{\lambda'_{2} } \right)} \right]  \nonumber \\
&&+ \frac{8 }{\hbar^{2}}  \sum_{\lambda_{1}} \sum_{\lambda_{2}} \sum_{m=1} \frac{1}{m!} V_{ \bar{\mu}_{m} \bar{ \nu}_{m}} \left(\lambda_{1};\lambda_{2}\right) \nonumber \\
&&\times \sum_{\lambda'_{1}} \sum_{\lambda'_{2}} \sum_{m'=1} \frac{1}{m'!} V_{ \bar{\mu}'_{m'}  \bar{ \nu}'_{m'}} \left(\lambda'_{1};\lambda'_{2}\right) \sum_{n''=3} \nonumber \\
&&\times  \sum_{\bar{\lambda}''_{n''-2}} V\left(-\lambda_{1};-\lambda_{2};-\lambda'_{1};-\lambda'_{2};\bar{\lambda}''_{n''-2};-\bar{\lambda}''_{n''-2}\right)  \nonumber \\
&&\times n'' \left(n'' -1 \right) \prod^{2n'' - 1}_{k'=1 \wedge k' = odd} k'  \nonumber \\
&&\times 2^{n''} \left(\bar{n}_{\lambda''_{1}} + \frac{1}{2}\right) \left(\bar{n}_{\lambda''_{2}} + \frac{1}{2}\right) \cdots \left(\bar{n}_{\lambda''_{n''-2}} + \frac{1}{2}\right) \nonumber \\
&&\times   \frac{ \omega_{\lambda_{1}} \left( \bar{n}_{\lambda_{2}} +\frac{1}{2}\right) - \omega_{\lambda_{2}} \left( \bar{n}_{\lambda_{1}} +\frac{1}{2}\right) }{ \omega^{2}_{\lambda_{1}} - \omega^{2}_{\lambda_{2}} } \nonumber \\
&&\times \frac{ \omega_{\lambda'_{1}} \left( \bar{n}_{\lambda'_{2}} +\frac{1}{2}\right) - \omega_{\lambda'_{2}} \left( \bar{n}_{\lambda'_{1}} +\frac{1}{2} \right) }{ \omega^{2}_{\lambda'_{1}} - \omega^{2}_{\lambda'_{2}} },
\label{eq:HarkThermPerturbThirdOrderResultsIsothermalElasticConstantsAndStressAppendixEq_6}
\end{eqnarray}
\begin{eqnarray} 
\tilde{c}&&^{T\left(3\right),7}_{ \bar{\mu}_{k} \bar{\nu}_{k} } \nonumber \\
&&= \frac{4}{ \hbar^{2}  }   \sum_{\lambda_{1}} \sum_{m=1} \frac{1}{m!} V_{ \bar{\mu}_{m} \bar{ \nu}_{m}} \left(\lambda_{1}\right) \nonumber \\
&&\times  \sum_{\lambda'_{2}} \sum_{m'=1} \frac{1}{m'!} V_{ \bar{\mu}'_{m'}  \bar{ \nu}'_{m'}} \left(-\lambda_{1};\lambda'_{2}\right) \nonumber \\
&&\times \sum_{n''=5 \wedge n''=odd } \sum_{\lambda''_{2}} \sum_{\lambda''_{4}} \cdots \sum_{\lambda''_{n''-1}}  \nonumber \\
&&\times V\left(-\lambda'_{2};\lambda''_{2};-\lambda''_{2};\lambda''_{4};-\lambda''_{4};\ldots;\lambda''_{n''-1};-\lambda''_{n''-1}\right)  \nonumber \\
&&\times 2 n'' 2^{n''-2} \prod^{n''-3}_{k'=1 \wedge k'=odd} k'  \nonumber \\
&&\times	\frac{\left( \bar{n}_{\lambda''_{2}} + \frac{1}{2}\right) \left( \bar{n}_{\lambda''_{4}} +  \frac{1}{2}\right) \cdots \left( \bar{n}_{\lambda''_{n''-1}} +  \frac{1}{2} \right)}{ \omega_{\lambda_{1}} \omega_{\lambda'_{2}} } \nonumber \\
&&+ \frac{8 }{ \hbar^{2} }   \sum_{\lambda_{1}} \sum_{m=1} \frac{1}{m!} V_{ \bar{\mu}_{m} \bar{ \nu}_{m}} \left(\lambda_{1}\right) \nonumber \\
&&\times \sum_{\lambda'_{1}}  \sum_{\lambda'_{2}} \sum_{m'=1} \frac{1}{m'!} V_{ \bar{\mu}'_{m'}  \bar{ \nu}'_{m'}} \left(\lambda'_{1};\lambda'_{2}\right) \nonumber \\
&&\times \sum_{n''=5 \wedge n''=odd} \sum_{\lambda''_{4}} \cdots \sum_{\lambda''_{n''-1}}  \nonumber \\
&&\times V\left(-\lambda_{1};-\lambda'_{1};-\lambda'_{2};\lambda''_{4};-\lambda''_{4};\ldots;\lambda''_{n''-1};-\lambda''_{n''-1}\right)  \nonumber \\
&&\times n'' \left( n'' - 1\right) \left( n'' - 2\right) \prod^{n''-5}_{k'=1 \wedge k'=odd} k'   \nonumber \\
&&\times 2 \left( \bar{n}_{\lambda''_{4}} + \frac{1}{2}\right) 2  \left( \bar{n}_{\lambda''_{6}} + \frac{1}{2}\right) \cdots 2 \left( \bar{n}_{\lambda''_{n''-1}} + \frac{1}{2}\right) \nonumber \\
&&\times \frac{ 1 }{  \omega_{\lambda_{1}} } \frac{ \omega_{\lambda'_{1}} \left( \bar{n}_{\lambda'_{2}} +\frac{1}{2}\right) - \omega_{\lambda'_{2}} \left( \bar{n}_{\lambda'_{1}} +\frac{1}{2} \right) }{ \omega^{2}_{\lambda'_{1}} - \omega^{2}_{\lambda'_{2}} }.
\label{eq:HarkThermPerturbThirdOrderResultsIsothermalElasticConstantsAndStressAppendixEq_7}
\end{eqnarray}
In Eqs. \ref{eq:HarkThermPerturbThirdOrderResultsIsothermalElasticConstantsAndStressAppendixEq_8}-\ref{eq:HarkThermPerturbThirdOrderResultsIsothermalElasticConstantsAndStressAppendixEq_14}, $m+m'+m''=k$
\begin{eqnarray} 
\tilde{c}^{T\left(3\right),8}_{ \bar{\mu}_{k} \bar{\nu}_{k} } =&& \frac{4 }{ \hbar^{2}  } \sum_{\lambda_{1}}  \sum_{m=1} \frac{1}{m!} V_{ \bar{\mu}_{m} \bar{ \nu}_{m}} \left(\lambda_{1}\right) \nonumber \\
&&\times \sum_{\lambda'_{1}} \sum_{m'=1} \frac{1}{m'!} V_{ \bar{\mu}'_{m'}  \bar{ \nu}'_{m'}} \left(\lambda'_{1}\right) \nonumber \\
&&\times \sum_{m''=1} \frac{1}{m''!} \frac{V_{ \bar{\mu}''_{m''}  \bar{ \nu}''_{m''}} \left(-\lambda_{1};-\lambda'_{1}\right) }{ \omega_{\lambda_{1}} \omega_{\lambda'_{1}} }, 
\label{eq:HarkThermPerturbThirdOrderResultsIsothermalElasticConstantsAndStressAppendixEq_8}
\end{eqnarray}
\begin{eqnarray} 
\tilde{c}&&^{T\left(3\right),9}_{ \bar{\mu}_{k} \bar{\nu}_{k} } \nonumber \\
&&= \frac{ 16 }{ 3 \hbar^{2}} \sum_{\lambda_{1}} \sum_{\lambda_{2}} \sum_{m=1} \frac{1}{m!} V_{ \bar{\mu}_{m} \bar{ \nu}_{m}} \left(\lambda_{1};\lambda_{2}\right) \nonumber \\
&&\times  \sum_{\lambda'_{2}} \sum_{m'=1} \frac{1}{m'!} V_{ \bar{\mu}'_{m'}  \bar{ \nu}'_{m'}} \left(-\lambda_{1};\lambda'_{2}\right) \nonumber \\
&&\times  \sum_{m''=1} \frac{1}{m''!} V_{ \bar{\mu}''_{m''}  \bar{ \nu}''_{m''}} \left(-\lambda_{2};-\lambda'_{2}\right) \nonumber \\
&&\times  \left[\frac{ \omega_{\lambda_{1}}  \left( \bar{n}_{\lambda_{2}} - \bar{n}_{\lambda'_{2}} \right)  }{\left(\omega_{\lambda_{1}} -\omega_{\lambda_{2}}\right) \left(\omega_{\lambda_{1}} +\omega_{\lambda_{2}}\right) \left(\omega_{\lambda'_{2}}-\omega_{\lambda_{2}}\right)} \right. \nonumber \\
&&+ \frac{ \omega_{\lambda_{1}}  \left( \bar{n}_{\lambda_{2}} +\bar{n}_{\lambda'_{2}}  + 1\right)  }{\left(\omega_{\lambda_{1}}  -\omega_{\lambda_{2}}\right) \left(\omega_{\lambda_{1}} +\omega_{\lambda_{2}}\right) \left(\omega_{\lambda_{2}}+\omega_{\lambda'_{2}}\right)} \nonumber \\
&&+\frac{ \omega_{\lambda_{2}} \left( \bar{n}_{\lambda_{1}} - \bar{n}_{\lambda'_{2}} \right) }{ \left( \omega_{\lambda_{2}}-\omega_{\lambda_{1}} \right) \left( \omega_{\lambda_{1}} +\omega_{\lambda_{2}}  \right)\left( \omega_{\lambda'_{2}}-\omega_{\lambda_{1}} \right)} \nonumber \\
&&+ \left. \frac{ \omega_{\lambda_{2}} \left( \bar{n}_{\lambda_{1}} + \bar{n}_{\lambda'_{2}} + 1  \right) }{\left( \omega_{\lambda_{2}}-\omega_{\lambda_{1}} \right) \left( \omega_{\lambda_{1}} +\omega_{\lambda_{2}} \right) \left( \omega_{\lambda_{1}} +\omega_{\lambda'_{2}} \right)} \right],
\label{eq:HarkThermPerturbThirdOrderResultsIsothermalElasticConstantsAndStressAppendixEq_9}
\end{eqnarray}
\begin{eqnarray} 
\tilde{c}&&^{T\left(3\right),10}_{ \bar{\mu}_{k} \bar{\nu}_{k} }  \nonumber \\
&&= \frac{192 }{\hbar^{2}}  \sum_{\lambda_{1}} \sum_{\lambda_{2}} \sum_{m=1} \frac{1}{m!} V_{ \bar{\mu}_{m} \bar{ \nu}_{m}} \left(\lambda_{1};\lambda_{2}\right) \nonumber \\
&&\times \sum_{\lambda'_{1}} \sum_{\lambda'_{2}} \sum_{m'=1} \frac{1}{m'!} V_{ \bar{\mu}'_{m'}  \bar{ \nu}'_{m'}} \left(\lambda'_{1};\lambda'_{2}\right) \nonumber \\
&&\times \sum_{m''=1} \frac{1}{m''!} V_{ \bar{\mu}''_{m''}  \bar{ \nu}''_{m''}}\left(-\lambda_{1};-\lambda_{2};-\lambda'_{1};-\lambda'_{2}\right)  \nonumber \\
&&\times   \frac{ \omega_{\lambda_{1}} \left( \bar{n}_{\lambda_{2}} +\frac{1}{2}\right) - \omega_{\lambda_{2}} \left( \bar{n}_{\lambda_{1}} +\frac{1}{2}\right) }{ \omega^{2}_{\lambda_{1}} - \omega^{2}_{\lambda_{2}} } \nonumber \\
&&\times \frac{ \omega_{\lambda'_{1}} \left( \bar{n}_{\lambda'_{2}} +\frac{1}{2}\right) - \omega_{\lambda'_{2}} \left( \bar{n}_{\lambda'_{1}} +\frac{1}{2} \right) }{ \omega^{2}_{\lambda'_{1}} - \omega^{2}_{\lambda'_{2}} },
\label{eq:HarkThermPerturbThirdOrderResultsIsothermalElasticConstantsAndStressAppendixEq_10}
\end{eqnarray}
\begin{eqnarray} 
\tilde{c}&&^{T\left(3\right),11}_{ \bar{\mu}_{k} \bar{\nu}_{k} } \nonumber \\
&&= \frac{4 }{ \hbar^{2}  } \sum_{\lambda_{1}} \sum_{m=1} \frac{1}{m!} V_{ \bar{\mu}_{m} \bar{ \nu}_{m}} \left(\lambda_{1}\right) \nonumber \\
&&\times  \sum_{\lambda'_{1}} \sum_{m'=1} \frac{1}{m'!} V_{ \bar{\mu}'_{m'}  \bar{ \nu}'_{m'}} \left(\lambda'_{1}\right) \nonumber \\
&&\times \sum_{n''=1} \sum_{\lambda''_{1}} \sum_{\lambda''_{2}} \cdots \sum_{\lambda''_{n''}}  \sum_{m''=1} \frac{1}{m''!} \nonumber \\
&&\times \frac{ V_{ \bar{\mu}''_{m''}  \bar{ \nu}''_{m''}}\left(-\lambda_{1};-\lambda'_{1};\left\{\lambda''_{ n'' }\right\}\right) }{ \omega_{\lambda_{1}} \omega_{\lambda'_{1}} } \nonumber \\
&&\times 2^{n''} \left( n'' + 1 \right) \prod^{2n'' + 1}_{k'=1 \wedge k'=odd} k' \nonumber \\
&&\times \left(\bar{n}_{\lambda''_{1}} + \frac{1}{2}\right) \left(\bar{n}_{\lambda''_{2}} + \frac{1}{2}\right) \cdots  \left(\bar{n}_{\lambda''_{n''}} + \frac{1}{2}\right),  \nonumber \\
&&\left\{\lambda''_{ n'' }\right\} \equiv \lambda''_{1};-\lambda''_{1};\ldots;\lambda''_{n''};-\lambda''_{ n'' },
\label{eq:HarkThermPerturbThirdOrderResultsIsothermalElasticConstantsAndStressAppendixEq_11}
\end{eqnarray}
\begin{eqnarray} 
\tilde{c}&&^{T\left(3\right),12}_{ \bar{\mu}_{k} \bar{\nu}_{k} }  \nonumber \\
&&= \frac{48 }{ \hbar^{2}  }   \sum_{\lambda_{1}} \sum_{m=1} \frac{1}{m!} V_{ \bar{\mu}_{m} \bar{ \nu}_{m}} \left(\lambda_{1}\right) \nonumber \\
&&\times  \sum_{\lambda'_{2}} \sum_{m'=1} \frac{1}{m'!}  V_{ \bar{\mu}'_{m'}  \bar{ \nu}'_{m'}} \left(-\lambda_{1};\lambda'_{2}\right)  \nonumber \\
&&\times \sum_{\lambda''_{2}} \sum_{m''=1} \frac{1}{m''!} V_{ \bar{\mu}''_{m''}  \bar{ \nu}''_{m''}}\left(-\lambda'_{2};\lambda''_{2};-\lambda''_{2}\right) \nonumber \\
&&\times  \frac{ \bar{n}_{\lambda''_{2}} + \frac{1}{2} }{ \omega_{\lambda_{1}} \omega_{\lambda'_{2}} } \nonumber \\
&&+ \frac{48 }{ \hbar^{2} }   \sum_{\lambda_{1}} \sum_{m=1} \frac{1}{m!} V_{ \bar{\mu}_{m} \bar{ \nu}_{m}} \left(\lambda_{1}\right) \nonumber \\
&&\times \sum_{\lambda'_{1}}  \sum_{\lambda'_{2}} \sum_{m'=1} \frac{1}{m'!} V_{ \bar{\mu}'_{m'}  \bar{ \nu}'_{m'}} \left(\lambda'_{1};\lambda'_{2}\right) \nonumber \\
&&\times \sum_{m''=1} \frac{1}{m''!} \frac{ V_{ \bar{\mu}''_{m''}  \bar{ \nu}''_{m''}}\left(-\lambda_{1};-\lambda'_{1};-\lambda'_{2}\right) }{  \omega_{\lambda_{1}} } \nonumber \\
&&\times \frac{ \omega_{\lambda'_{1}} \left( \bar{n}_{\lambda'_{2}} +\frac{1}{2}\right) - \omega_{\lambda'_{2}} \left( \bar{n}_{\lambda'_{1}} +\frac{1}{2} \right) }{ \omega^{2}_{\lambda'_{1}} - \omega^{2}_{\lambda'_{2}} }, 
\label{eq:HarkThermPerturbThirdOrderResultsIsothermalElasticConstantsAndStressAppendixEq_12}
\end{eqnarray}
\begin{eqnarray} 
\tilde{c}&&^{T\left(3\right),13}_{ \bar{\mu}_{k} \bar{\nu}_{k} } \nonumber \\
&&= \frac{4 }{ \hbar^{2}  }   \sum_{\lambda_{1}} \sum_{m=1} \frac{1}{m!} V_{ \bar{\mu}_{m} \bar{ \nu}_{m}} \left(\lambda_{1}\right) \nonumber \\
&&\times  \sum_{\lambda'_{2}} \sum_{m'=1} \frac{1}{m'!} V_{ \bar{\mu}'_{m'}  \bar{ \nu}'_{m'}} \left(-\lambda_{1};\lambda'_{2}\right) \nonumber \\
&&\times \sum_{n''=5 \wedge n''=odd } \sum_{\lambda''_{2}} \cdots \sum_{\lambda''_{n''-1}} \sum_{m''=1} \frac{1}{m''!} \nonumber \\
&&\times  V_{ \bar{\mu}''_{m''}  \bar{ \nu}''_{m''}}\left(-\lambda'_{2};\lambda''_{2};-\lambda''_{2};\ldots;\lambda''_{n''-1};-\lambda''_{n''-1}\right)  \nonumber \\
&&\times 2 n'' 2^{n''-2} \prod^{n''-3}_{k'=1 \wedge k'=odd} k'  \nonumber \\
&&\times \frac{\left( \bar{n}_{\lambda''_{2}} + \frac{1}{2}\right) \left( \bar{n}_{\lambda''_{4}} +  \frac{1}{2}\right) \cdots \left( \bar{n}_{\lambda''_{n''-1}} +  \frac{1}{2} \right)}{ \omega_{\lambda_{1}} \omega_{\lambda'_{2}} } \nonumber \\
&&+ \frac{8 }{ \hbar^{2} }   \sum_{\lambda_{1}} \sum_{m=1} \frac{1}{m!} V_{ \bar{\mu}_{m} \bar{ \nu}_{m}} \left(\lambda_{1}\right) \nonumber \\
&&\times \sum_{\lambda'_{1}}  \sum_{\lambda'_{2}} \sum_{m'=1} \frac{1}{m'!} V_{ \bar{\mu}'_{m'}  \bar{ \nu}'_{m'}} \left(\lambda'_{1};\lambda'_{2}\right) \nonumber \\
&&\times \sum_{n''=5 \wedge n''=odd} \sum_{\lambda''_{4}} \sum_{\lambda''_{6}} \cdots \sum_{\lambda''_{n''-1}} \sum_{m''=1} \frac{1}{m''!}  \nonumber \\
&&\times V_{ \bar{\mu}''_{m''}  \bar{ \nu}''_{m''}}\left(-\lambda_{1};-\lambda'_{1};-\lambda'_{2};\left\{\lambda''_{ n''-1 }\right\}\right)  \nonumber \\
&&\times n'' \left( n'' - 1\right) \left( n'' - 2\right) \prod^{n''-5}_{k'=1 \wedge k'=odd} k'   \nonumber \\
&&\times 2 \left( \bar{n}_{\lambda''_{4}} + \frac{1}{2}\right) 2  \left( \bar{n}_{\lambda''_{6}} + \frac{1}{2}\right) \cdots 2 \left( \bar{n}_{\lambda''_{n''-1}} + \frac{1}{2}\right) \nonumber \\
&&\times \frac{ 1 }{  \omega_{\lambda_{1}} } \frac{ \omega_{\lambda'_{1}} \left( \bar{n}_{\lambda'_{2}} +\frac{1}{2}\right) - \omega_{\lambda'_{2}} \left( \bar{n}_{\lambda'_{1}} +\frac{1}{2} \right) }{ \omega^{2}_{\lambda'_{1}} - \omega^{2}_{\lambda'_{2}} },  \nonumber \\
&&\left\{\lambda''_{ n''-1 }\right\} \equiv \lambda''_{4};-\lambda''_{4};\ldots;\lambda''_{n''-1};-\lambda''_{n''-1},
\label{eq:HarkThermPerturbThirdOrderResultsIsothermalElasticConstantsAndStressAppendixEq_13}
\end{eqnarray}
\begin{eqnarray} 
\tilde{c}&&^{T\left(3\right),14}_{ \bar{\mu}_{k} \bar{\nu}_{k} } \nonumber \\
&&= \frac{16 }{ \hbar^{2}} \sum_{\lambda_{1}} \sum_{\lambda_{2}} \sum_{m=1} \frac{1}{m!} V_{ \bar{\mu}_{m} \bar{ \nu}_{m}} \left(\lambda_{1};\lambda_{2}\right) \nonumber \\
&&\times  \sum_{\lambda'_{2}} \sum_{m'=1} \frac{1}{m'!} V_{ \bar{\mu}'_{m'}  \bar{ \nu}'_{m'}} \left(-\lambda_{1};\lambda'_{2}\right) \nonumber \\
&&\times \sum_{n''=2} \sum_{\lambda''_{3}} \sum_{\lambda''_{5}} \cdots \sum_{\lambda''_{2 n''-1}} \sum_{m''=1} \frac{1}{m''!} \nonumber \\
&&\times V_{ \bar{\mu}''_{m''}  \bar{ \nu}''_{m''}}\left(-\lambda_{2};-\lambda'_{2};\left\{\lambda''_{ 2n''-1 }\right\}^{\left(1\right)}\right)    \nonumber \\
&&\times n'' \prod^{2n'' - 1}_{k'=1 \wedge k' = odd} k' \nonumber \\
&&\times \left(2\bar{n}_{\lambda''_{3}} + 1\right) \left(2\bar{n}_{\lambda''_{5}} + 1\right) \cdots \left(2\bar{n}_{\lambda''_{2n''-1}} + 1\right)   \nonumber \\
&&\times \left[\frac{ \omega_{\lambda_{1}}  \left( \bar{n}_{\lambda_{2}} - \bar{n}_{\lambda'_{2}} \right)  }{\left(\omega_{\lambda_{1}} -\omega_{\lambda_{2}}\right) \left(\omega_{\lambda_{1}} +\omega_{\lambda_{2}}\right) \left(\omega_{\lambda'_{2}}-\omega_{\lambda_{2}}\right)} \right. \nonumber \\
&&+ \frac{ \omega_{\lambda_{1}}  \left( \bar{n}_{\lambda_{2}} +\bar{n}_{\lambda'_{2}}  + 1\right)  }{\left(\omega_{\lambda_{1}}  -\omega_{\lambda_{2}}\right) \left(\omega_{\lambda_{1}} +\omega_{\lambda_{2}}\right) \left(\omega_{\lambda_{2}}+\omega_{\lambda'_{2}}\right)} \nonumber \\
&&+\frac{ \omega_{\lambda_{2}} \left( \bar{n}_{\lambda_{1}} - \bar{n}_{\lambda'_{2}} \right) }{ \left( \omega_{\lambda_{2}}-\omega_{\lambda_{1}} \right) \left( \omega_{\lambda_{1}} +\omega_{\lambda_{2}}  \right)\left( \omega_{\lambda'_{2} }-\omega_{\lambda_{1}} \right)} \nonumber \\
&&+ \left. \frac{ \omega_{\lambda_{2}} \left( \bar{n}_{\lambda_{1}} + \bar{n}_{\lambda'_{2} } + 1  \right) }{\left( \omega_{\lambda_{2}}-\omega_{\lambda_{1}} \right) \left( \omega_{\lambda_{1}} +\omega_{\lambda_{2}} \right) \left( \omega_{\lambda_{1}} +\omega_{\lambda'_{2} } \right)} \right]  \nonumber \\
&&+ \frac{32 }{\hbar^{2}}  \sum_{\lambda_{1}} \sum_{\lambda_{2}} \sum_{m=1} \frac{1}{m!} V_{ \bar{\mu}_{m} \bar{ \nu}_{m}} \left(\lambda_{1};\lambda_{2}\right) \nonumber \\
&&\times \sum_{\lambda'_{1}} \sum_{\lambda'_{2}} \sum_{m'=1} \frac{1}{m'!} V_{ \bar{\mu}'_{m'}  \bar{ \nu}'_{m'}} \left(\lambda'_{1};\lambda'_{2}\right) \nonumber \\
&&\times \sum_{n''=3}  \sum_{\lambda''_{5}} \sum_{\lambda''_{7}} \cdots \sum_{\lambda''_{2 n''-1}} \sum_{m''=1} \frac{1}{m''!} \nonumber \\
&&\times V_{ \bar{\mu}''_{m''}  \bar{ \nu}''_{m''}}\left(-\lambda_{1};-\lambda_{2};-\lambda'_{1};-\lambda'_{2};\left\{\lambda''_{ 2n''-1 }\right\}\right)  \nonumber \\
&&\times n'' \left(n'' -1 \right) \prod^{2n'' - 1}_{k'=1 \wedge k' = odd} k' \nonumber \\
&&\times 2\left(\bar{n}_{\lambda''_{5}} + \frac{1}{2}\right) 2 \left( \bar{n}_{\lambda''_{7}} + \frac{1}{2}\right) \cdots 2 \left( \bar{n}_{\lambda''_{2 n''-1}} + \frac{1}{2}\right) \nonumber \\
&&\times   \frac{ \omega_{\lambda_{1}} \left( \bar{n}_{\lambda_{2}} +\frac{1}{2}\right) - \omega_{\lambda_{2}} \left( \bar{n}_{\lambda_{1}} +\frac{1}{2}\right) }{ \omega^{2}_{\lambda_{1}} - \omega^{2}_{\lambda_{2}} } \nonumber \\
&&\times \frac{ \omega_{\lambda'_{1}} \left( \bar{n}_{\lambda'_{2}} +\frac{1}{2}\right) - \omega_{\lambda'_{2}} \left( \bar{n}_{\lambda'_{1}} +\frac{1}{2} \right) }{ \omega^{2}_{\lambda'_{1}} - \omega^{2}_{\lambda'_{2}} },  \nonumber \\
&&\left\{\lambda''_{ 2n''-1 }\right\}^{\left(1\right)} \equiv \lambda''_{3};-\lambda''_{3};\ldots;\lambda''_{2n''-1};-\lambda''_{2n''-1},  \nonumber \\
&&\left\{\lambda''_{ 2n''-1 }\right\} \equiv \lambda''_{5};-\lambda''_{5};\ldots;\lambda''_{2n''-1};-\lambda''_{2n''-1}.
\label{eq:HarkThermPerturbThirdOrderResultsIsothermalElasticConstantsAndStressAppendixEq_14}
\end{eqnarray}
In Eq. \ref{eq:HarkThermPerturbThirdOrderResultsIsothermalElasticConstantsAndStressAppendixEq_1} (see Eq. \ref{eq:HarkThermPerturbEvaluationOfPerturbationExpansionEq_4} for the notation)
\begin{eqnarray} 
&&\int^{\beta}_{0} d\tau_{1} \int^{\beta}_{0} d\tau_{2} \int^{\beta}_{0} d\tau_{3} G_{0}\left(\lambda_{1} \tau_{1}|\lambda_{1} \tau_{2}\right)  \nonumber \\
&&\times  G_{0}\left(\lambda_{2} \tau_{1}|\lambda_{2} \tau_{3}\right) G_{0}\left(\lambda_{3} \tau_{1}|\lambda_{3} \tau_{2}\right) G_{0}\left(\lambda'_{3} \tau_{2}|\lambda'_{3} \tau_{3}\right) \nonumber \\
&&=  \frac{ 2 \beta }{ \hbar^{2}}  \left[  \frac{ \left( \bar{n}_{\lambda_{3}} +  \bar{n}_{\lambda_{1}} +1 \right) \left[\omega_{\lambda'_{3}} \left( \bar{n}_{\lambda_{2}} + \frac{1}{2} \right)  - \omega_{\lambda_{2}} \left( \bar{n}_{\lambda'_{3}} + \frac{1}{2} \right) \right] }{  (\omega_{\lambda_{2}}-\omega_{\lambda'_{3}}) (\omega_{\lambda_{2}}+\omega_{\lambda'_{3}}) (\omega_{\lambda_{1}}+\omega_{\lambda_{2}}+\omega_{\lambda_{3}})}  \right. \nonumber \\
&&+  	\frac{ \left( \bar{n}_{\lambda_{3}} +  \bar{n}_{\lambda_{1}} +1 \right) \left[\omega_{\lambda'_{3}} \left( \bar{n}_{\lambda_{2}} + \frac{1}{2} \right) - \omega_{\lambda_{2}} \left( \bar{n}_{\lambda'_{3}} + \frac{1}{2} \right) \right] }{ (\omega_{\lambda_{2}}-\omega_{\lambda'_{3}}) (\omega_{\lambda_{2}}+\omega_{\lambda'_{3}}) (\omega_{\lambda_{1}}-\omega_{\lambda_{2}}+\omega_{\lambda_{3}})}\nonumber \\
&&-  \frac{1}{2} \frac{  \omega_{\lambda_{2}} \omega_{\lambda'_{3}} \left[ 2 \bar{n}_{\lambda_{1}} \bar{n}_{\lambda_{3}} + \left( \bar{n}_{\lambda_{3}} +  \bar{n}_{\lambda_{1}} +1 \right)\right] }{ (\omega_{\lambda_{1}}-\omega_{\lambda_{2}}+\omega_{\lambda_{3}}) (\omega_{\lambda_{1}}+\omega_{\lambda_{2}}+\omega_{\lambda_{3}}) } \nonumber \\
&&\times \frac{ 1 }{ (\omega_{\lambda_{1}}+\omega_{\lambda_{3}}-\omega_{\lambda'_{3}}) (\omega_{\lambda_{1}}+\omega_{\lambda_{3}}+\omega_{\lambda'_{3}})} \nonumber \\
&&-  \frac{2 \omega_{\lambda'_{3}} (\omega_{\lambda_{1}}+\omega_{\lambda_{3}}) \left(\bar{n}_{\lambda_{1}} +  \bar{n}_{\lambda_{3}} +1\right) }{ (\omega_{\lambda_{2}}-\omega_{\lambda'_{3}}) (\omega_{\lambda_{2}}+\omega_{\lambda'_{3}})} \nonumber \\
&&\times  \frac{ \left( \bar{n}_{\lambda_{2}} + \frac{1}{2} \right) }{ (\omega_{\lambda_{1}}-\omega_{\lambda_{2}}+\omega_{\lambda_{3}}) (\omega_{\lambda_{1}}+\omega_{\lambda_{2}}+\omega_{\lambda_{3}})} \nonumber \\
&&+   \frac{2 (\omega_{\lambda_{1}}+\omega_{\lambda_{3}}) \left(\bar{n}_{\lambda_{1}} +  \bar{n}_{\lambda_{3}} +1\right) }{ (\omega_{\lambda'_{3}}-\omega_{\lambda_{2}}) (\omega_{\lambda_{2}}+\omega_{\lambda'_{3}}) } \nonumber \\
&&\times \frac{ \left( \bar{n}_{\lambda'_{3}} + \frac{1}{2} \right) }{ (\omega_{\lambda_{1}}+\omega_{\lambda_{3}}-\omega_{\lambda'_{3}}) (\omega_{\lambda_{1}}+\omega_{\lambda_{3}}+\omega_{\lambda'_{3}})} \nonumber \\
&&- \frac{ \left(\bar{n}_{\lambda_{1}} - \bar{n}_{\lambda_{3}}\right) \left[ \omega_{\lambda'_{3}}  \left( \bar{n}_{\lambda_{2}} + \frac{1}{2} \right) - \omega_{\lambda_{2}}  \left( \bar{n}_{\lambda'_{3}} + \frac{1}{2} \right) \right] }{  (\omega_{\lambda_{2}}-\omega_{\lambda'_{3}})  (\omega_{\lambda_{2}}+\omega_{\lambda'_{3}}) (\omega_{\lambda_{1}}-\omega_{\lambda_{2}}-\omega_{\lambda_{3}})} \nonumber \\
&&- \frac{ \left(\bar{n}_{\lambda_{1}} - \bar{n}_{\lambda_{3}}\right) \left[ \omega_{\lambda'_{3}} \left( \bar{n}_{\lambda_{2}} + \frac{1}{2} \right) - \omega_{\lambda_{2}} \left( \bar{n}_{\lambda'_{3}} + \frac{1}{2} \right) \right] }{  (\omega_{\lambda_{2}}-\omega_{\lambda'_{3}}) (\omega_{\lambda_{2}}+\omega_{\lambda'_{3}}) (\omega_{\lambda_{1}}+\omega_{\lambda_{2}}-\omega_{\lambda_{3}})} \nonumber \\
&&- \frac{2 \omega_{\lambda_{2}} \omega_{\lambda'_{3}} \left[  \bar{n}_{\lambda_{3}} \left(\bar{n}_{\lambda_{1}} + 1 \right) +  \bar{n}_{\lambda_{1}} \left(\bar{n}_{\lambda_{3}} + 1 \right)\right] 
}{ (\omega_{\lambda_{1}}+\omega_{\lambda_{2}}-\omega_{\lambda_{3}}) (-\omega_{\lambda_{1}}+\omega_{\lambda_{2}}+\omega_{\lambda_{3}}) } \nonumber \\
&&\times \frac{ 1 }{ (\omega_{\lambda_{1}}-\omega_{\lambda_{3}}-\omega_{\lambda'_{3}}) (\omega_{\lambda_{1}}-\omega_{\lambda_{3}}+\omega_{\lambda'_{3}})} \nonumber \\
&&- \frac{ \left( \bar{n}_{\lambda_{1}}  -  \bar{n}_{\lambda_{3}}
 \right) \left[\omega_{\lambda'_{3}} \left( \bar{n}_{\lambda_{2}} + \frac{1}{2} \right) - \omega_{\lambda_{2}} \left( \bar{n}_{\lambda'_{3}} + \frac{1}{2} \right) \right] }{  (\omega_{\lambda_{2}}-\omega_{\lambda'_{3}}) (\omega_{\lambda_{2}}+\omega_{\lambda'_{3}}) (\omega_{\lambda_{1}}+\omega_{\lambda_{2}}-\omega_{\lambda_{3}})} \nonumber \\
&&- \left. \frac{ \left( \bar{n}_{\lambda_{1}}  -  \bar{n}_{\lambda_{3}}
 \right) \left[ \omega_{\lambda'_{3}} \left( \bar{n}_{\lambda_{2}} + \frac{1}{2} \right) - \omega_{\lambda_{2}} \left( \bar{n}_{\lambda'_{3}} + \frac{1}{2} \right)\right] }{ (\omega_{\lambda_{2}}-\omega_{\lambda'_{3}}) (\omega_{\lambda_{2}}+\omega_{\lambda'_{3}}) (\omega_{\lambda_{1}}-\omega_{\lambda_{2}}-\omega_{\lambda_{3}})} \right].
\label{eq:HarkThermPerturbThirdOrderResultsIsothermalElasticConstantsAndStressAppendixEq_15}
\end{eqnarray}

\section{Expressions for thermal quantities}
\label{ExpressionsForThermalQuantities}
The static second-order contribution can be written as
\begin{eqnarray} 
U^{\left(2\right)}_{aa,n=n'=3} =&& - \frac{ 36 }{  \hbar } \sum_{\lambda,\lambda',\lambda''} V\left(\lambda;-\lambda;\lambda'\right)  \nonumber \\
&&\times V\left(-\lambda';\lambda'';-\lambda''\right)  \frac{ \left( \bar{n}_{\lambda} + \frac{1}{2}\right) \left( \bar{n}_{\lambda''} + \frac{1}{2}\right) }{ \omega_{\lambda'}}  \nonumber \\
&&- \frac{6 }{ \hbar }  \sum_{\lambda,\lambda',\lambda''} \left|V\left(\lambda;\lambda';\lambda''\right)\right|^{2}  G^{\left(3\right)}\left(\lambda;\lambda';\lambda''\right)  \nonumber \\
&&+ \frac{72}{\beta }   \sum_{\lambda,\lambda',\lambda''} V\left(\lambda;-\lambda;\lambda'\right) V\left(-\lambda';\lambda'';-\lambda''\right) \nonumber \\
&&\times  \frac{ \omega_{\lambda}  \bar{n}_{\lambda} \left(\bar{n}_{\lambda} + 1\right) \left(\bar{n}_{\lambda''} +\frac{1}{2}\right) }{ \omega_{\lambda'}}  \nonumber \\
&&- 18 \beta \sum_{\lambda,\lambda',\lambda''} \left|V\left(\lambda;\lambda';\lambda''\right)\right|^{2} G^{\left(3,\beta\right)}\left(\lambda;\lambda';\lambda''\right). \nonumber \\
\label{eq:ExpressionsForThermalQuantitiesEq_1} 
\end{eqnarray} 
In Eq. \ref{eq:ExpressionsForThermalQuantitiesEq_1}, $G^{\left(3\right)}\left(\lambda;\lambda';\lambda''\right)$ and $G^{\left(3,\beta\right)}\left(\lambda;\lambda';\lambda''\right)$ are given by Eqs. \ref{eq:HarkThermPerturbSecondOrderResultsFreeEnergyEq_3} and \ref{eq:HarkThermPerturbSecondOrderResultsAdiabaticElasticConstantsAndStressEq_13}, respectively. The static second-order contribution, to lowest-order in the IFCs, may be written as
\begin{eqnarray} 
S&&^{\left(2\right)}_{aa,n=n'=3} \nonumber \\
&&= \frac{72}{\beta T }   \sum_{\lambda} \sum_{\lambda'} \sum_{\lambda''} V\left(\lambda;-\lambda;\lambda'\right)  \nonumber \\
&&\times V\left(-\lambda';\lambda'';-\lambda''\right) \frac{ \omega_{\lambda}  \bar{n}_{\lambda} \left(\bar{n}_{\lambda} + 1\right) \left(\bar{n}_{\lambda''} +\frac{1}{2}\right) }{ \omega_{\lambda'}}  \nonumber \\
&&- \frac{ 18 \beta  }{T} \sum_{\lambda} \sum_{\lambda'} \sum_{\lambda''} \left|V\left(\lambda;\lambda';\lambda''\right)\right|^{2} G^{\left(3,\beta\right)}\left(\lambda;\lambda';\lambda''\right). \nonumber \\
\label{eq:ExpressionsForThermalQuantitiesEq_2} 
\end{eqnarray}


\begin{acknowledgments}
The author gratefully acknowledge funding from the Foundation for Research of Natural Resources in Finland (grant 17591/13). The author thanks Professor Robert van Leeuwen and Dr. Gerrit Groenhof (University of Jyv\"{a}skyl\"{a}) for useful discussions on various aspects of the present work.
\end{acknowledgments}
\bibliography{bibfile}
\end{document}